\documentclass[a4paper,11pt]{article}
\pdfoutput=1 % if your are submitting a pdflatex (i.e. if you have
\usepackage{jheppub} % for details on the use of the package, please
\usepackage[T1]{fontenc} % if needed

\usepackage[latin1]{inputenc}
\usepackage[english]{babel}

\usepackage{amssymb,amsthm,cancel,hyperref,graphicx,xcolor}
\usepackage{picinpar,graphicx,xypic}
\usepackage{booktabs}
\usepackage{mathrsfs}
\usepackage{amsfonts}
\usepackage{latexsym}
\usepackage{booktabs,graphicx,hyperref,epsfig,multirow}
\usepackage{bm}

\usepackage{array}
\newcolumntype{C}[1]{>{\centering\let\newline\\\arraybackslash\hspace{0pt}}m{#1}}

\allowdisplaybreaks
\def\beq{\begin{equation}}
\def\eeq{\end{equation}}
\def\bea#1\eea{\begin{align}#1\end{align}}
\def\be{\begin{equation}}
\def\ee{\end{equation}}
\newcommand{\MSbar}{\overline{\text{MS}}}
\newcommand{\muf}{\mu_{\rm\sss F}}
\newcommand{\mur}{\mu_{\rm\sss R}}
\newcommand{\mt}{m_t}

\newcommand{\Lum}{\mathcal{L}}

\newcommand{\Ord}{\mathcal{O}}
\newcommand{\gsim}{\gtrsim}
\newcommand{\lsim}{\lesssim}
\newcommand{\sss}{\scriptscriptstyle\rm}
\newcommand{\as}{\alpha_s}
\newcommand{\troll}{\texttt{TROLL}}
\newcommand{\toppp}{\texttt{Top++}}
\newcommand{\gbar}{\mbox{$\bar g_0$}}

\newcommand{\lc}{\left[}
\newcommand{\rc}{\right]}

\def\({\left(}
\def\){\right)}
\def\[{\left[}
\def\]{\right]}
\def    \hepph  #1 {{\tt hep-ph/#1}}
\def    \hepex  #1 {{\tt hep-ex/#1}}
\long\def\symbolfootnote[#1]#2{\begingroup%
\def\thefootnote{\fnsymbol{footnote}}\footnote[#1]{#2}\endgroup}

\numberwithin{equation}{section}
\title{\boldmath Parton Distributions with Threshold Resummation}

\author[a]{Marco~Bonvini,}
\author[b]{Simone~Marzani,}
\author[a]{Juan~Rojo,}
\author[a]{Luca~Rottoli,}
\author[c]{Maria~Ubiali,}
\author[d,e]{Richard~D.~Ball,}
\author[e]{Valerio~Bertone,}
\author[f]{Stefano~Carrazza,}
\author[a]{Nathan~P.~Hartland}

\affiliation[a]{Rudolf Peierls Centre for Theoretical Physics,\\
  1 Keble Road, University of Oxford, OX1 3NP, Oxford, UK}
\affiliation[b]{Center for Theoretical Physics, Massachusetts Institute
  of Technology,\\ 77 Massachusetts Ave, Cambridge, MA 02139, USA}
\affiliation[c]{Cavendish Laboratory, HEP group,\\
JJ Thomson Avenue, University of Cambridge, CB3 0HE, Cambridge, UK}
\affiliation[d]{ The Higgs Centre for Theoretical Physics, University of Edinburgh,\\
  JCMB, KB, Mayfield Rd, Edinburgh EH9 3JZ, Scotland}
\affiliation[e]{PH Department, TH Unit, CERN, CH-1211 Geneva 23, Switzerland}
\affiliation[f]{Dipartimento di Fisica, Universit\`a di Milano and
INFN, Sezione di Milano,\\ Via Celoria 16, I-20133 Milano, Italy}

\emailAdd{marco.bonvini@physics.ox.ac.uk}

\preprint{
\begin{flushright}
OUTP-15-03P\\
MIT-CTP 4657\\
Cavendish-HEP-15/02\\
CERN-PH-TH-2015-060\\
Edinburgh 2015/04\\
TIF-UNIMI-2015-6\\
\end{flushright}
}

\abstract{We construct a set of parton distribution
  functions (PDFs) in which fixed-order NLO and NNLO calculations
  are supplemented with soft-gluon (threshold) resummation
  up to NLL and NNLL accuracy respectively, suitable for use in
  conjunction with any QCD calculation in which threshold
  resummation is included at the level of partonic cross sections.
  These resummed PDF sets, based on
  the NNPDF3.0 analysis, are extracted from
  deep-inelastic scattering, Drell-Yan,
  and top quark pair production data, for which resummed
  calculations can be consistently used.
  We find that, close
  to threshold, the inclusion
  of resummed PDFs can partially compensate the enhancement
  in resummed matrix elements, leading to resummed hadronic
  cross-sections closer to the
  fixed-order calculations.
  On the other hand, far from threshold,
  resummed PDFs reduce to their fixed-order counterparts.
  Our results demonstrate the need for a consistent use of resummed
  PDFs in resummed calculations.
}

\begin{document} 
\maketitle
\flushbottom

\section{Introduction}

The accurate determination of the parton distribution functions (PDFs)
of the proton is
an essential ingredient of the LHC physics program~\cite{Forte:2013wc,Ball:2012wy,Perez:2012um,Rojo:2015xta,Watt:2011kp}.
In order to reduce theoretical uncertainties, it is crucial to incorporate in
global PDF fits higher-order
perturbative QCD corrections,  both to the hard partonic cross sections and to the
parton evolution.
While recent progress in fixed-order NLO (see e.g.\ \cite{Butterworth:2014efa} for a recent review),
NNLO (e.g.\ \cite{Czakon:2013goa,Catani:2011qz,Currie:2013dwa,Chen:2014gva,Boughezal:2015dra,
  Boughezal:2013uia,Brucherseifer:2014ama,deFlorian:2013jea,Bolzoni:2010xr,Cacciari:2015jma,Grazzini:2015nwa,
  Gehrmann:2014fva,Ferrera:2014lca,Cascioli:2014yka,Grazzini:2013bna,Boughezal:2015dva,Boughezal:2015aha})
and even N$^3$LO~\cite{Anastasiou:2015vya} calculations for different processes in hadron-hadron collisions has been impressive,
it is also well-known that fixed-order
perturbative calculations display classes of logarithmic contributions that become large in some kinematic regions, thus spoiling
the perturbative expansion in the strong coupling constant $\as$.
The importance of these contributions varies significantly with both the 
type and the kinematic regime of the processes which enter PDF fits. 
Therefore, their omission can lead to a significant distortion of the PDFs, 
thereby reducing their theoretical accuracy (see Ref.~\cite{Dittmar:2005ed} for a detailed discussion).
In order to avoid this problem, it is necessary to supplement fixed-order calculations with
all-order resummations of these large logarithms.

Logarithmic enhancements of higher-order perturbative contributions originate 
from a number of different kinematic regions and require, in general, different 
resummation techniques (see e.g.\ Ref.~\cite{Luisoni:2015xha} for a recent review). 
For instance, enhancements may take place when the centre-of-mass
energy of the partonic collision is much higher than the hard scale of
the process: this corresponds to the small-$x$ region of the PDFs, and
the resummation of such terms is known as high-energy or small-$x$
resummation, see e.g.~\cite{Ciafaloni:2007gf,Ball:2007ra,Altarelli:2008aj,Marzani:2008uh}.
Small-$x$ resummation is certainly relevant for PDF determination and might be
needed to describe the most recent HERA data, where some tensions with fixed-order DGLAP
have been reported~\cite{Caola:2009iy,Caola:2010cy,H1:2015mha}.
A study of small-$x$ resummation in PDF fits will be presented elsewhere.
In this paper we concentrate instead on another type
of logarithmic enhancement of higher order perturbative contributions
which appear close to threshold for the production of the final states: this
is the large-$x$ kinematic region, and the resummation
of logarithms from this region is known as large-$x$, soft gluon,
or threshold resummation.

All-order threshold resummations
exist for many of the processes which play a central 
role in the exploration of the electroweak scale being pursued at the LHC. 
For instance, the current Higgs Cross Section Working Group recommendation 
for the gluon-fusion cross section includes threshold resummation~\cite{Catani:2003zt,deFlorian:2009hc}, 
and resummed Higgs cross sections
in this channel are available up to N$^{3}$LL~\cite{Bonvini:2014joa,Bonvini:2014tea,Catani:2014uta}.
Additional resummed calculations for Higgs physics exist, for example
for gluon-induced Higgs Strahlung~\cite{Harlander:2014wda} and for Higgs-pair production~\cite{deFlorian:2015moa}.
The basis for threshold resummation in QCD were laid in Refs.~\cite{Sterman:1986aj,Catani:1989ne,Catani:1990rp}, 
in which explicit resummed expressions to NLL accuracy were given for
processes relevant for global fits of parton distributions, such as
deep-inelastic structure functions and Drell-Yan total cross section.
%
%Concerning the 
Subsequently, higher-order resummed
calculations have been made available for
deep-inelastic scattering structure functions~\cite{Vogt:2000ci,Moch:2005ba,Becher:2006mr,Manohar:2003vb,Idilbi:2006dg},
invariant mass distributions~\cite{Manohar:2003vb,Idilbi:2006dg,Moch:2005ky,Laenen:2005uz,Ahmed:2014cla,Catani:2014uta} 
and rapidity distributions~\cite{Laenen:1992ey, Bolzoni:2006ky,Mukherjee:2006uu,Ravindran:2006bu,Ravindran:2007sv,Becher:2007ty,Bonvini:2010tp}
in Drell-Yan production, and top quark pair production, both inclusive~\cite{Czakon:2009zw,Beneke:2011mq}
and differentially~\cite{Kidonakis:2001nj,Ahrens:2011mw,Ferroglia:2013awa}.
As far as processes relevant for New Physics searches are concerned, resummed
calculations exist for squark and gluino 
production~\cite{Beenakker:2011fu,Beenakker:2011sf,Falgari:2012hx,Beenakker:2014sma},
stop quark pair production~\cite{Beenakker:2010nq,Broggio:2013cia}, 
slepton and gaugino pair production~\cite{Fuks:2013vua,Fuks:2013lya,Bozzi:2007qr,Broggio:2011bd}
among many others.

Moreover, it is well known~\cite{Korchemsky:1988si,Albino:2000cp} that in the commonly
used $\MSbar$ scheme, threshold resummation affects only partonic
coefficient functions, while the singular part of the DGLAP
splitting function is given, to any order in perturbation theory, by
the cusp contribution: $P(x,\as)\sim \Gamma_{\rm cusp}(\as)/(1-x)$, as $x\to1$.
Therefore, to perform a resummed PDF fit it is only necessary to modify the partonic
cross sections, while the NLO~\cite{gNLOf,gNLOd,gNLOa,gNLOb} and
NNLO~\cite{gNNLOa,gNNLOb} DGLAP evolution kernels remain unchanged.

Despite the wide range of resummed calculations available, a complete global PDF fit 
including the effects of threshold resummation
has never been produced (although some preliminary results were presented in Ref.~\cite{Westmark:2013vea}).
A first study, restricted to non-singlet DIS structure functions, was performed
in Ref.~\cite{Corcella:2005us}, finding that at NLO resummation could suppress the large-$x$ valence quark
PDFs by as much as ten percent.
The impact of
threshold resummation in direct photon production and its
implications on the large-$x$ gluon
was studied in Ref.~\cite{Sato:2013wea}.
More recently, threshold resummation has been studied in the context of the
CJ fits~\cite{Accardi:2014qda}, with emphasis on the description of the large-$x$
JLAB data.
Given the impressive theoretical developments in the resummation of hard-scattering cross sections, it
is clear that a state-of-the-art resummed global PDF fit is most timely. This is what we plan to achieve in this paper.

To this end we will produce for the first time
NLO+NLL and NNLO+NNLL threshold-resummed fits
based on the NNPDF methodology~\cite{DelDebbio:2004qj,DelDebbio:2007ee,Ball:2008by,Ball:2009mk,Ball:2010de,Ball:2011mu,Ball:2011uy,Ball:2012cx}. Since NNPDF fitting is free from theoretical bias, due to the very flexible PDF parametrisation, it is sufficiently precise to be able to detect even small changes in PDFs due to threshold resummation of the various processes that go into the global fit.

A major obstacle to producing a truly global resummed fit is that for a number
of important processes, in particular inclusive jet production and 
$W$ production at the leptonic
level, threshold resummation is not readily available.
For inclusive jets, resummed calculations have been used to
determine approximate expressions~\cite{Kidonakis:2000gi,deFlorian:2013qia,Carrazza:2014hra}
for the yet unknown NNLO contributions, but codes that provide all-order results are not publicly available.
For $W$ production, resummation is available only at the level of reconstructed $W$ but
not for the measured lepton-level distributions.

For this reason, in this work we have begun by producing variants of the
NNPDF3.0 global NLO and NNLO fits~\cite{Ball:2014uwa} based only on those processes which
can be consistently resummed: fixed-target and collider
neutral and charged current
deep-inelastic structure functions,
fixed-target and collider neutral current
Drell-Yan production, and inclusive top-quark pair production.
These DIS+DY+top fits then provide a suitable baseline to compare with the NLL and NNLL resummed
fits.
One important drawback is that the resulting fits will be affected by larger PDF
uncertainties as compared to the global NNPDF3.0 set,
due to the missing experiments, affecting in particular
gluon-initiated processes.
In this respect, it will be important to produce updated resummed fits as soon as 
the missing resummed calculations become available.

An important goal of this paper is to quantify the inaccuracies that affect current
resummed calculations due to the inconsistent use of a fixed-order PDF with resummed partonic
cross sections.
As we will show, for final states with large invariant mass, close to the hadronic threshold,
the main effect of the resummed PDFs is to bring the resummed hadronic calculation closer to the fixed-order
result, thereby canceling partially the effect of the resummation in the matrix elements.
On the other hand, for final states far below threshold, such as inclusive Higgs production at the LHC, the effect of the resummation on the PDFs can be small compared
with the resummation in the matrix elements.
We also find that, unsurprisingly, resummed and unresummed PDFs are much closer at NNLO than at NLO.
Our results emphasise the need for a consistent use of resummed PDFs in resummed calculations: the use
of fixed-order PDFs with resummed matrix elements can lead to misleading results, particularly at NLO.

The outline of this paper is as follows.
In Sect.~\ref{sec:resummation} we review some basic concepts and results in threshold resummation, as well as their implementation. 
In Sect.~\ref{sec:fitsettings}, we discuss the settings of
the global PDF fit used here to include threshold resummation effects, which is a variant
of the recent NNPDF3.0 global fit.
The results of the resummed fits are then discussed in
Sect.~\ref{sec:results}, where we compare resummed with fixed-order
PDFs at NLO(+NLL) and NNLO(+NNLL).
Then in Sect.~\ref{sec:lhcpheno} we discuss the implications of the resummed PDFs for LHC phenomenology,
with emphasis on the mismatch that can arise if fixed-order PDFs are used in resummed calculations.
Finally we summarise in Sect.~\ref{sec:delivery} and discuss the delivery
of the resummed PDFs produced in this work.
%
%A number of technical aspects are collected in the Appendices.

\section{Threshold resummation}
\label{sec:resummation}

In this section we review the theoretical formalism of threshold resummation,
and then we discuss its practical implementation in order to be able
to use it in the resummed NNPDF fits.  We work in the traditional framework of 
perturbative QCD (see e.g.\ \cite{Catani:1998tm});
alternative results can be obtained using the methods of Soft-Collinear Effective Theory
(see Ref.~\cite{Becher:2014oda} for a recent review and Refs.~\cite{Bonvini:2012az,Bonvini:2013td,Bonvini:2014qga,Sterman:2013nya}
for more detailed comparisons between the two formalisms).

\subsection{Theoretical framework}
\label{sec:framework}

We start by considering a hadron-level cross section
\beq\label{eq:xsec_def}
\sigma(x,Q^2) = 
x\sum_{a,b}\int_x^1 \frac{dz}{z}\,\Lum_{ab}\(\frac{x}{z},\muf^2\)
\frac1z\hat\sigma_{ab}\(z, Q^2, \as(\mur^2),\frac{Q^2}{\muf^2},\frac{Q^2}{\mur^2}\),
\eeq
where $a,b$ run over parton flavors, $Q^2$ is the hard scale of the process,
$x$ is a dimensionless variable and $x\to1$ defines the threshold limit.
For the resummed fit we are going to consider three processes: deep-inelastic scattering (DIS)
of a lepton off a hadron, the Drell-Yan process (DY) and top--anti-top production ($t \bar t$). 
In DIS, $Q^2$ is the off-shellness of the exchanged boson $Q^2=-q^2$ and $x=\frac{Q^2}{2 p\cdot q}$, where $p$ is the hadron momentum. In DY, $Q$ is the invariant mass of the lepton pair and $x=\frac{Q^2}{s}$, being $\sqrt{s}$ the 
collider centre-of-mass energy. Finally, for $t \bar t$, $Q^2=4 \mt^2$ and $x=\frac{Q^2}{s}$. 
In Eq.~(\ref{eq:xsec_def}),  $\Lum_{ab}(z,\mu^2)$ is a parton luminosity,
defined as
\beq\label{eq:lum}
\Lum_{ab}(z,\mu^2) = \int_z^1 \frac{dw}w\, f_a\(\frac zw,\mu^2\) f_b(w,\mu^2) \, ,
\eeq
in the hadron-hadron collision case, while in the case of DIS it is just a single PDF.
In the following we are going to set $\mur=\muf=Q$.

In order to diagonalise the convolution integral, we take Mellin moments of Eq.~(\ref{eq:xsec_def}): 
\begin{equation}\label{eq:mellin1}
\sigma(N,Q^2)=\int_0^1 d x \, x^{N-2} \sigma(x,Q^2) = \sum_{a,b}\Lum_{ab}(N,Q^2) \hat\sigma_{ab}\(N, Q^2,\as\),
\end{equation}
where $\as=\as(Q^2)$ and 
\begin{subequations}
\begin{align}
\Lum_{ab}(N,Q^2) &=\int_0^1 d z \, z^{N-1} \Lum_{ab}(z,Q^2),\\
 \hat\sigma_{ab}\(N, Q^2,\as\) &=\int_0^1 d z \, z^{N-2} \hat\sigma_{ab}\(z, Q^2,\as\).
\end{align}
\end{subequations}
In Mellin space the threshold limit corresponds to $N\to \infty$ and the aim of threshold resummation is to obtain a more reliable estimate of the
hadron-level cross section by resumming to all orders in the strong coupling $\as$ the logarithmically enhanced contributions
to the partonic cross section $\hat\sigma_{ab}$ at large $N$.
The resummed partonic cross section can be written as the product of a Born contribution and an all-order coefficient function:
\begin{equation}\label{eq:gen_resum_start}
\hat \sigma^{(\text{res})}_{ab}(N,Q^2,\as)= \sigma_{ab}^{(\text{born})}(N,Q^2, \as) \, C_{ab}^{(\text{res})}(N,\as),
\end{equation}
where
\begin{align}\label{eq:gen_resum}
C^{(\text{res})}_{ab}(N,\as)&=\sum_{\mathbf I} \gbar_{ab}^{({\mathbf I})}(\as)\exp \bar{\mathcal{S}}^{({\mathbf I})}(N,\as),
\nonumber \\
 \bar{\mathcal{S}}^{({\mathbf I})}(N,\as)&= \ln \Delta_a+\ln \Delta_b+\ln J_c+\ln J_d+\ln \Delta^{({\mathbf I})}_{ab\to c d}.
\end{align}
The notation $ab\to c d$ has been chosen to accommodate all the processes that enter our fit. For $t \bar t$ production, we have to consider the resummation of two Born-level processes, namely $q \bar q \to t \bar t $ and $g g \to t \bar t$. For DIS instead we have $V^* q \to q$ and for DY $q \bar q \to V^*$. Moreover, while in DIS and DY we have one color structure, in the $t \bar t$ case we have two contributions, i.e.\ $\mathbf{I}= \text{singlet, octet}$.

Let us now examine the different contributions to the resummed exponent.
If $i$ is a color-singlet, then $\Delta_i=J_i=1$. For each initial-state QCD parton, we have an initial-state jet function
\beq \label{eq:beam_func}
\ln \Delta_i=\int_0^1 d z \frac{z^{N-1}-1}{1-z}\int_{\muf^2}^{(1-z)^2Q^2} \frac{d q^2}{q^2}A_i \(\as \(q^2 \) \), \quad i=a,b.
\eeq	
For each \emph{massless} final-state QCD parton we have a final-state jet function
\beq \label{eq:massless_jet_func}
\ln J_i=\int_0^1 d z \frac{z^{N-1}-1}{1-z} \[ \int_{(1-z)^2Q^2}^{(1-z)Q^2} \frac{d q^2}{q^2}A_i  \(\as \(q^2 \) \)+ \frac{1}{2}B_i  \(\as \(Q^2(1-z) \) \) \], \quad i=c,d,
\eeq	
while there is no jet-function for $t$ or $\bar t$. Finally we also have a large-angle soft contribution, which depends in principle on both the process and the color flow: 
\beq \label{eq:soft1}
\ln \Delta^{({\mathbf I})}_{ab\to c d}=\int_0^1 d z \frac{z^{N-1}-1}{1-z} D^{({\mathbf I})}_{ab\to c d} \(\as \(Q^2(1-z)^2 \) \).
\eeq
The functions $A_i \(\as \)$, $B_i \(\as \)$, $D^{({\mathbf I})}_i \(\as \)$, and obviously $\gbar_{ab}^{({\mathbf I})} (\as)$, are free of large logarithms and can be computed in fixed-order perturbation theory.
The accuracy of their determination fixes the logarithmic accuracy of the resummation.
In particular, (N)NLL requires $A_i$ to second (third) order in the strong coupling $\as$, and $B_i $, $D^{({\mathbf I})}_i $, and $\gbar_{ab}$ to first (second) order.\footnote{This accuracy is sometimes referred to as (N)NLL$^\prime$. For a precise definition of all possible accuracies and their nomenclature in threshold resummation, see Tab.~1 of Ref.~\cite{Bonvini:2014joa}.}
Threshold resummation is actually known to N$^3$LL for DIS~\cite{Moch:2005ba} and DY~\cite{Moch:2005ky,Laenen:2005uz,Ahmed:2014cla,Catani:2014uta}, and to NNLL
for $t \bar t$ production~\cite{Czakon:2009zw,Cacciari:2011hy}.

We have left out of our discussion inclusive jet production.\footnote{We acknowledge discussions with Mrinal Dasgupta and Werner Vogelsang on this topic.} The general framework to perform this resummation has been worked out long ago~\cite{Kidonakis:1998bk}.
However, different treatments of the jet kinematics at threshold can lead to substantially different results, see e.g.\ Ref.~\cite{Kidonakis:2000gi} and~\cite{deFlorian:2005yj}.
Moreover, depending on the way the threshold limit is defined, NLL resummation can be affected by non-global logarithms~\cite{Dasgupta:2001sh} and the result may acquire a non-trivial dependence on the jet algorithm~\cite{Banfi:2005gj,Delenda:2006nf}.
In addition, as previously mentioned, computer programs that implement threshold resummation for jet production are not, to the best of our knowledge, publicly available.
On the other hand, recent progress~\cite{Catani:2013vaa,Broggio:2014hoa,Hinderer:2014qta} has shown that NNLL accuracy is perhaps achievable in the near future. We leave a detailed phenomenological analysis of jet production and its inclusion in a PDF fit to future work.

The Mellin integrals in the resummed expression Eq.~(\ref{eq:gen_resum}) are often evaluated in the $N\to \infty$ limit,
thereby keeping only those contributions that do not vanish at large $N$ and behave as powers of $\ln N$.
In this approximation, which we refer to as $N$-soft in the following, the resummed coefficient function becomes
\begin{align}\label{eq:gen_resum_Nsoft}
C^{(\text{$N$-soft})}(N,\as)&= {g_0}(\as)\exp {\mathcal{S}}(\ln N,\as), \nonumber\\
{\mathcal{S}}(\ln N,\as)&= \left[\frac{1}{\as} g_1(\as\ln N)+ g_2(\as\ln N)+ \as g_3(\as\ln N)+\dots\],
\end{align}
where, in order to simplify our notation, we henceforth drop all the flavor and color-flow indices and it is understood that all the modifications we discuss are applied to each partonic subprocess and each color-flow.
The functions $g_i(\as\ln N)$ with $i\geq1$ resum $\as^n \ln^n N$ contributions to all orders in perturbation theory. They can be derived directly by the integral representations
Eqs.~\eqref{eq:beam_func}--\eqref{eq:soft1} by computing the integrals as an expansion in powers
of $\as$ at fixed $\as\ln N$, in the large $N$ limit.

Other resummation schemes, which are equivalent to $N$-soft at large $N$, but preserve the analytic structure of fixed-order coefficient functions at finite $N$, have been considered in the context of Higgs production~\cite{Ball:2013bra,Bonvini:2014joa,Bonvini:2014jma} and heavy quark production~\cite{Muselli:2015kba}. The extension of these resummation schemes to DIS and DY, and their application to the determination of PDFs, will be considered elsewhere.

Threshold resummation can be extended to rapidity distributions, see e.g.~\cite{Laenen:1992ey, Bolzoni:2006ky,Mukherjee:2006uu,Ravindran:2006bu,Ravindran:2007sv,Becher:2007ty,Bonvini:2010tp}. In this work we follow the approach of Ref.~\cite{Bonvini:2010tp}. The basic observation is that the resummed partonic rapidity distribution coincides with the rapidity integrated one up to terms which are power-suppressed in the threshold limit. Therefore, in order to obtain the hadron-level resummed rapidity distribution, we have only to modify the parton luminosity. While we refer the Reader to Ref.~\cite{Bonvini:2010tp} for an explicit derivation, here we limit ourselves to note that the resummed rapidity distribution is constructed in such a way that the integral over rapidity gives back the resummation of the rapidity-integrated cross section.

Finally, we mention that the calculation of hadron-level cross sections and distributions from resummed results in $N$ space requires a prescription because of the presence of a logarithmic branch-cut for real $N>N_L$, originating from the Landau pole of the running coupling.
As a consequence, the resummed result does not admit an inverse-Mellin transform.
Different solutions to this problem exists, such as the Minimal Prescription~\cite{Catani:1996yz}, which consists on a simple modification of the Mellin inversion integral, and the Borel prescription~\cite{Forte:2006mi, Abbate:2007qv,Bonvini:2008ei,Bonvini:2010tp,Bonvini:2012sh}, which relies on a Borel summation of the divergent series of the order-by-order inverse Mellin transform of  the resummed coefficient function.
In this paper we adopt the Minimal Prescription, but we stress that from a practical point of view, differences between these prescriptions become only relevant at extremely large values $x$~\cite{Bonvini:2012sh}, a region where no experimental data is available.

\subsection{Numerical implementation}\label{sec:troll}

In this section we discuss the numerical implementation of the $N$-soft
threshold resummation described above.
For the PDF fits performed in this work,
the processes that we are interested in  are
DIS (both neutral and charged currents), lepton-pair invariant mass and rapidity distribution for Drell-Yan production, and inclusive top pair production cross section.
For DIS and Drell-Yan, we use a
new version of the public code \texttt{ResHiggs},
written originally~\cite{Bonvini:2014joa} to perform threshold resummation
of Higgs inclusive cross section, including several improvements with respect to 
standard $N$-soft resummation,
and later extended~\cite{Bonvini:2014tea} to also perform (improved) resummation in the Soft-Collinear Effective Theory formalism.
Because of the inclusion of additional processes, the new version of this code changes name from \texttt{ResHiggs} to \troll,
standing for \texttt{TROLL Resums Only Large-x Logarithms}, publicly available at the webpage~\cite{gghiggs}.
To give continuity with the original code \texttt{ResHiggs}, the first version of \troll\ is \texttt{v3.0}.
For top pair production we use the public code \toppp~\cite{Czakon:2013goa}.

The code \troll\ is designed to provide only the contribution of the resummation, while the fixed-order calculation is obtained from a separate code (in our case, the same {\tt FKgenerator} code used for the NNPDF3.0 fits).
More specifically, the output of \troll\ is $\Delta_jK_{\text{N$^k$LL}}$, defined as the
difference between a resummed $K$-factor at N$^j$LO+N$^k$LL and a fixed-order $K$-factor at N$^j$LO, such that
\beq \label{eq:deltaK}
\sigma_{\text{N$^j$LO+N$^k$LL}} = \sigma_{\text{N$^j$LO}} + \sigma_{\rm LO}\times \Delta_jK_{\text{N$^k$LL}}\, ,
\eeq
where all the cross sections appearing in the above equation are evaluated
with a
common N$^j$LO+N$^k$LL PDF set.
Internally, $\Delta_jK_{\text{N$^k$LL}}$ is computed by subtracting off the expansion of the resummed coefficient
up to $\Ord(\as^j)$ from the coefficient itself, multiplying this by the parton luminosity, computing the inverse Mellin
transform and finally dividing by the parton luminosity.

There are several advantages in using the $\Delta K$-factors
defined in Eq.~(\ref{eq:deltaK}): the fixed-order normalisation
is irrelevant, the $K$-factor is much less sensitive to the input
PDFs than the cross section itself, and
finally, since the resummed contribution has the same kinematic structure as the Born cross section
(soft radiation does not change the kinematics), the effect of phase space constraints like kinematic
cuts are correctly taken into account if they are applied to the LO cross section in Eq.~\eqref{eq:deltaK}.

%%%%%%%%%%%%%%%%%%%%%%%%%%%%%%%
\begin{figure}[t]
  \centering
  \includegraphics[width=0.495\textwidth,page=1]{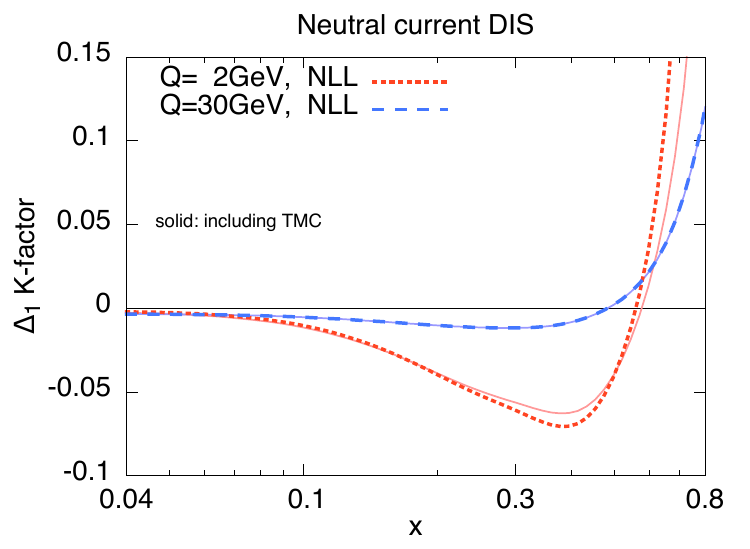}
  \includegraphics[width=0.495\textwidth,page=2]{plots/paper_DIS_deltaK_NNPDF30.pdf}
  \caption{\small $\Delta K$-factors Eq.~(\ref{eq:deltaK}) for the
    neutral current DIS structure function $F_2(x,Q)$,
    as a function of $x$, for $Q=2$~GeV and $Q=30$~GeV.
    The plot on the left corresponds to $j=1$, $k=1$ in Eq.~(\ref{eq:deltaK}), i.e.\ NLO and NLL,
    while the one on the right to $j=2$, $k=2$, i.e.\ NNLO and NNLL.
    The effect of adding TMCs is shown as a thin solid line.}
  \label{fig:DIS-Kfactor}
\end{figure}
%%%%%%%%%%%%%%%%%%%%%%%%%%%%%

We note that in DIS, Target Mass Corrections (TMCs) at next-to-leading twist are included in the resummation
according to the same prescription used in the NNPDF fitting code~\cite{Ball:2008by}, which amounts to multiplying the Mellin transforms of the partonic coefficient functions by an $N$-dependent factor.
No TMCs are included for the fixed-target DY data.

In Fig.~\ref{fig:DIS-Kfactor} we show the $\Delta K$-factors for the neutral current
DIS structure function $F_2(x,Q)$,
as a function of $x$, for $Q=2$~GeV and $Q=30$~GeV.
The plot on the left corresponds to $j=1$, $k=1$ in Eq.~(\ref{eq:deltaK}), i.e.\ NLL to be matched to NLO,
while the one on the right to $j=2$, $k=2$, i.e.\ NNLL to be matched to NNLO.
We note that the resummation enhances the cross section at large $x$, while it gives a very small contribution at small $x$, as it should.
We also note a dip in the region of intermediate $x$, which is also present in fixed-order calculations~\cite{Vermaseren:2005qc}.

TMC effects are also shown as light shadows to the actual curves: as expected, at large scales
they are negligible, while at smaller scales their effect is non-negligible,
in particular at large $x$, where they reduce the effect of the resummation.
Note that in the definition of $\Delta K$ we use the
fact that the same TMCs are already included
in the LO cross section, so much of their effect cancels out.

%%%%%%%%%%%%%%%%%%%%%%%%%%%
\begin{figure}[t]
  \centering
  \includegraphics[width=0.495\textwidth,page=1]{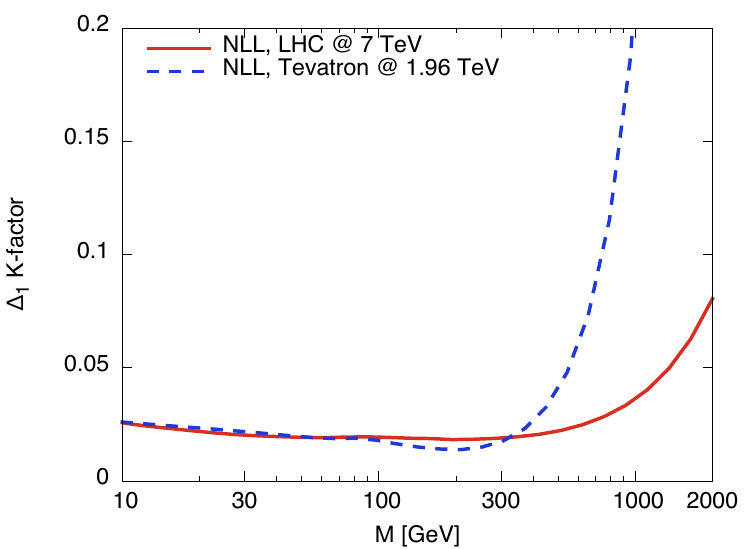}
  \includegraphics[width=0.495\textwidth,page=1]{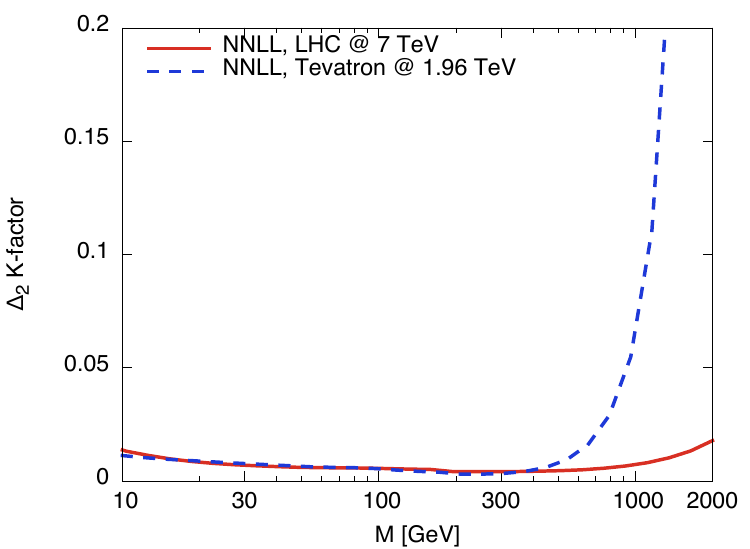}
  \caption{\small Same as Fig.~\ref{fig:DIS-Kfactor}
    for the NLL (left plot)
and NNLL (right plot)
    resummation
    of neutral-current Drell-Yan invariant mass
    distribution at the Tevatron and at the LHC.
 }
  \label{fig:DY-Kfactor}
\end{figure}
%%%%%%%%%%%%%%%%%%

%%%%%%%%%%%%%%%%%%%%%%%%%
\begin{figure}[t]
  \centering
  \includegraphics[width=0.495\textwidth]{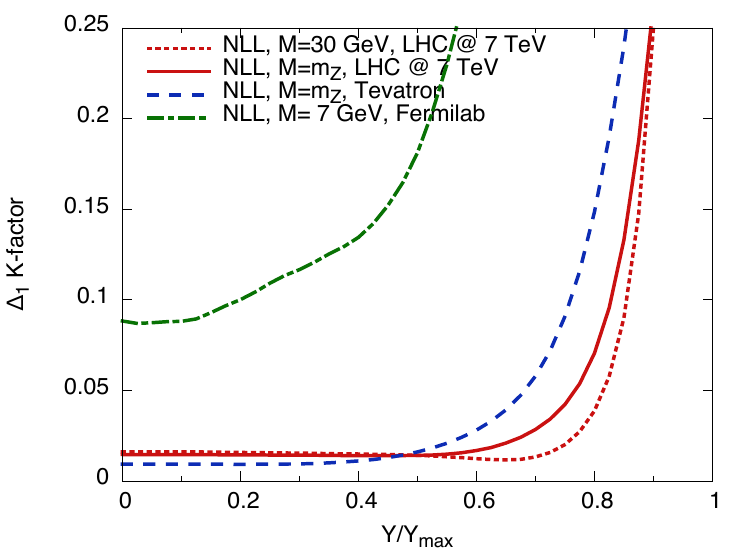}
  \includegraphics[width=0.495\textwidth]{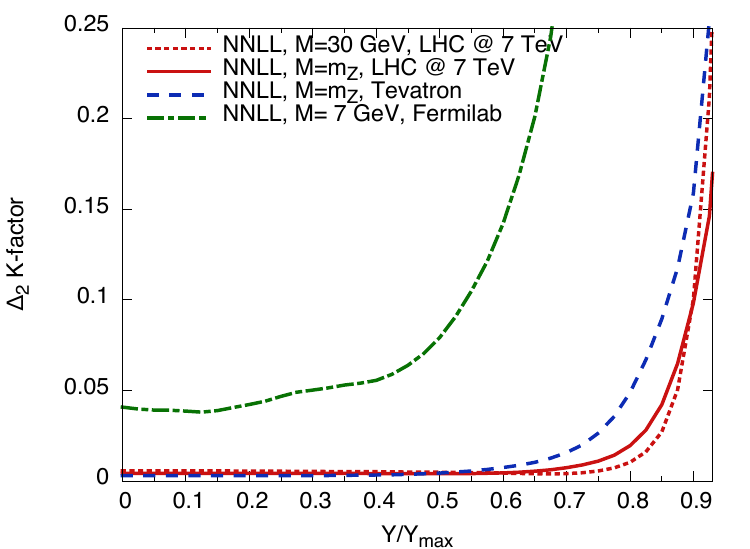}
  \caption{\small Same as Fig.~\ref{fig:DIS-Kfactor} for the neutral-current
    Drell-Yan rapidity distribution,
    for different experiments and different values of the lepton invariant mass.}
  \label{fig:DY-Kfactor_Rap}
\end{figure}
%%%%%%%%%%%%%%%%%%%%%%%%%%%

In Fig.~\ref{fig:DY-Kfactor} we show the corresponding $\Delta K$-factors
this time for Drell-Yan invariant mass distributions,
for LHC at $\sqrt{s}=7$~TeV and for the Tevatron at $\sqrt{s}=1.96$~TeV.
In Fig.~\ref{fig:DY-Kfactor_Rap} we also present the results for the lepton pair rapidity distribution in DY processes
as a function of $Y/Y_{\rm max}$, with $Y_{\rm max}=\frac12\ln(s/M^2)$
being the maximum rapidity of the lepton pairs allowed by kinematics.
The experiments cover different kinematic regimes: close to threshold (Fermilab's fixed-target Drell-Yan experiments at $M=7$~GeV),
an intermediate region (Tevatron and LHC at the $Z$ pole) and away from threshold (low mass DY at LHC).
We note that threshold resummation always gives a significant correction at large rapidities.
For the fixed-target kinematics and $M=7$ GeV, the effect of threshold
resummation is important even at central rapidities.

%%%%%%%%%%%%%%%%%%%%%%%%%%%%%%%%%%
\begin{table}[t]
  \centering
  \begin{tabular}{c|cc}
    & LHC $7$ TeV &  LHC $8$ TeV \\[2ex]
    \midrule
    $\sigma_{\rm NLO+NLL}/\sigma_{\rm NLO}$ & 1.086 & 1.081 \\
    $\sigma_{\rm NNLO+NNLL}/\sigma_{\rm NNLO}$ & 1.031 & 1.029 \\
    $\sigma_{\rm NNLO}/\sigma_{\rm NLO}$ & 1.123 & 1.122
  \end{tabular}
  \caption{\small $K$-factors for $t\bar t$ production at LHC at $7$ and $8$~TeV. The first line is obtained using NLO PDFs,
    while the second and third lines are obtained with NNLO PDFs
    (in all cases, both numerator and denominator are computed with the same PDFs).}
  \label{tab:top}
\end{table}
%%%%%%%%%%%%%%%%%%%%%%%%%%%%%%%%%%

%
Finally, in Table~\ref{tab:top} we collect the $K$-factors for $t\bar t$ production obtained using \toppp.
In this case we provide directly the $K$-factors for the
(N)NLO+(N)NLL over (N)NLO cross sections,
with both numerator and denominator computed with the same (N)NLO PDFs.
We can see from the table that the impact of the resummation is non-negligible, and in fact rather important especially at NLO+NLL,
where the correction is about $9\%$ of the fixed-order NLO result, comparable to the NNLO correction.
Even at NNLO+NNLL, the effect of resummation is comparable to
other theory uncertainties like the values of $\alpha_s(m_Z^2)$ or
of the top quark
mass~\cite{Czakon:2013tha}.

\section{Settings of the resummed PDF fit}
\label{sec:fitsettings}

In this section we present the settings used in the
resummed PDF fits.
These are constructed
as variant of the recent NNPDF3.0
global fits~\cite{Ball:2014uwa}:
they use exactly the same fitting methodology, the same input
parameters
(strong coupling, heavy quark masses, etc.), and the same fixed-order 
theoretical calculations.
The experimental dataset is
also similar except that some
specific processes have been excluded.

In this section, we first review
the experimental data that can be consistently included in
a threshold resummed global PDF analysis, and then
we explain the procedure used to construct the
resummed $K$-factors that  will be used to include threshold
resummation in the NNPDF fits.
We also show the resulting resummed $K$-factors for a representative
subset of the experiments used in the fit.

\subsection{Experimental data}

\label{sec:exp_data}

In a PDF fit with threshold resummation,
as compared to  fixed-order 
fits, some datasets cannot be included since for
these processes threshold resummation is either unknown 
or not currently available in a format that can be used in a  fit.
In particular, when compared
to NNPDF3.0, in the present resummed fit we include
all the neutral and charged current DIS data, neutral current
DY production and top quark pair production data.
However
we exclude the DY charged current datasets,
for which data is provided in terms of the lepton kinematics,
and the construction of resummed expressions is more involved,
and inclusive jet production for the reasons discussed in Sect.~\ref{sec:framework}.

In Table~\ref{tab:completedataset} we list all the datasets
used in the NNPDF3.0 NLO and NNLO global analysis, and indicate whether or
not they are now included in the NLL and NNLL resummed NNPDF3.0 fits.
For each dataset we also display the corresponding measured
observable, and the relevant publication.
A more complete description of each of these datasets, as well as of
their impact in terms of PDF constraints in the global fit, can be
found in~\cite{Ball:2014uwa}.
%

%%%%%%%%%%%%%%%%%%%%%%%%%%%%%%%%%%%%%%%%%%%%%%%%%%%%%%%%%%%%%%%%%
%%%%%%%%%%%%%%%%%%%%%%%%%%%%%%%%%%%%%%%%%%%%%%%%%%%%%%%%%%%%%%%%%
\begin{table}[t]
\footnotesize
\begin{centering}
\begin{tabular}{|c|c|c|c|c|}
\hline
    {Experiment} & Observable & {Ref.} &
NNPDF3.0 global  & NNPDF3.0 DIS+DY+top\\
&&& (N)NLO & (N)NLO [+(N)NLL]
    \tabularnewline 
\hline
\hline
NMC & $\sigma^{\rm NC}_{\rm dis},F^d_2/F_2^p$ & \cite{Arneodo:1996kd,Arneodo:1996qe}&  Yes  &  Yes\\[0.06cm]
BCDMS & $F^d_2,F_2^p$ & \cite{Benvenuti:1989rh,Benvenuti:1989fm}&  Yes  &  Yes \\[0.06cm]
  SLAC & $F^d_2,F_2^p$  & \cite{Whitlow:1991uw}  & Yes & Yes \\[0.06cm]
  CHORUS & $\sigma^{\rm CC}_{\nu N}$ &  \cite{Onengut:2005kv}       & Yes & Yes \\[0.06cm]
  NuTeV &  $\sigma^{\rm CC,charm}_{\nu N}$ & \cite{Goncharov:2001qe}  & Yes & Yes \\[0.06cm]    
\hline
  HERA-I & $\sigma^{\rm NC}_{\rm dis},\sigma^{\rm CC}_{\rm dis}$ & \cite{Aaron:2009aa} & Yes & Yes \\[0.06cm]
ZEUS HERA-II & $\sigma^{\rm NC}_{\rm dis},\sigma^{\rm CC}_{\rm dis}$ & \cite{Chekanov:2009gm,Chekanov:2008aa,Abramowicz:2012bx,Collaboration:2010xc}
& Yes & Yes \\[0.06cm]
H1 HERA-II & $\sigma^{\rm NC}_{\rm dis},\sigma^{\rm CC}_{\rm dis}$ & \cite{Aaron:2012qi,Collaboration:2010ry}  & Yes & Yes \\[0.06cm]
HERA charm & $\sigma^{\rm NC,charm}_{\rm dis}$ &  \cite{Abramowicz:1900rp}& Yes & Yes \\[0.06cm]
\hline
  DY E866 & $\sigma^{\rm NC}_{\rm DY,p},\sigma^{\rm NC}_{\rm DY,d}/\sigma^{\rm NC}_{\rm DY,p}$  & \cite{Towell:2001nh,Webb:2003ps,Webb:2003bj}& Yes & Yes \\[0.06cm]
  DY E605 &  $\sigma^{\rm NC}_{\rm DY,p}$ &  \cite{Moreno:1990sf} & Yes & Yes \\[0.06cm]
\hline
 CDF $Z$ rap & $\sigma^{\rm NC}_{\rm DY,p}$ & \cite{Aaltonen:2010zza} & Yes & Yes \\[0.06cm]
 CDF Run-II $k_t$ jets & $\sigma_{\rm jet}$ & \cite{Abulencia:2007ez}& Yes & No \\[0.06cm]
 D0 $Z$ rap & $\sigma^{\rm NC}_{\rm DY,p}$ &  \cite{Abazov:2007jy} & Yes & Yes \\[0.06cm]
 \hline
ATLAS $Z$ 2010 & $\sigma^{\rm NC}_{\rm DY,p}$  & \cite{Aad:2011dm}  & Yes & Yes \\[0.06cm]
 ATLAS $W$ 2010 & $\sigma^{\rm CC}_{\rm DY,p}$ & \cite{Aad:2011dm}  & Yes & No \\[0.06cm]
    ATLAS 7 TeV jets 2010 & $\sigma_{\rm jet}$&  \cite{Aad:2011fc} & Yes & No \\[0.06cm]
    ATLAS 2.76 TeV jets  &  $\sigma_{\rm jet}$ &\cite{Aad:2013lpa} & Yes & No \\[0.06cm]
       ATLAS high-mass DY  & $\sigma^{\rm NC}_{\rm DY,p}$  & \cite{Aad:2013iua} & Yes & Yes \\[0.06cm]
       ATLAS $W$ $p_T$  &  $\sigma^{\rm CC}_{\rm DY,p}$  & \cite{Aad:2011fp} & Yes & No \\[0.06cm]
       \hline
 CMS $W$ electron asy &  $\sigma^{\rm CC}_{\rm DY,p}$& \cite{Chatrchyan:2012xt} & Yes & No \\[0.06cm]
   CMS $W$ muon asy  & $\sigma^{\rm CC}_{\rm DY,p}$  & \cite{Chatrchyan:2013mza}& Yes & No \\[0.06cm]
   CMS jets 2011     &   $\sigma_{\rm jet}$ & \cite{Chatrchyan:2012bja}& Yes & No \\[0.06cm]
   CMS $W+c$ total  & $\sigma^{\rm NC,charm}_{\rm DY,p}$  & \cite{Chatrchyan:2013uja}& Yes & No \\[0.06cm]
   CMS 2D DY 2011   &  $\sigma^{\rm NC}_{\rm DY,p}$ & \cite{Chatrchyan:2013tia}  & Yes & Yes \\[0.06cm]
   \hline
    LHCb $W$ rapidity  & $\sigma^{\rm CC}_{\rm DY,p}$  &\cite{Aaij:2012vn} & Yes & No \\[0.06cm]
      LHCb $Z$ rapidity & $\sigma^{\rm NC}_{\rm DY,p}$  & \cite{Aaij:2012mda} &  Yes & Yes \\[0.06cm]
\hline
ATLAS CMS top prod &  $\sigma(t\bar{t})$  &  \cite{ATLAS:2012aa,ATLAS:2011xha,TheATLAScollaboration:2013dja,Chatrchyan:2013faa,Chatrchyan:2012bra,Chatrchyan:2012ria}&
Yes & Yes \\[0.06cm]
\hline
\end{tabular}
\par\end{centering}
\caption{List of all the experiments
  that were used in the NNPDF3.0 global analysis, and whether or not
  they are now included in the present (N)NLL resummed fits
  (and in the corresponding baseline fixed-order fits).
  For each dataset we also provide the type
  of cross section that has been measured and 
  the corresponding
  publication references.
}
\label{tab:completedataset}
\end{table}
%%%%%%%%%%%%%%%%%%%%%%%%%%%%%%%%%%%%%%%%%%%%%%%%%%%%%%%%%%%%%%%%%%%%%%%%%
%%%%%%%%%%%%%%%%%%%%%%%%%%%%%%%%%%%%%%%%%%%%%%%%%%%%%%%%%%%%%%%%%%%%%%%%%

From Table~\ref{tab:completedataset}, we infer that, when compared to the
global fit, the resummed fits lose experimental constraints on the medium and
large-$x$ gluon (due to the exclusion of the jet data) and on the
quark-flavor separation (due to the exclusion of the $W$ data).
Still, given that we include in the resummed fit more than 3000
data points, the loss
of accuracy due to the exclusion of these datasets is not dramatic, 
as we will show in Sect.~\ref{sec:results}.
In future studies, we aim to include the missing processes
once the corresponding
resummed calculations become available.

The kinematic cuts applied in the present fits closely follow the ones of
the NNPDF3.0 fixed-order analysis. 
In particular, a cut on the final-state invariant mass of DIS data
$W^2 \ge 12.5$ GeV$^2$ is applied, in order to reduce the dependence
on higher-twists at large $x$.
It would be interesting to loosen this cut in future analyses,
in order to test the
stability of the leading-twist PDF determination once the large-$x$
resummation is included; this might also allow us to include additional
large-$x$, low-$Q^2$ DIS
measurements, such as for example JLAB data~\cite{Tkachenko:2014byy}.

In addition, a stability analysis of the calculations for neutral-current
Drell-Yan production indicates that our results, even when supplemented with resummation, 
become unstable for data points
too close to the production threshold, either because the invariant mass $M_{ll}$ of
the Drell-Yan pairs is too large, or because the rapidity $Y$ is too close
to the kinematic boundary $Y_{\rm max}$.
Therefore, we have supplemented the
NNPDF3.0 kinematic cuts with two additional cuts
for the fixed-target Drell-Yan experiments, as summarised in Table~\ref{tab:kincuts}.
For the collider Drell-Yan data, the cuts are the same as in
NNPDF3.0.

%%%%%%%%%%%%%%%%%%%%%%%%%%%%%%%%%%%%%%%%%%%%%%%%%%%%%%%%%%%
%%%%%%%%%%%%%%%%%%%%%%%%%%%%%%%%%%%%%%%%%%%%%%%%%%%%%%%%%%%
\begin{table}[t]
\centering
\begin{tabular}{|c|c|}
\hline     
Experiment   &  Kinematic cuts \\
\hline
\hline
DIS         &  $Q^2 \ge Q^2_{\rm min}=3.5$ GeV$^2$ \\
         &  $W^2 \ge W^2_{\rm min}=12.5$ GeV$^2$ \\
         \hline
Fixed target Drell-Yan  &  $ \tau\le 0.08 $ \\
  &   $|Y|/Y_{\rm max}\le 0.663$ \\
\hline
\end{tabular}
\caption{\small Kinematic cuts applied to the DIS and fixed-target Drell-Yan
data in the baseline and resummed fits.
For the collider Drell-Yan data, the cuts are the same as in
NNPDF3.0.}
\label{tab:kincuts}
\end{table}
%%%%%%%%%%%%%%%%%%%%%%%%%%%%%%%%%%%%%%%%%%%%%%%%%%%%%%%%%%%
%%%%%%%%%%%%%%%%%%%%%%%%%%%%%%%%%

%%%%%%%%%%%%%%%%%%%%%%%%%%%%%%
\begin{figure}[t]
  \centering
  \includegraphics[width=0.495\textwidth]{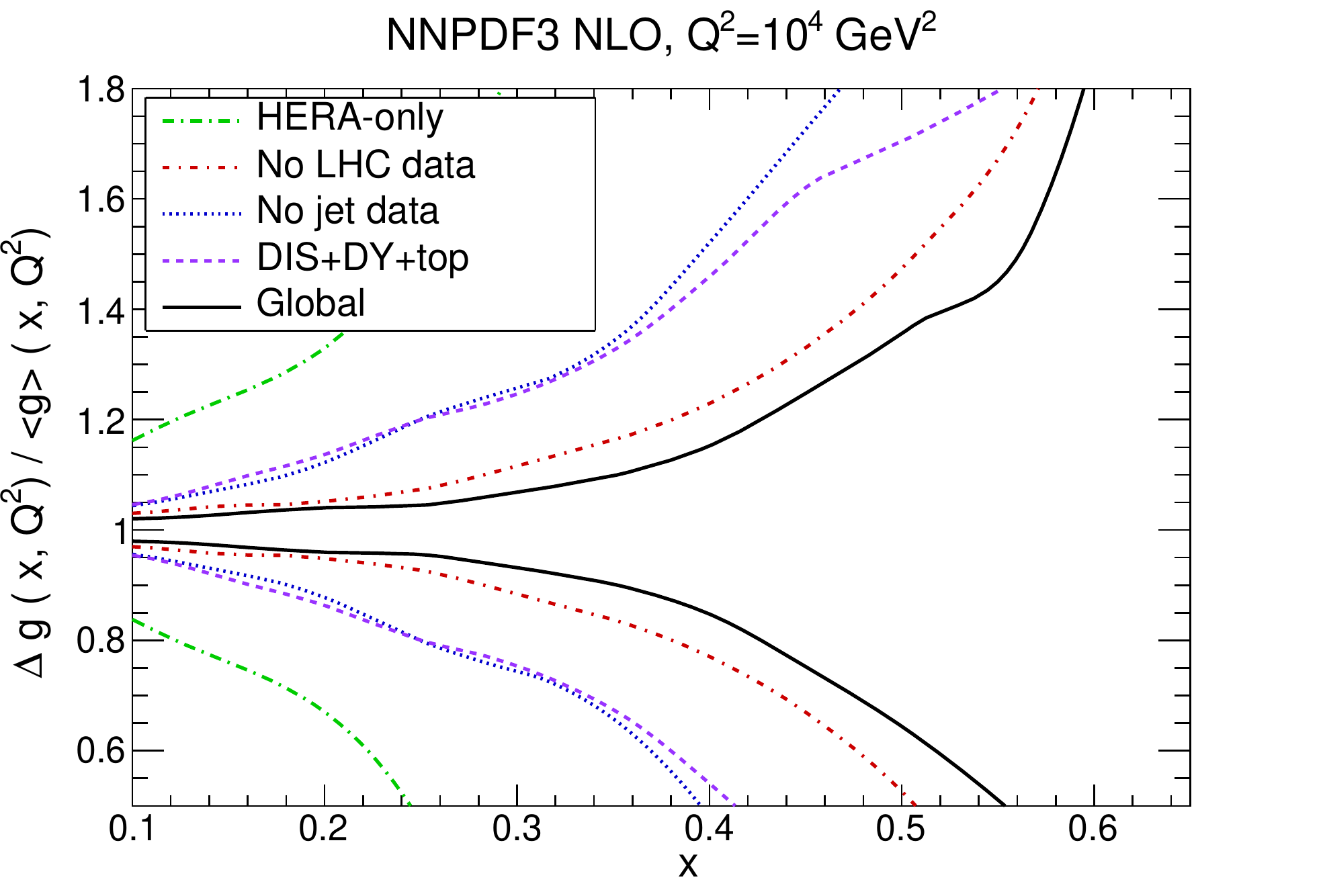}
  \includegraphics[width=0.495\textwidth]{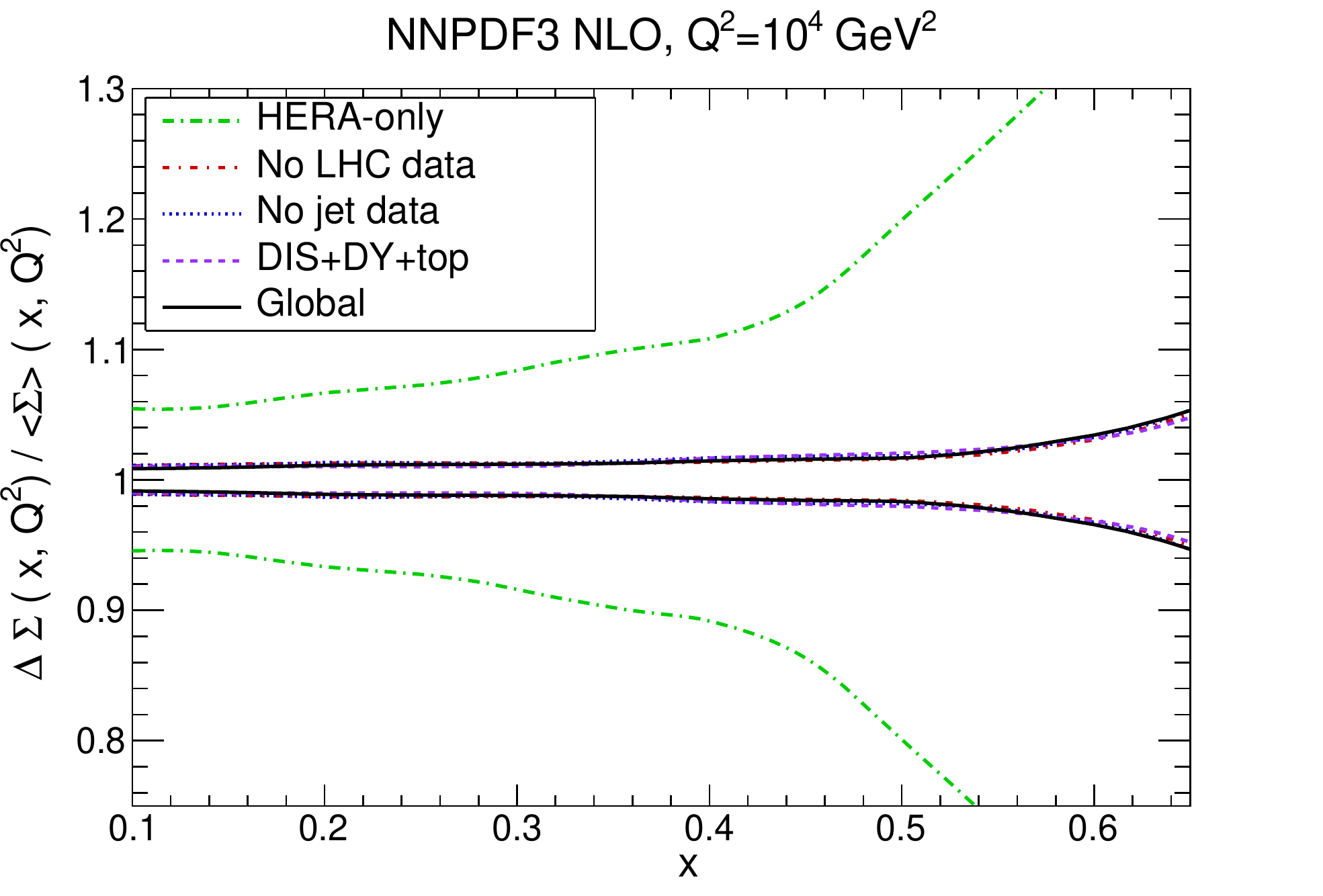}
  \includegraphics[width=0.495\textwidth]{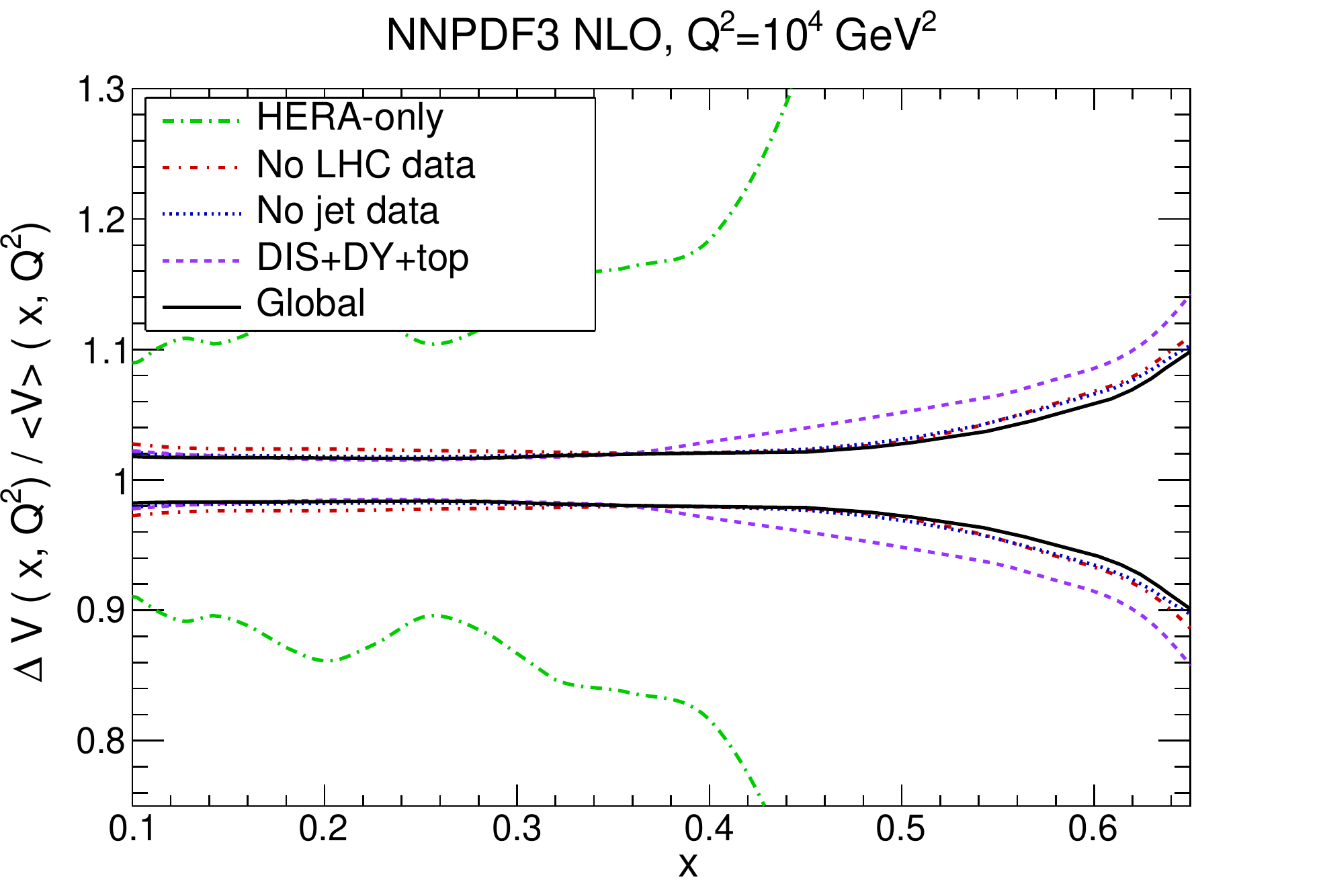}
  \includegraphics[width=0.495\textwidth]{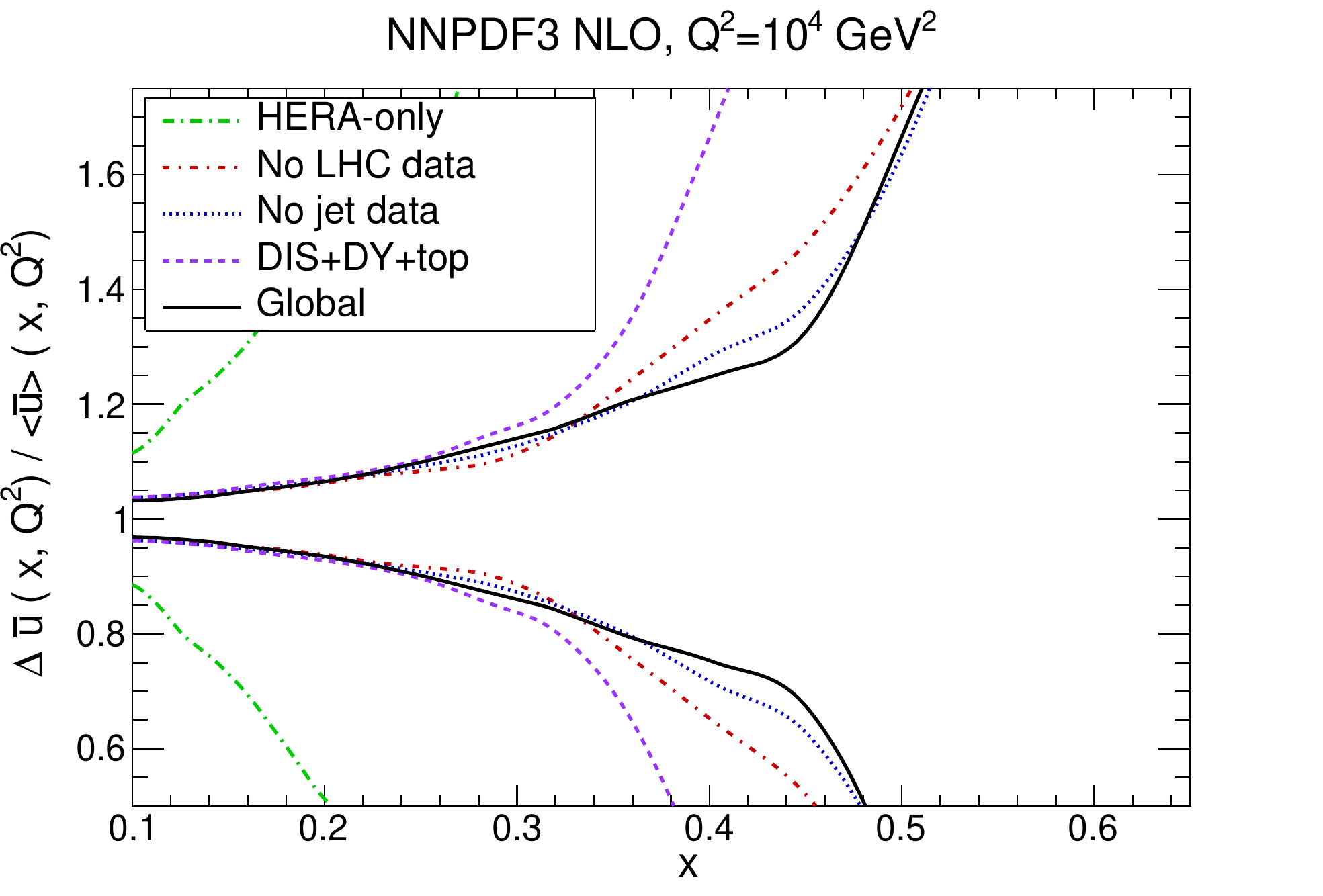}  
  \caption{\small Comparison of the relative PDF uncertainty at large-$x$
    between NNPDF3.0 NLO fits based on different input datasets.
    We show the results for the gluon, total quark singlet, total valence
    and $\bar{u}$ quark PDFs, at a typical LHC scale of $Q^{2}=10^{4}$ GeV$^2$.
    The fits shown are the HERA-only, no LHC data, no jet data, DIS+DY+top (our
    baseline for the resummed fits) and finally the global fit.
  }
  \label{fig:datasetdep}
\end{figure}
%%%%%%%%%%%%%%%%%%%%%%%%%%%%%%%%

It is useful to quantify which experiments determine the behaviour
of the large-$x$ PDFs in the global analysis.
In Fig.~\ref{fig:datasetdep} we compare
the relative PDF uncertainties in variants of the NNPDF3.0 NLO fit based
on different input datasets: HERA-only, no LHC data, no jet data,
a DIS+DY+top fit (our baseline for the resummed fits) and the
global fit.
We focus 
in the large-$x$ region for $Q^2=10^4$ GeV, a typical
scale for LHC phenomenology.
We show the gluon, the total quark
singlet, the total valence and the $\bar{u}$ quark PDFs.
From this
comparison we see that the PDF that at large-$x$ is most
dependent on the choice of input dataset is the gluon.
For the total
valence and singlet quark PDFs, the bulk of the constraints are
provided by the DIS and fixed target Drell-Yan data, which are common
datasets in all these fits (except for the HERA-only fit).
For the sea
quarks, in this case the $\bar{u}$ quark, the information on both jet
data and LHC data are necessary to achieve the best possible accuracy.
The baseline DIS+DY+top fit is slightly less accurate at large-$x$ for
the quark flavor separation due to the missing charged-current Drell-Yan data.

\subsection{Calculation of resummed $K$-factors}

As mentioned in the introduction, in the $\MSbar$ scheme
all the effects of threshold resummation are encoded in
the partonic cross sections, and thus parton evolution
is the same as in fixed-order calculations.
Therefore, apart from the modification of the hard-scattering cross sections,
all theoretical settings in the resummed fit will be the same as
those of the NNPDF3.0 fixed-order analysis, including
the use of the FONLL general-mass VFN scheme~\cite{Forte:2010ta},
the values of the heavy quark masses, and so on.
We will produce results for single a value of the strong coupling,
$\as(m_Z^2)=0.118$.

As discussed in Ref.~\cite{Ball:2014uwa}, the NNPDF3.0 global
analysis for hadronic observables always uses fast NLO
calculations~\cite{Bertone:2014zva,Carli:2010rw,Wobisch:2011ij}
supplemented with NNLO/NLO $K$-factors when required.
These are defined as the ratio of the NNLO over the NLO
bin-by-bin cross sections, using a common PDF luminosity computed with
a NNLO PDF set.
For the resummed fits, we follow exactly the same procedure: we include the effect of resummation
supplementing the fixed-order computation with a $K$-factor.
Since the $K$-factor is computed externally using a fixed set of PDFs, the fit
is re-iterated several times, recomputing each time the resummed $K$-factor using as input PDFs
those obtained from the previous iteration.

As discussed in Sect.~\ref{sec:troll}, the resummed contributions for DIS and
DY processes is obtained using the 
program \troll\ in the form of $\Delta K$-factors, Eq.~\eqref{eq:deltaK}, and hence must be converted into actual $K$-factors.
For DIS cross sections, since the NNLO calculation is implemented exactly in the NNPDF fitting code,
this is done according to
\be
\label{eq:cfact1}
K^{\text{N$^k$LO+N$^k$LL}}_{\rm DIS} \equiv \frac{\sigma^{\text{N$^k$LO+N$^k$LL}}}{\sigma^{\text{N$^k$LO}}}
    = 1+ \Delta_{k} K_{\text{N$^k$LL}} \cdot \frac{\sigma^{\rm LO}}{\sigma^{\text{N$^k$LO}}},
\ee
with $k=1,2$ for NLO+NLL and NNLO+NNLL respectively.
For hadronic processes we use a similar expression, but (at NNLO) also including the NNLO/NLO $K$-factor,
\bea
\label{eq:cfact3}
K^{\rm NLO+NLL}_{\rm hadr} &\equiv \frac{\sigma^{\rm NLO+NLL}}{\sigma^{\rm NLO}}
= 1+ \Delta_{1} K_{\rm NLL} \cdot \frac{\sigma^{\rm LO}}{\sigma^{\rm NLO}}\, , \\
K^{\rm NNLO+NNLL}_{\rm hadr} &\equiv \frac{\sigma^{\rm NNLO+NNLL}}{\sigma^{\rm NLO}}
= K^{\rm NNLO} + \Delta_{2} K_{\rm NNLL} \cdot \frac{\sigma^{\rm LO}}{\sigma^{\rm NLO}} \, ,
\label{eq:cfact2}
\eea
where $K^{\rm NNLO}=\sigma^{\rm NNLO}/\sigma^{\rm NLO}$.
In the above expressions, all contributions are meant to be computed with the same N$^k$LO+N$^k$LL PDF set.
For the leading-order
cross section $\sigma^{\rm LO}$, dedicated
{\tt FK} tables with LO coefficient functions but NLO and NNLO
PDF evolution have been produced 
using the {\tt APFEL} program~\cite{Bertone:2013vaa}, validated with the
same {\tt FKgenerator} internal code used in the NNPDF3.0 fits.

In principle, all these contributions should be recomputed at each iteration of the fit;
in practice, the computation of $K^{\rm NNLO}$ is time consuming, so for this contribution
we use a fixed value.
Specifically, in the present work, these fixed-order NNLO/NLO $K$-factors are the same
as in the NNPDF3.0 fits, with the exception of those for fixed-target Drell-Yan experiments, which
have been recomputed using {\tt Vrap}~\cite{Anastasiou:2003ds}
with the NNPDF3.0 global PDF set
as input.

For the computation of the
resummed $K$-factors Eq.~(\ref{eq:cfact1})--(\ref{eq:cfact2}),
we find that two iterations of the fit
are enough to reach a satisfactory convergence,
meaning that these $K$-factors
are essentially unchanged if we use
resummed PDFs from the last or from the next-to-last iteration of the fit.

It is now interesting to illustrate the
effect of the (N)NLL resummation for some of the datasets used in the present resummed PDF fit.
To this purpose, we plot the DIS, Eq.~\eqref{eq:cfact1}, and hadronic,
Eqs.~\eqref{eq:cfact3} and~\eqref{eq:cfact2}, resummed $K$-factors
for representative experimental datasets with exactly
the same kinematics as for the data points to be used in the fit.
In these calculations, we have consistently used the
NNPDF3.0 DIS+DY+top NLO+NLL and NNLO+NNLL PDF sets discussed
in the next Section, with $\alpha_s(m_Z^2)=0.118$, in both the fixed-order and resummed cross sections.
To isolate the effect of the resummation, in these
comparison plots we will factor out
$K^{\rm NNLO}$ from the hadronic NNLO+NNLL resummed
$K$-factor Eq.~\eqref{eq:cfact2}.
Note that in these plots we will only include those data points
that satisfy the kinematic cuts imposed in the fit,
summarised in Table~\ref{tab:kincuts}.

%%%%%%%%%%%%%%
\begin{figure}[t]
\begin{center}
\epsfig{width=0.49\textwidth,figure=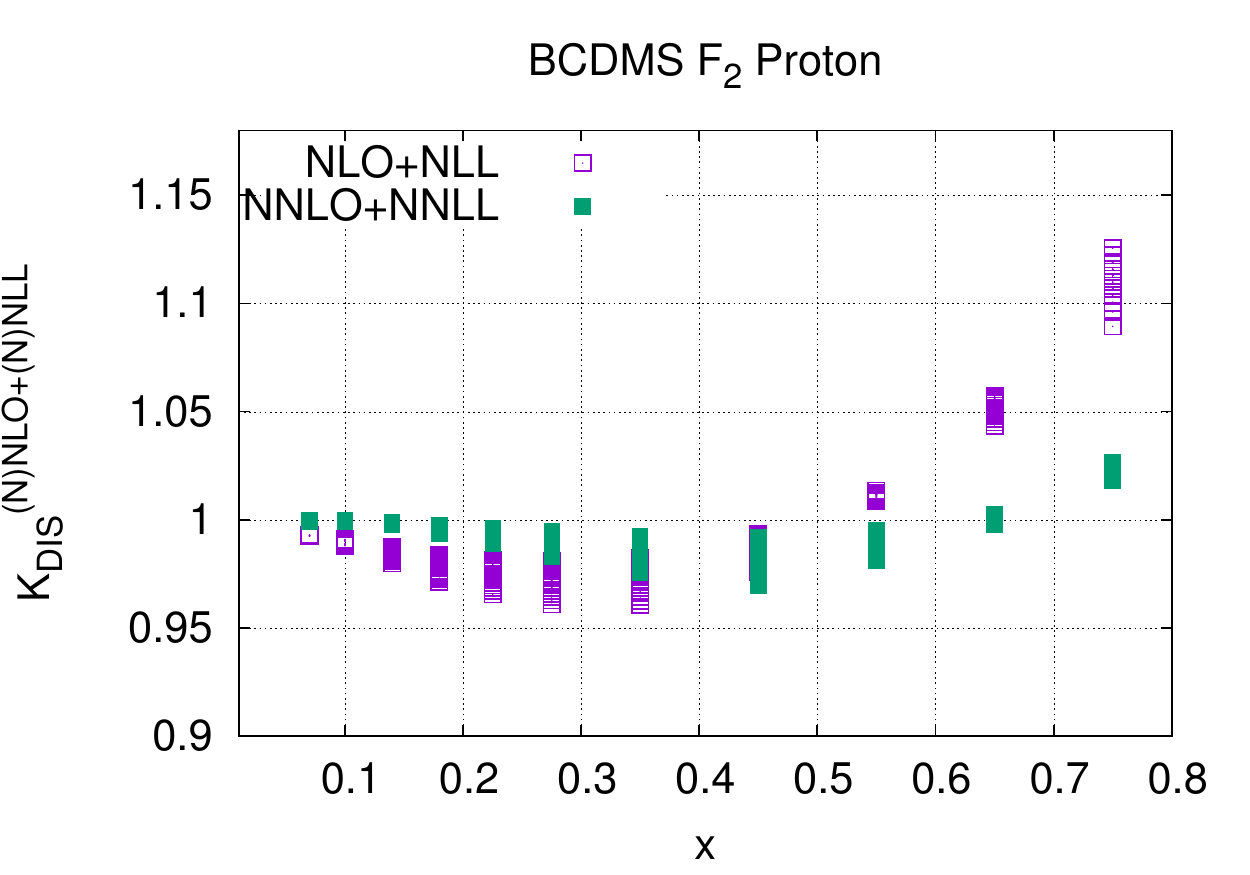}
\epsfig{width=0.49\textwidth,figure=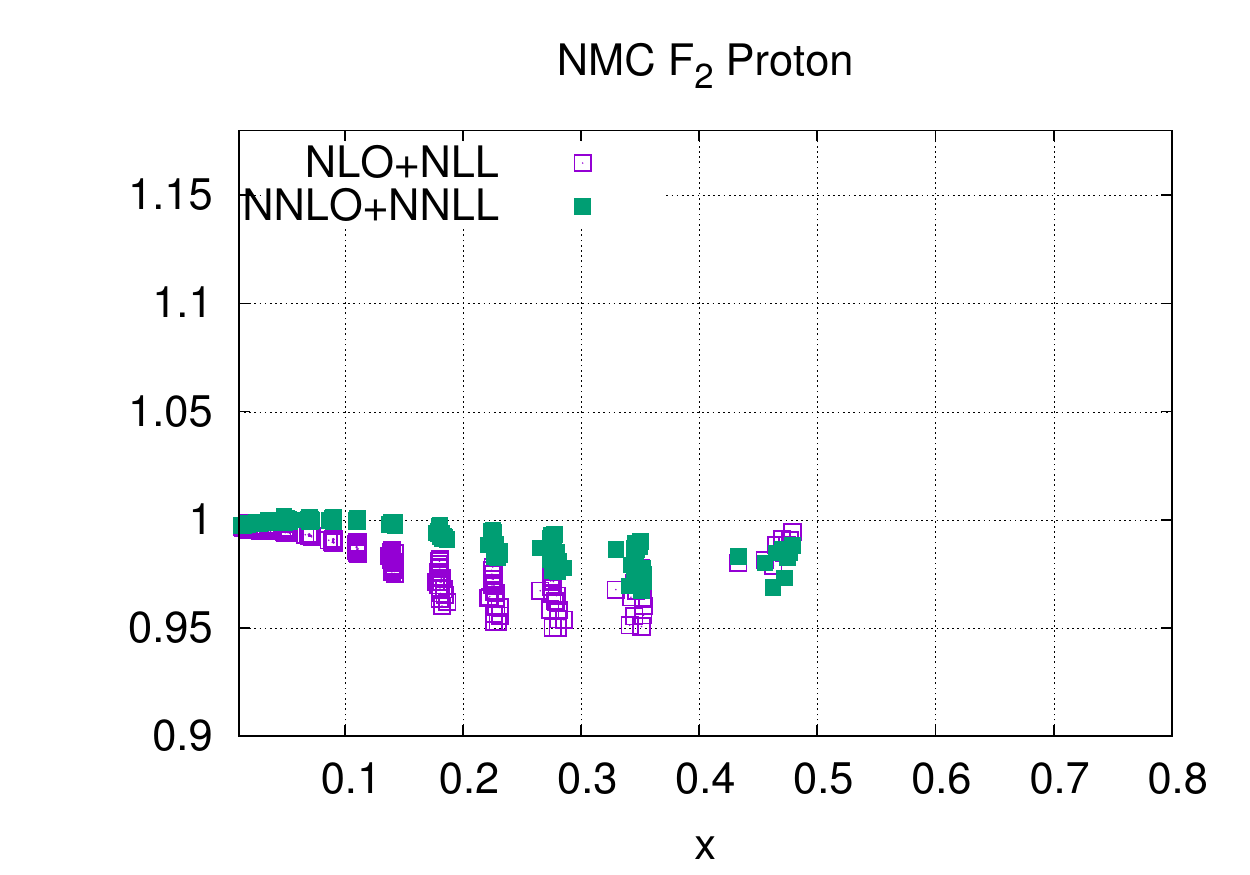}
\epsfig{width=0.49\textwidth,figure=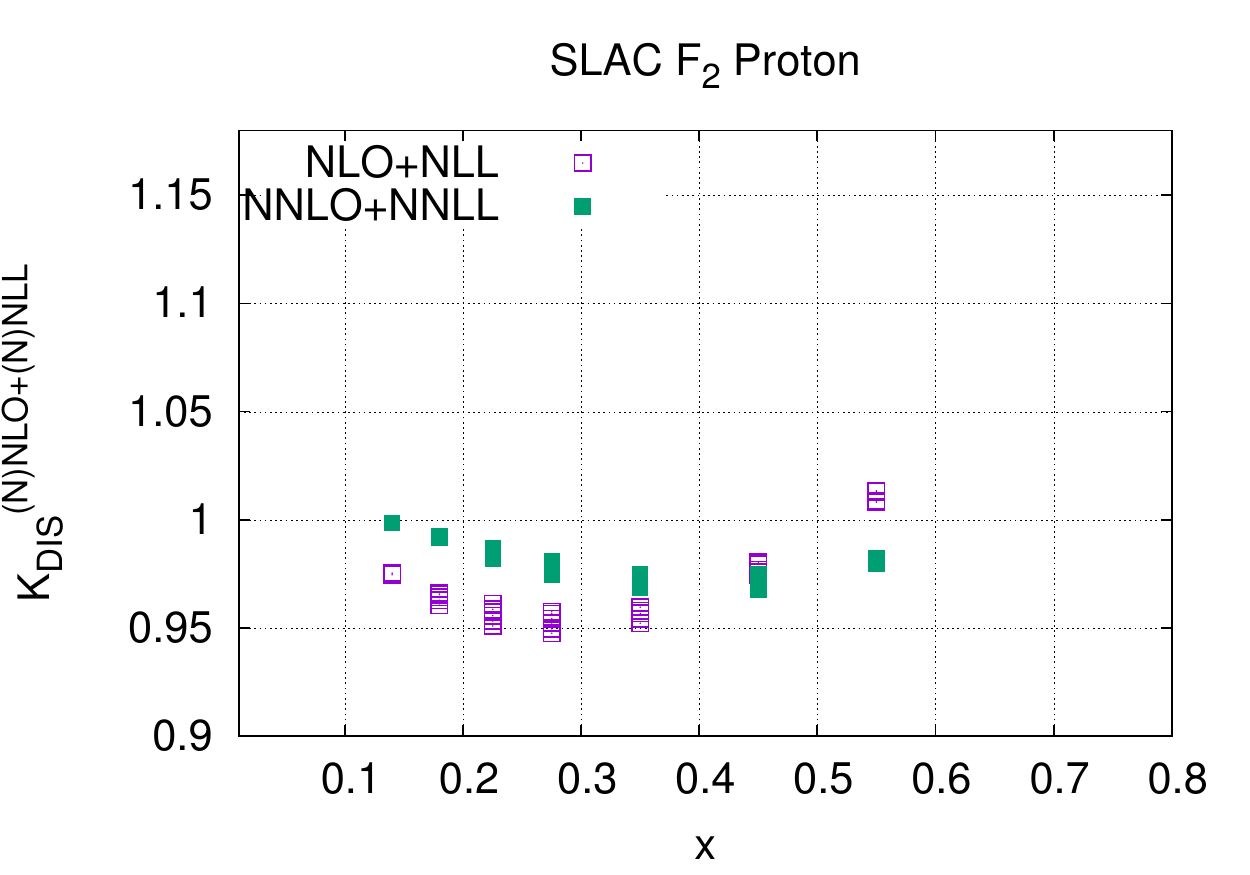}
\epsfig{width=0.49\textwidth,figure=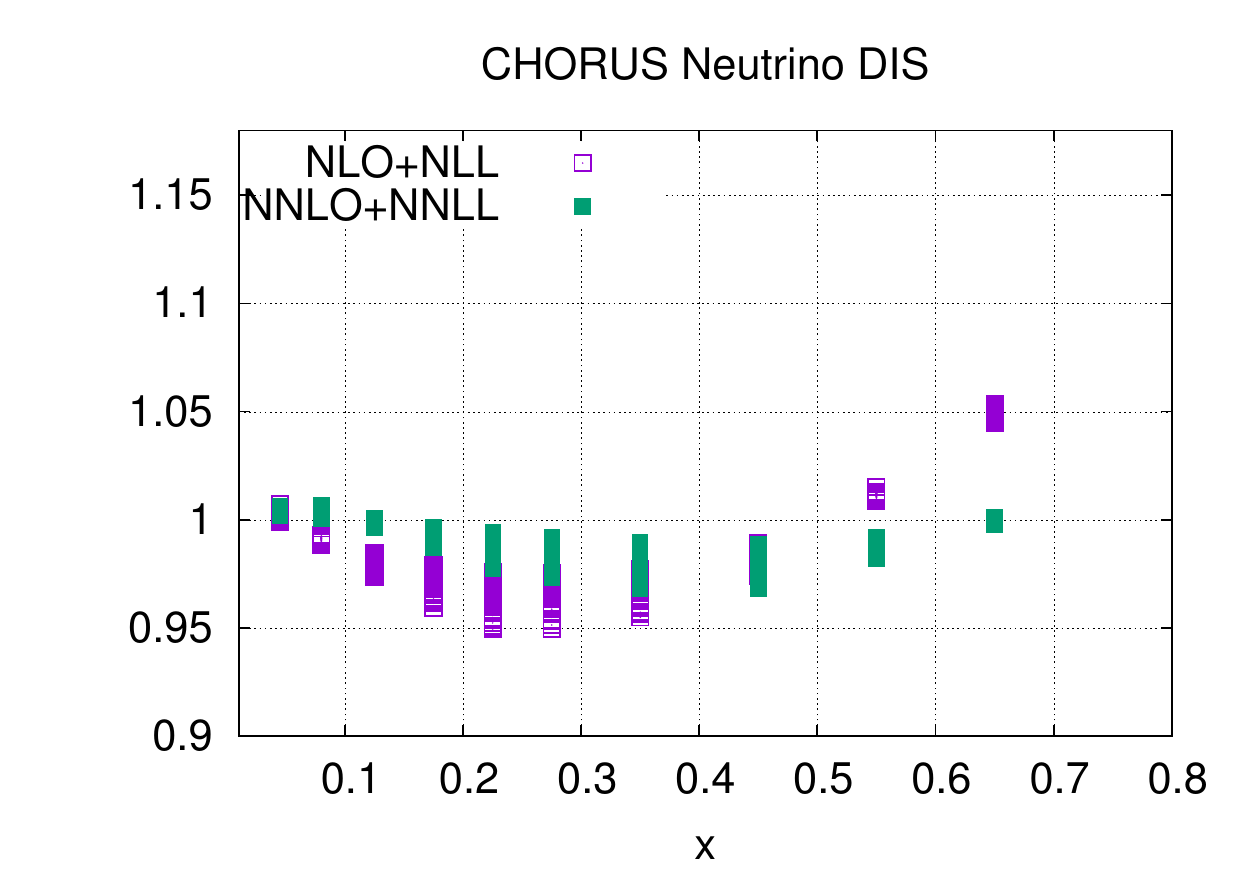}
\caption{The resummed $K$-factors for DIS, Eq.~(\ref{eq:cfact1}), for a representative subset
  of the experiments included in the resummed fit: BCDMS $F_2^p$, NMC $\sigma_{\rm NC}^p$,
  SLAC $F_2^p$ and CHORUS $\sigma_{\nu N}$.
  We show both the results corresponding to NLL and to NNLL resummation.
  The DIS kinematics $(x,Q^2,y)$ is that of the corresponding experimental data included in the fit,
  plotted just as a function of $x$, so for each value of $x$ there are measurements
  at different values of $Q^2$ and $y$.
  }
   \label{fig:nnpdf_cfact_1nsoft}
  \end{center}
\end{figure}
%%%%%%%%%%%%%%%%%%

The results for the DIS case
are shown in Fig.~\ref{fig:nnpdf_cfact_1nsoft}.
For each experiment, we show both the results corresponding to NLL and to NNLL resummation.
The DIS kinematics $(x,Q^2,y)$ is that of the
associated experimental measurements, so for each value of $x$ there are measurements
at different values of $Q^2$ and $y$.
We do not show the results for any of the HERA datasets, for which the
effect of the resummation turns out to be
negligible since the data is either at small-$x$ or at high scales.
As expected, the impact of the resummation is only relevant
at large $x$, and of course the impact of the resummation
decreases when more fixed perturbative orders are included
in the calculation.

From the results of Fig.~\ref{fig:nnpdf_cfact_1nsoft} we see that
the effect of threshold resummation
is most important for the BCDMS data, while it is milder for
the other experiments.
Effects of NLL resummation reach up to 15\% at the highest values
of $x$ available, which is reduced to up to a few percent for
NNLL resummation (since part of the effects at NLL are now included
in the fixed-order NNLO calculation).
Note also that the cut in $W^2$ removes most of the large-$x$
SLAC data, where resummation effects are very large.
We also note that the effect of resummation is comparable to
the experimental uncertainties, and thus
we should expect to see an impact on the resulting
large-$x$ parton distributions.

%%%%%%%%%%%%%%
\begin{figure}[t]
  \begin{center}
    \epsfig{width=0.49\textwidth,figure=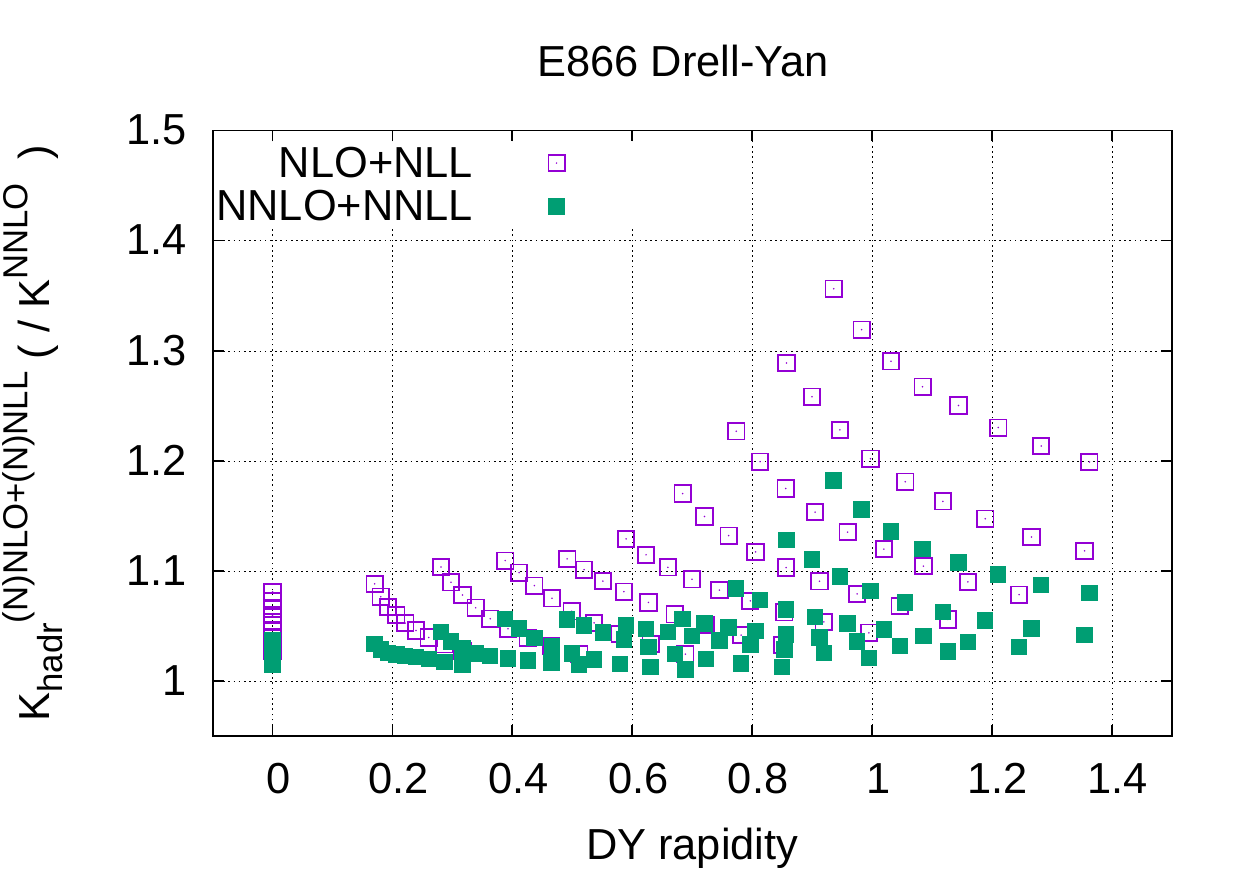}
    \epsfig{width=0.49\textwidth,figure=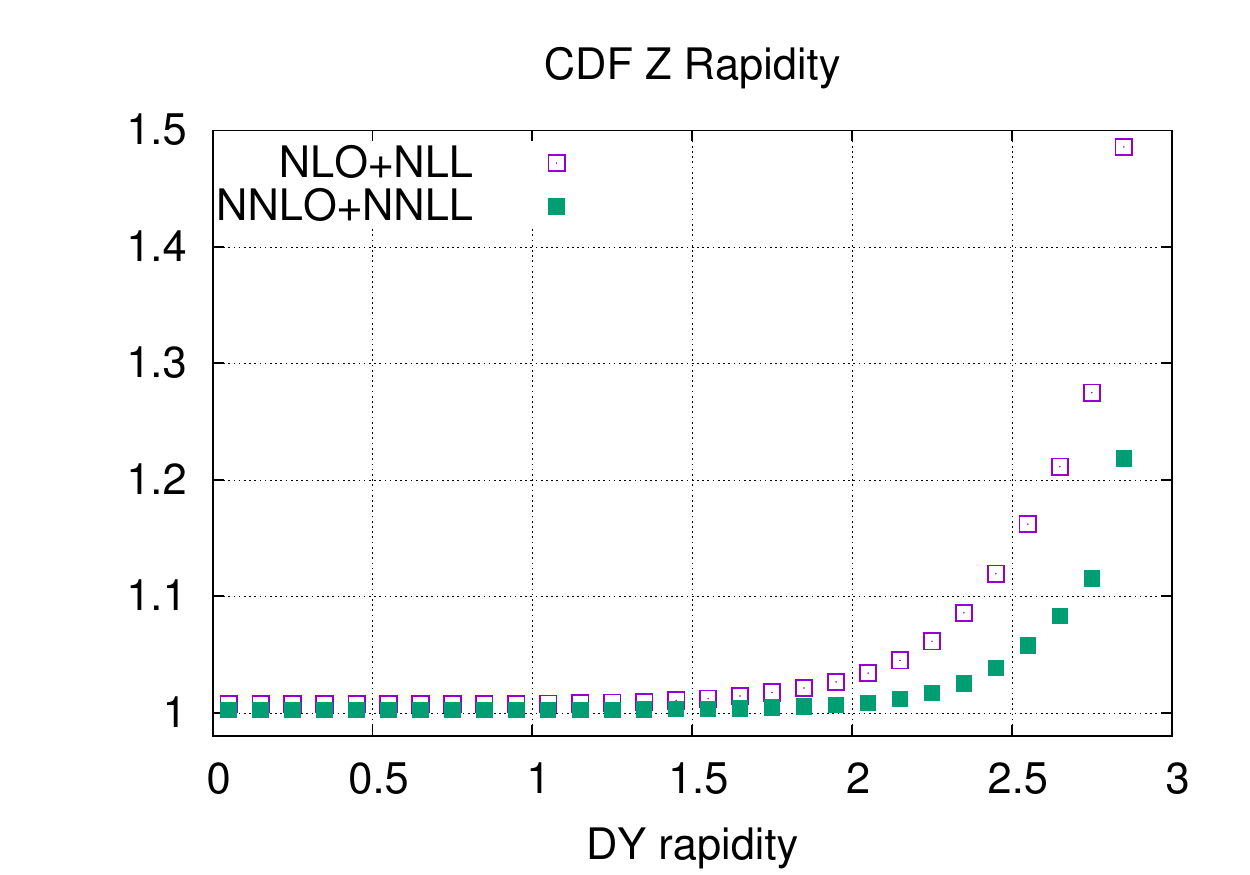}
    \epsfig{width=0.49\textwidth,figure=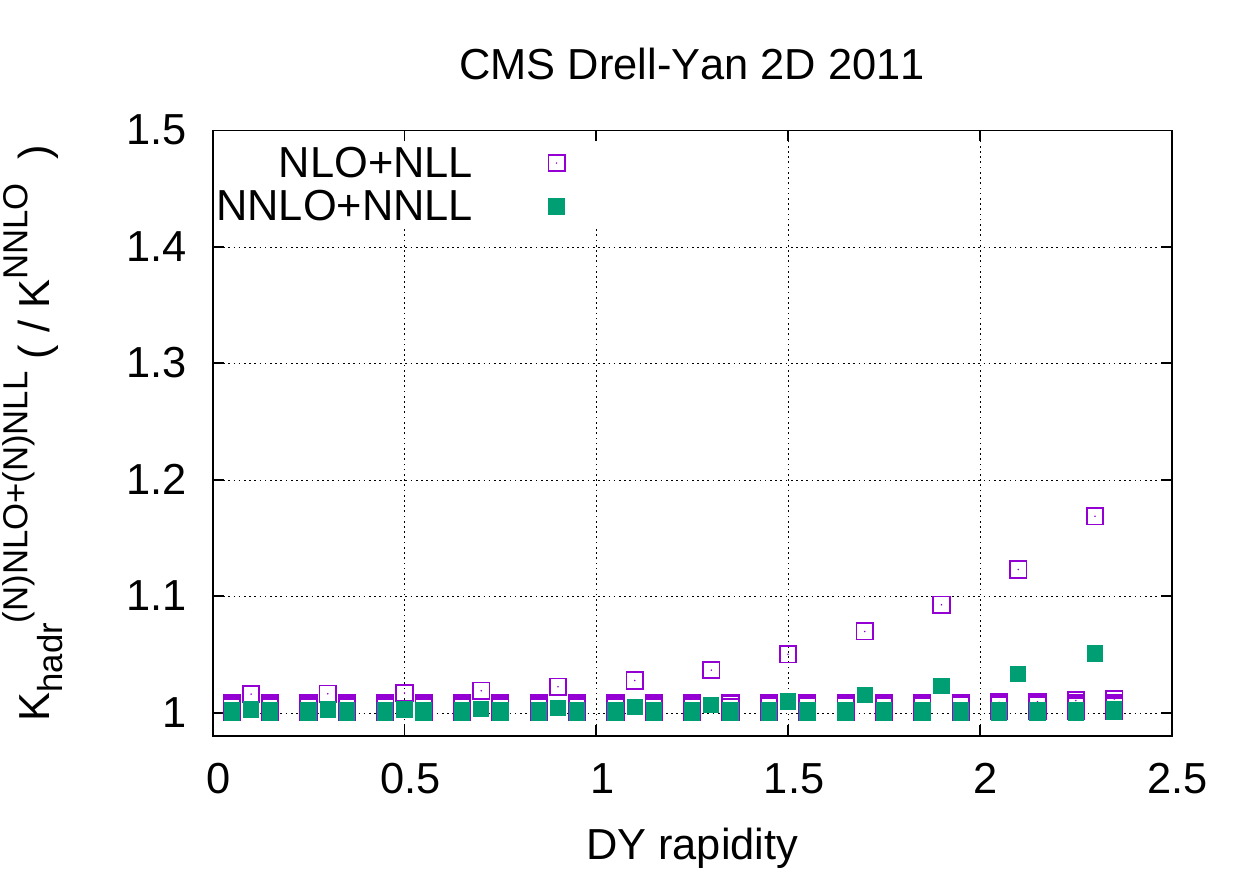}
    \epsfig{width=0.49\textwidth,figure=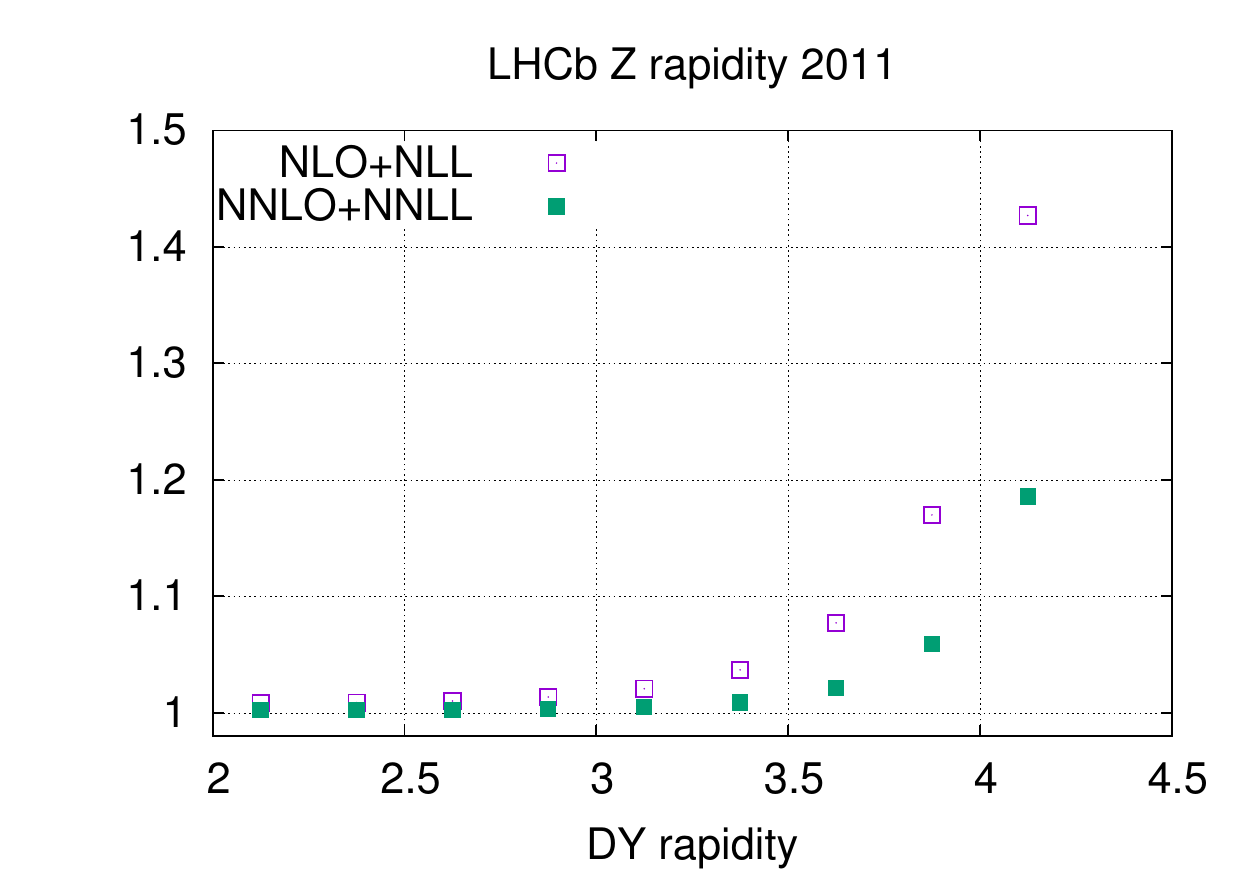}
\caption{Same as Fig.~\ref{fig:nnpdf_cfact_1nsoft} for selected Drell-Yan experimental datasets included in the fit:
the E866 $pp$ neutral Drell-Yan cross sections,
the CDF $Z$ rapidity distribution, the CMS double-differential
Drell-Yan distribution at 7 TeV and the LHCb $Z\to \mu\mu$ rapidity
distribution.
The resummed $K$-factors are now those defined in Eqs.~(\ref{eq:cfact3}) and~(\ref{eq:cfact2}),
but to isolate the effect of resummation from that of the fixed-order
NNLO corrections, in the NNLL case we divide Eq.~(\ref{eq:cfact2}) by $K^{\rm NNLO}$.
In the two left plots data points differ by the values of the rapidity and the invariant mass of the pair,
but only the dependence on the rapidity is shown.
} 
    \label{fig:nnpdf_cfact_2nsoft}
\end{center}
\end{figure}
%%%%%%%%%%%%%%%%%%

The corresponding results for the resummed
$K$-factors Eq.~({\ref{eq:cfact2}) for
  representative neutral-current DY
experiments are collected in Fig.~\ref{fig:nnpdf_cfact_2nsoft}.
As in the case of DIS, except for $Z$ peak measurements,
for each DY rapidity $Y$ value there are various data points
at different invariant mass $M$.
We show 
the Drell-Yan E866 $pp$ cross sections,
the CDF $Z$ rapidity distribution, the CMS double-differential
Drell-Yan distribution at 7 TeV and the LHCb $Z\to \mu\mu$ rapidity
distribution.

From Fig.~\ref{fig:nnpdf_cfact_2nsoft} we verify the expectation that the
impact of resummation is rather more important at NLO+NLL than at NNLO+NNLL,
and that it grows with the di-lepton rapidity (since in this case the
kinematic threshold is approached).
For those collider measurements differential in rapidity, the effect
of (N)NLL resummation can be as large as 50\% (20\%) at the
highest rapidities.
For the fixed-target DY experiments the effect of resummation
is substantial even at NNLL, since in this case many data points
have kinematics close to threshold.
For example, for the E866 $pp$ dataset, the effect of the resummation
results is an enhancement of the cross section that can be as large
as 35\% at NLL, and 20\% at NNLL.

Following this discussion on the settings used to
produce the threshold resummed fits, in the
the next section we turn to explore the actual effects that 
the inclusion of resummed calculations have on
the NNPDF3.0 PDFs.

%%%%%%%%%%%%%%%%%%%%%%%%%%%%%%%%%%%%%%%%%%%%%%%%%%%%%%%%%%%%%%%%%
%%%%%%%%%%%%%%%%%%%%%%%%%%%%%%%%%%%%%%%%%%%%%%%%%%%%%%%%%%%%%%%%%

\section{Parton distributions with threshold resummation}
\label{sec:results}

In this section we discuss the results of the NNPDF3.0 fits
with threshold resummation.
One important difference of the resummed fits as compared to the
NNPDF3.0 global fits is that the dataset is different, because we leave out the inclusive jet and $W$
production data, as discussed in Sect.~\ref{sec:fitsettings}.
Therefore, first we quantify the information loss due to the
reduced dataset by comparing the global NNPDF3.0 fits and the reduced dataset
fits obtained with fixed-order matrix elements (henceforth
denoted as the baseline fits).

Having established this, we move to quantify the impact of threshold
resummation on the fit quality and the resulting PDFs,
by comparing fits at NLO and NLO+NLL first,
and then at NNLO and NNLO+NNLL.
This is done both for DIS-only fits and for DIS+DY+top fits, and the
comparison is performed both at the level of PDFs and of $\chi^2$.
Finally, we assess the impact of threshold resummation
at the level of partonic luminosities.

The phenomenological implications of the resummed PDFs for
LHC applications will be discussed
in the next section.

\subsection{Baseline fixed-order fits}

%%%%%%%%%%%%%%%%%%%%
\begin{figure}[t]
\begin{center}
\includegraphics[width=0.49\textwidth]{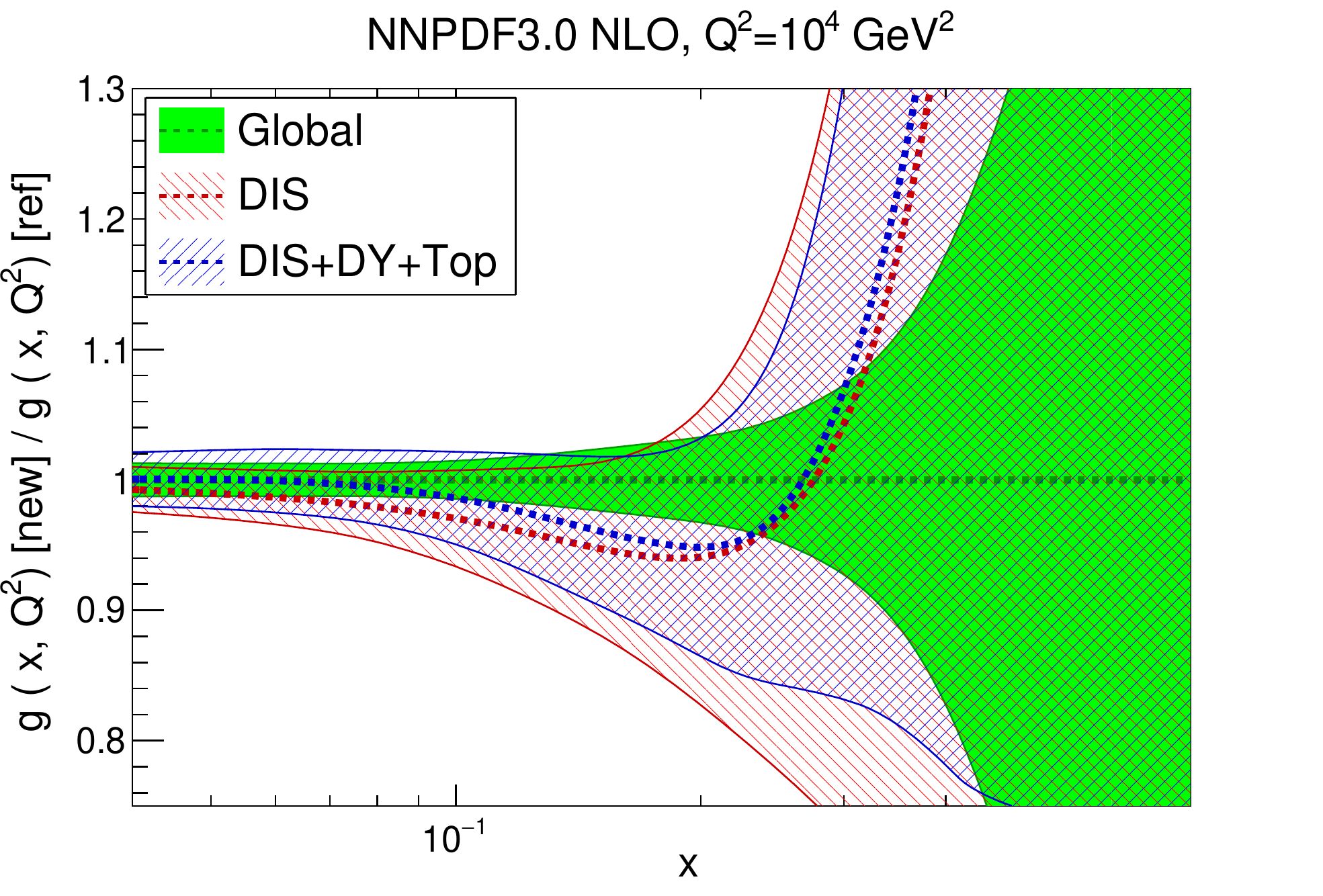}
\includegraphics[width=0.49\textwidth]{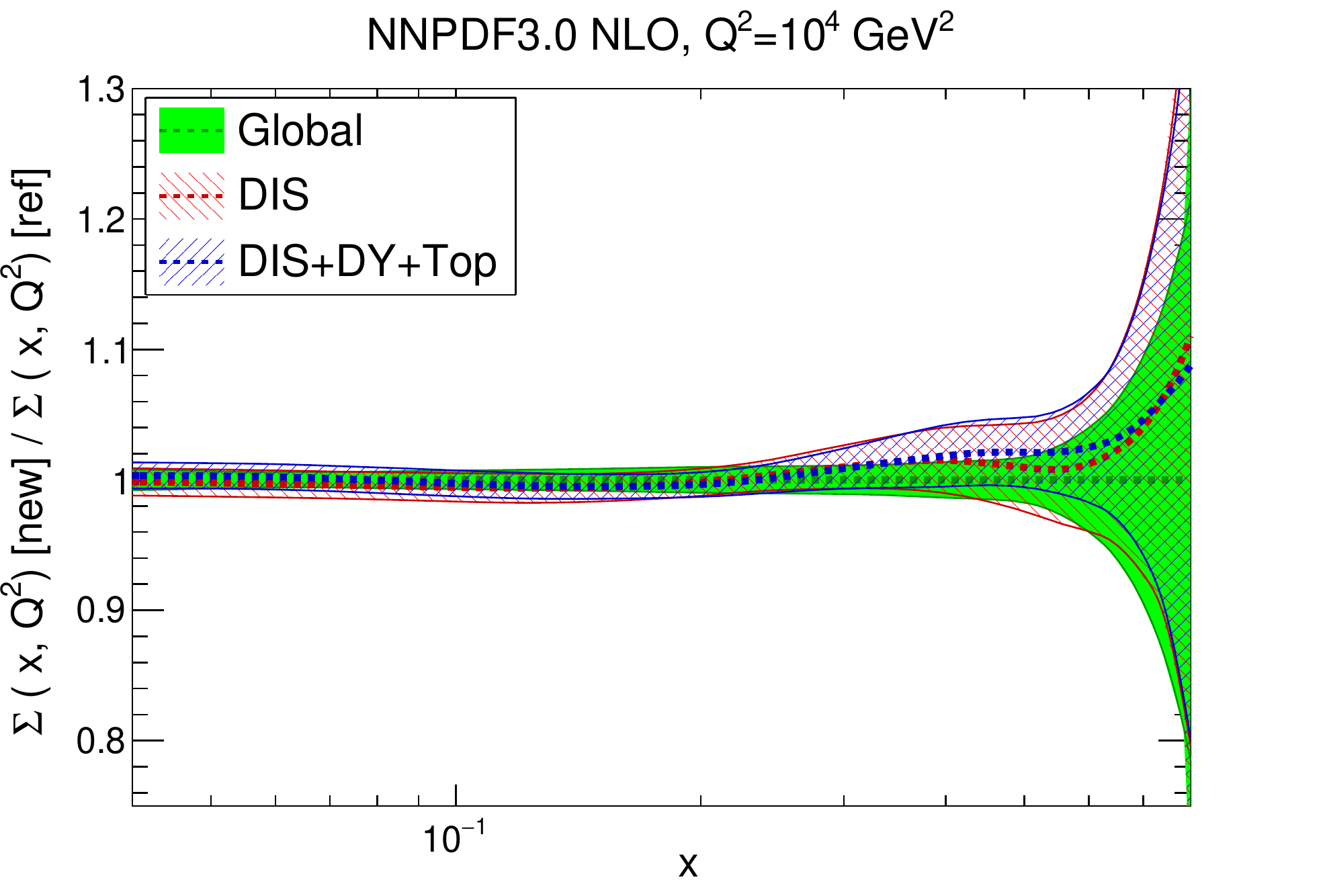}
\includegraphics[width=0.49\textwidth]{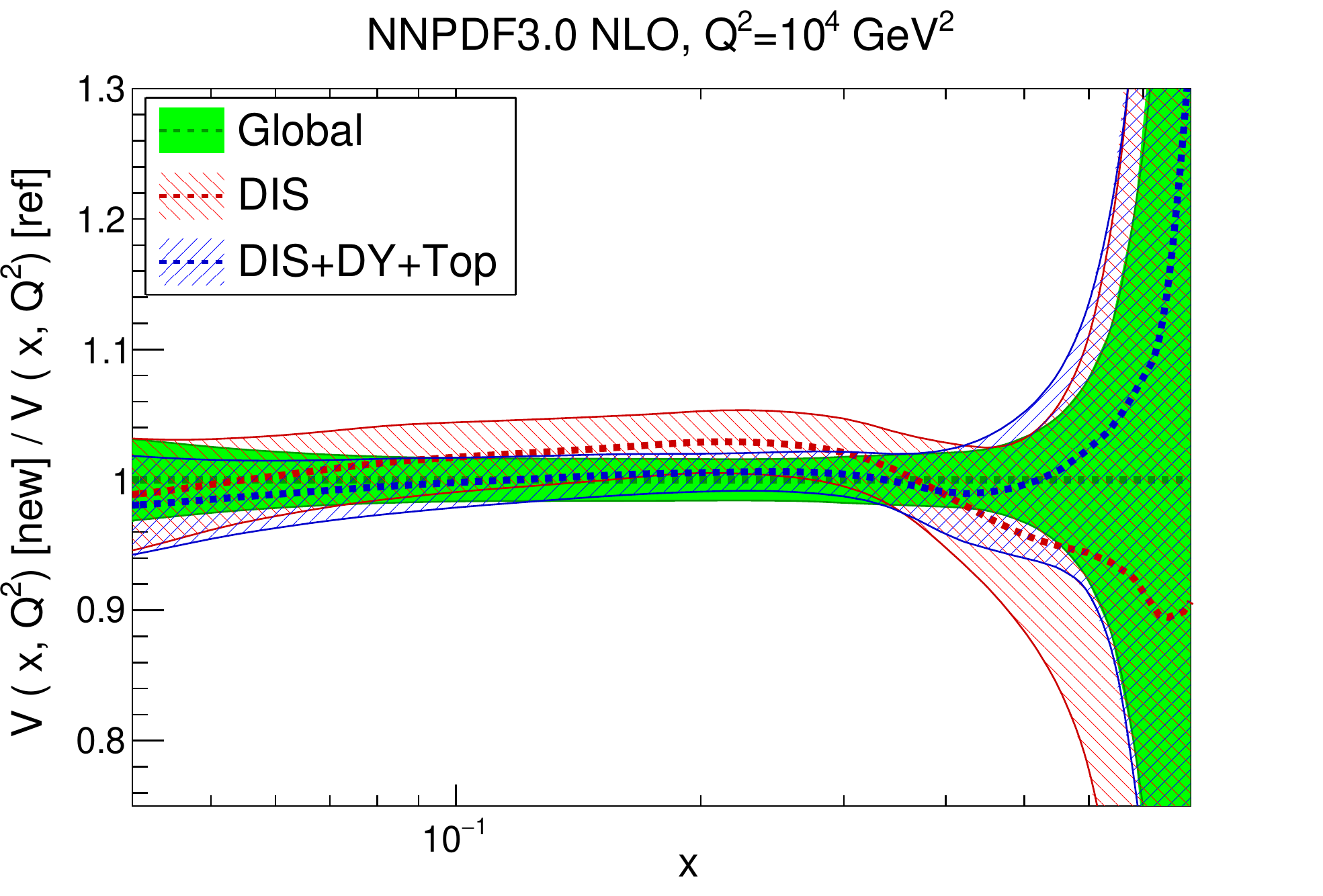}
\includegraphics[width=0.49\textwidth]{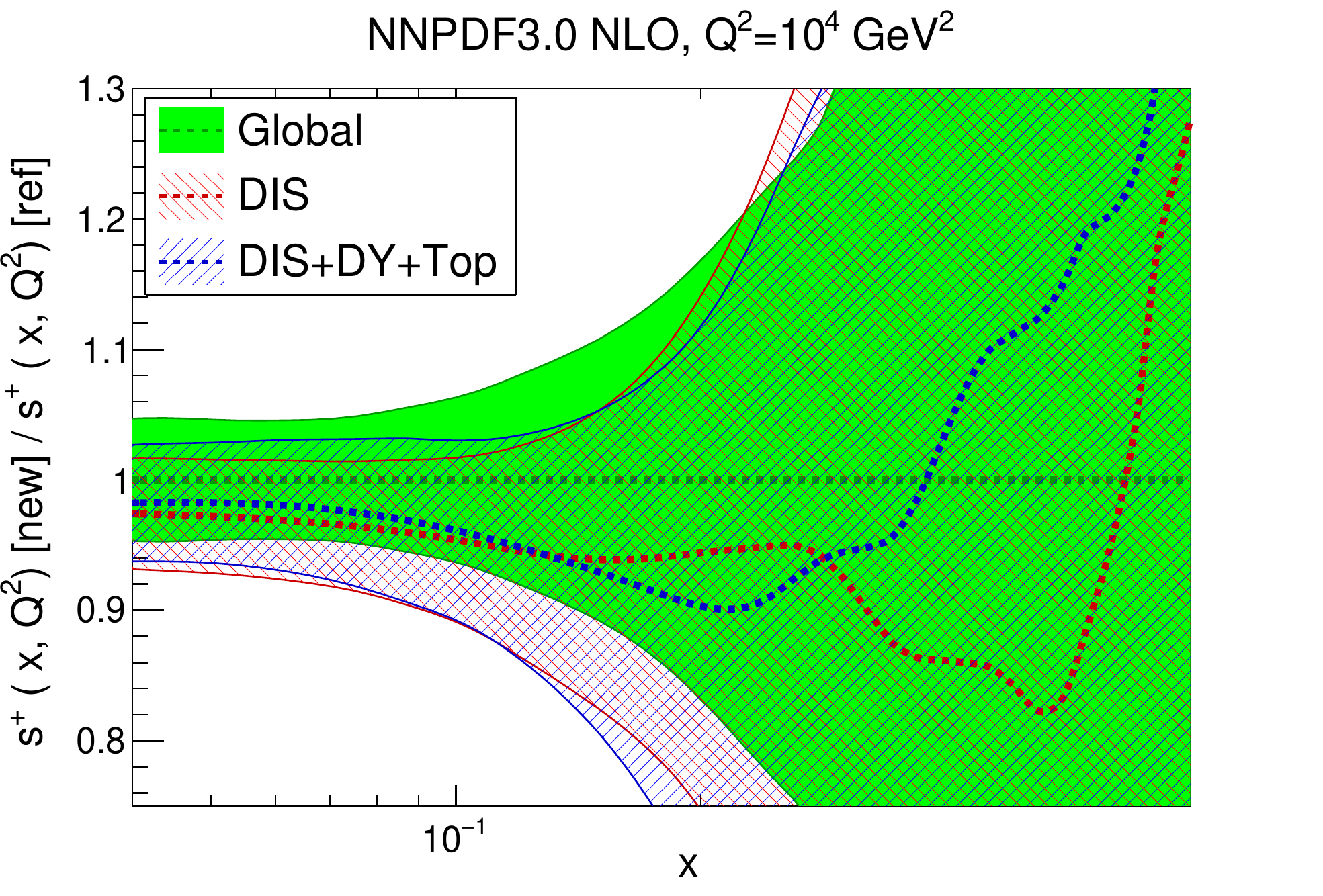}
\end{center}
\vspace{-0.3cm}
\caption{\small Comparison of the fixed-order NNPDF3.0 NLO
  fits based on different datasets: global, DIS-only and DIS+DY+top,
  for  $\as(m_Z^2)=0.118$, at a typical LHC scale
 of $Q^2=10^4$ GeV$^2$.
 Results are normalised to the central prediction of the
 NNPDF3.0 NLO global fit.
From left to right and from top to bottom, we show the gluon, the total
quark singlet, the total valence PDF and the total strangeness.
}
\label{fig:dis-vs-global-nlo-highQ}
\end{figure}
%%%%%%%%%%%%%%%%%%%%%%%

First of all, we compare the baseline fixed-order fits with the 
NNPDF3.0 global sets.
As mentioned in Sect.~\ref{sec:fitsettings}, in the
fixed-order  baseline fits all settings are identical
to those of  NNPDF3.0 with the only difference of
the use of a reduced dataset, see
Table~\ref{tab:completedataset}.
Therefore, we expect the two fits to be  consistent, with the
baseline fit affected by larger PDF uncertainties due
to the reduced dataset.
For simplicity, we restrict these comparisons to NLO, since the impact
of the reduced dataset is roughly independent of the perturbative order.

We have produced two baseline fits: one with
the all the data marked in the last column of Table~\ref{tab:completedataset},
and the other with only DIS-data included.
In Fig.~\ref{fig:dis-vs-global-nlo-highQ} we compare
  the NNPDF3.0 NLO DIS-only and DIS+DY+top set with $\as(m_Z^2)=0.118$, with the
  corresponding global set.\footnote{In the rest of this section, we concentrate
    only on the large-$x$ region of the PDFs, since as we will show the effects of
  threshold resummation are negligible at medium and small-$x$.}
In both cases we use  $N_{\rm rep}=100$ replicas, and the
comparison is performed at a typical LHC scale $Q=100$~GeV.
Results are normalised to the central prediction of the global fit.
From left to right and from top to bottom, we show the gluon, the total
quark singlet, the total valence PDF and the total strangeness.

As we can see, there is a reasonable agreement for most of the PDF
flavours and 
of momentum fraction $x$ between the three fits,
with as expected larger PDF uncertainties
  in the DIS-only and DIS+DY+top fits due to the reduced dataset.
  The DIS-only fit is relatively close to the global fit for the large-$x$ quarks,
  since these are well constrained by the DIS fixed-target data.
  On the other hand,
  the DIS-only fit is affected by rather larger uncertainties as compared
  to the global fit for the gluon (due to the missing jet data) and for
  the total valence (due to the missing Drell-Yan data that constrains
  flavor separation).
  In any case, the DIS-only and the global fit are always consistent at the
  one-sigma level.

  Concerning the DIS+DY+top fit, for quark PDFs (singlet, valence
  and strangeness) the
  results of the DIS+DY+top fit are quite
  close to the global fit.
Therefore, we can conclude that for quark-initiated processes, calculations done
with the DIS+DY+top fits are essentially equivalent to those performed using the
global PDFs.
The only differences are as expected related to the gluon PDF, where the missing
inclusive jet data cause a substantial increase in the large-$x$ gluon
PDF uncertainties compared to the global fit.
Note however that in the resummed DIS+DY+top fit a handle on the large-$x$ gluon
is still provided by the total top-quark pair production
cross section~\cite{Czakon:2013tha,Beneke:2012wb}.

After having established the impact of the reduced datasets on the baseline
fits that will be used for the resummation, in the following we concentrate
in quantifying the impact of resummation for fits based
on a common dataset, first for the
DIS-only fits, and then for the DIS+DY+top fits.

\subsection{DIS-only resummed PDFs}

Now we present the results of the resummed fits.
We begin with the DIS-only fits,
and compare the baseline NLO and NNLO fixed-order with the
corresponding NLO+NLL and NNLO+NNLL threshold resummed
fits.
%

%%%%%%%%%%%%%%%%%%%%%%%%%%%%%%%%%%%%%%%%%%%%%%%%%%%%%%%%%%%%%%%%%
%%%%%%%%%%%%%%%%%%%%%%%%%%%%%%%%%%%%%%%%%%%%%%%%%%%%%%%%%%%%%%%%%
\begin{table}[h]
\small
\begin{centering}
\begin{tabular}{|c|C{1.8cm}|C{1.8cm}|C{1.8cm}|c|}
\hline
  Experiment & 
    \multicolumn{4}{c|}{NNPDF3.0 DIS-only} \\
    \hline
    & NLO & NNLO & NLO+NLL & NNLO+NNLL  \\
    \hline
\hline
NMC &  1.36   & 1.39  & 1.36  &  1.32  \\
SLAC &  1.12  & 1.15  & 1.02  &  1.04  \\
BCDMS & 1.19  & 1.20  & 1.21  &  1.22   \\
CHORUS & 1.10 & 1.05  & 1.09  &  1.07  \\
  NuTeV & 0.52 & 0.46  & 0.55  &  0.51  \\
\hline
  HERA-I & 1.07 & 1.13 & 1.06  &  1.07  \\
ZEUS HERA-II & 1.40 & 1.42 & 1.42 & 1.43    \\
H1 HERA-II & 1.67 & 1.79 & 1.68   &  1.74  \\
HERA charm & 1.28 & 1.29 &  1.29  &  1.24  \\
\hline
\hline
Total   &   1.237  &  1.257    &  1.237  & 1.242    \\
\hline
\end{tabular}
\par\end{centering}
\caption{\small The  $\chi^2$ per data point
  for all experiments
  included in the DIS-only threshold resummed fits, at NLO and
  NNLO, compared with their resummed counterparts.}
\label{tab:chi2disfit}
\end{table}
%%%%%%%%%%%%%%%%%%%%%%%%%%%%%%%%%%%%%%%%%%%%%%%%%%%%%%%%%%%%%%%%%%%%%%%%%
%%%%%%%%%%%%%%%%%%%%%%%%%%%%%%%%%%%%%%%%%%%%%%%%%%%%%%%%%%%%%%%%%%%%%%%%%

In Table~\ref{tab:chi2disfit} we provide the
$\chi^2$ per data point for all experiments
included in the DIS-only threshold resummed fits, at NLO+NLL and
NNLO+NNLL, to be compared with their unresummed counterparts.\footnote{
  As in NNPDF3.0, the present fits used the $t_0$ definition of
  the covariance matrix for the $\chi^2$ minimisation, but then
  use the experimental definition to assess the consistency
  between theory and data.
See~\cite{Ball:2012wy,Ball:2009qv} for the explanation of the different
definitions of the $\chi^2$ estimators.
}
From Table~\ref{tab:chi2disfit} we see that, as expected, the impact of resummation
is moderate and restricted to the fixed-target DIS experiments. 
In the case of SLAC, there is a clear improvement in the $\chi^2$ due to the inclusion
of threshold resummation, both at NLO and at NNLO.
For other experiments the change in $\chi^2$ is not significant, meaning that the
small effect of threshold resummation can be
absorbed in the fitted PDFs.
The total $\chi^2$ is slightly improved when going from the NNLO to the NNLO+NNLL
fit, while it is essentially unaffected in the NLO+NLL case.

%%%%%%%%%%%%%%%%%%%
\begin{figure}[t]
\begin{center}
\includegraphics[width=0.49\textwidth]{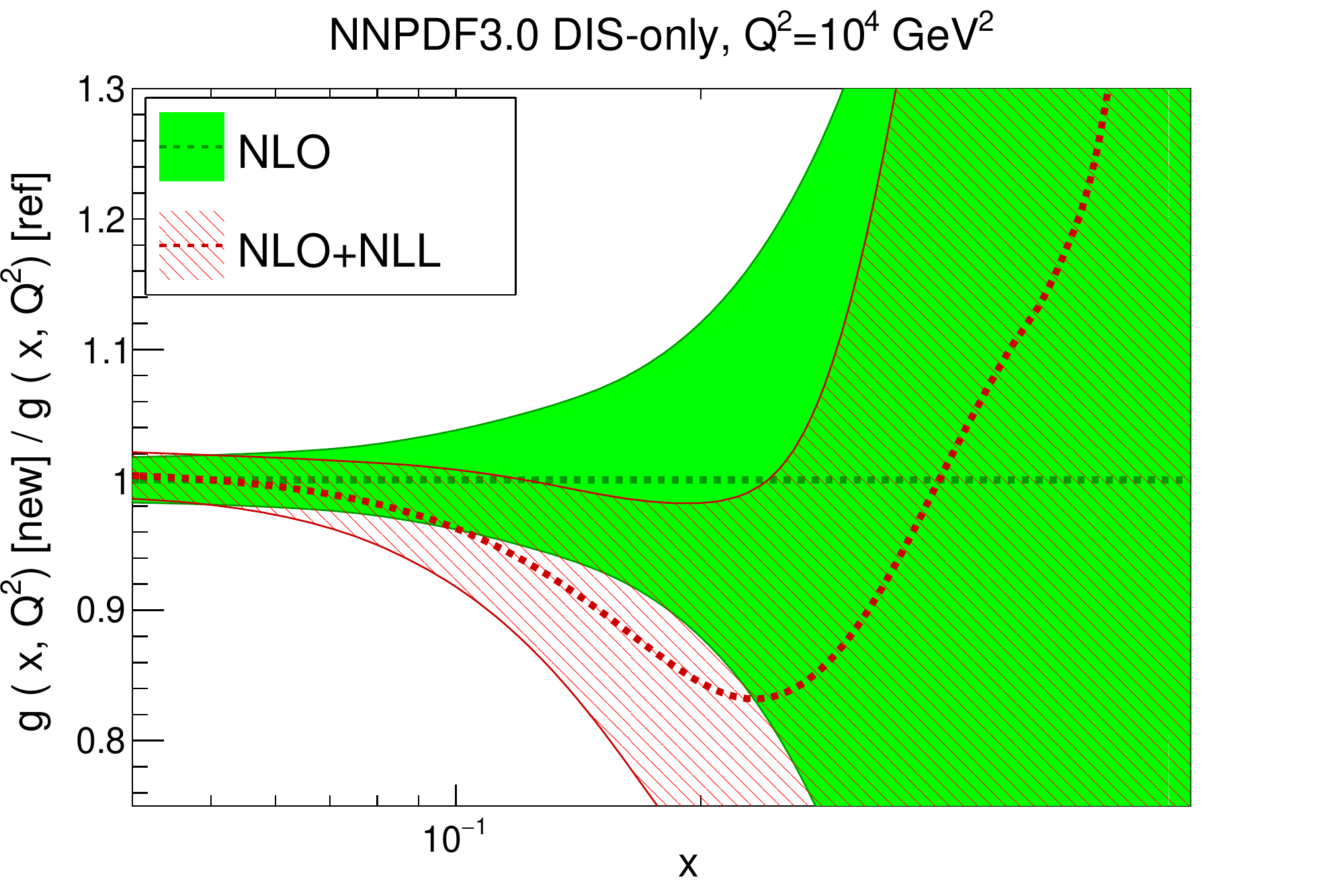}
\includegraphics[width=0.49\textwidth]{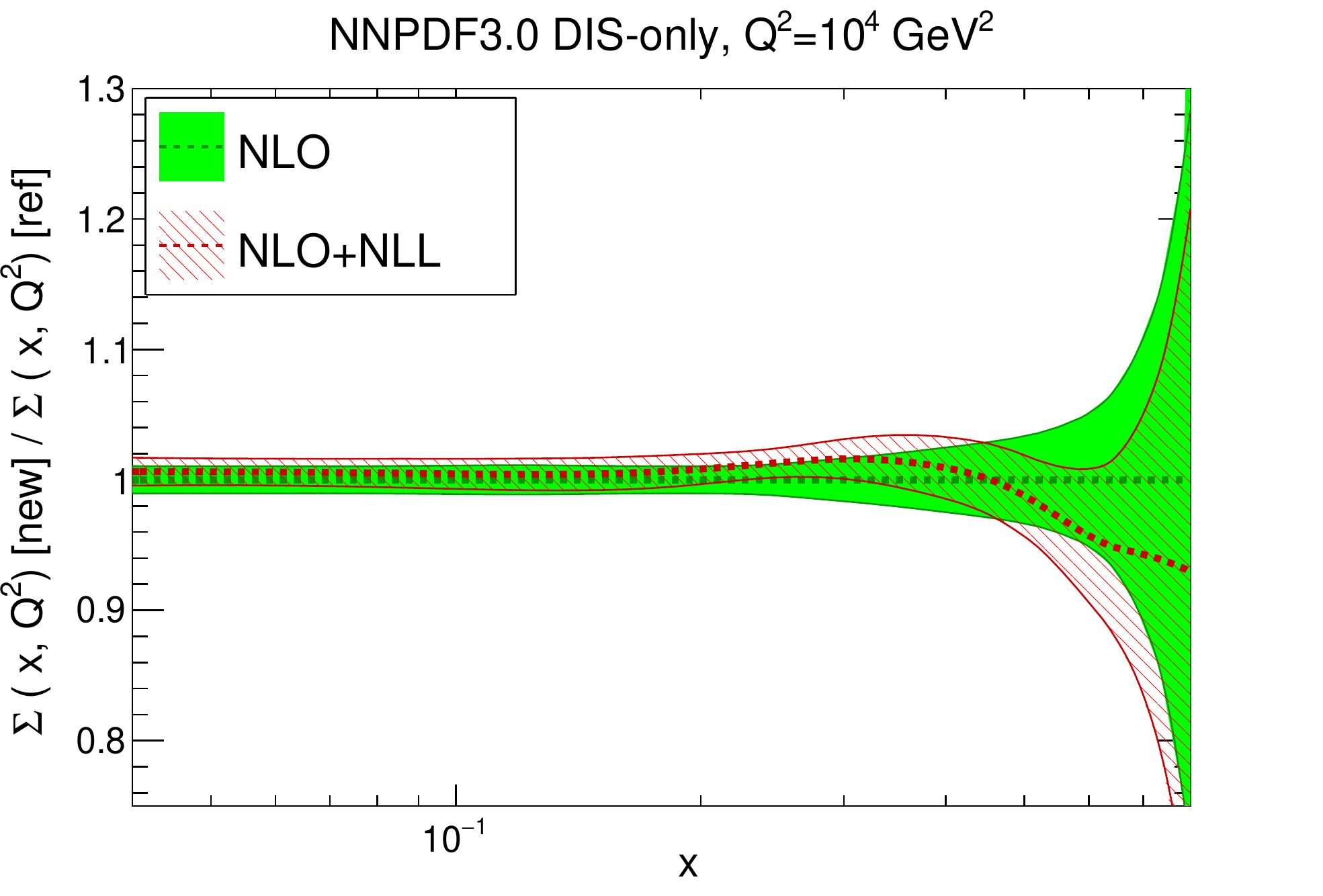}
\includegraphics[width=0.49\textwidth]{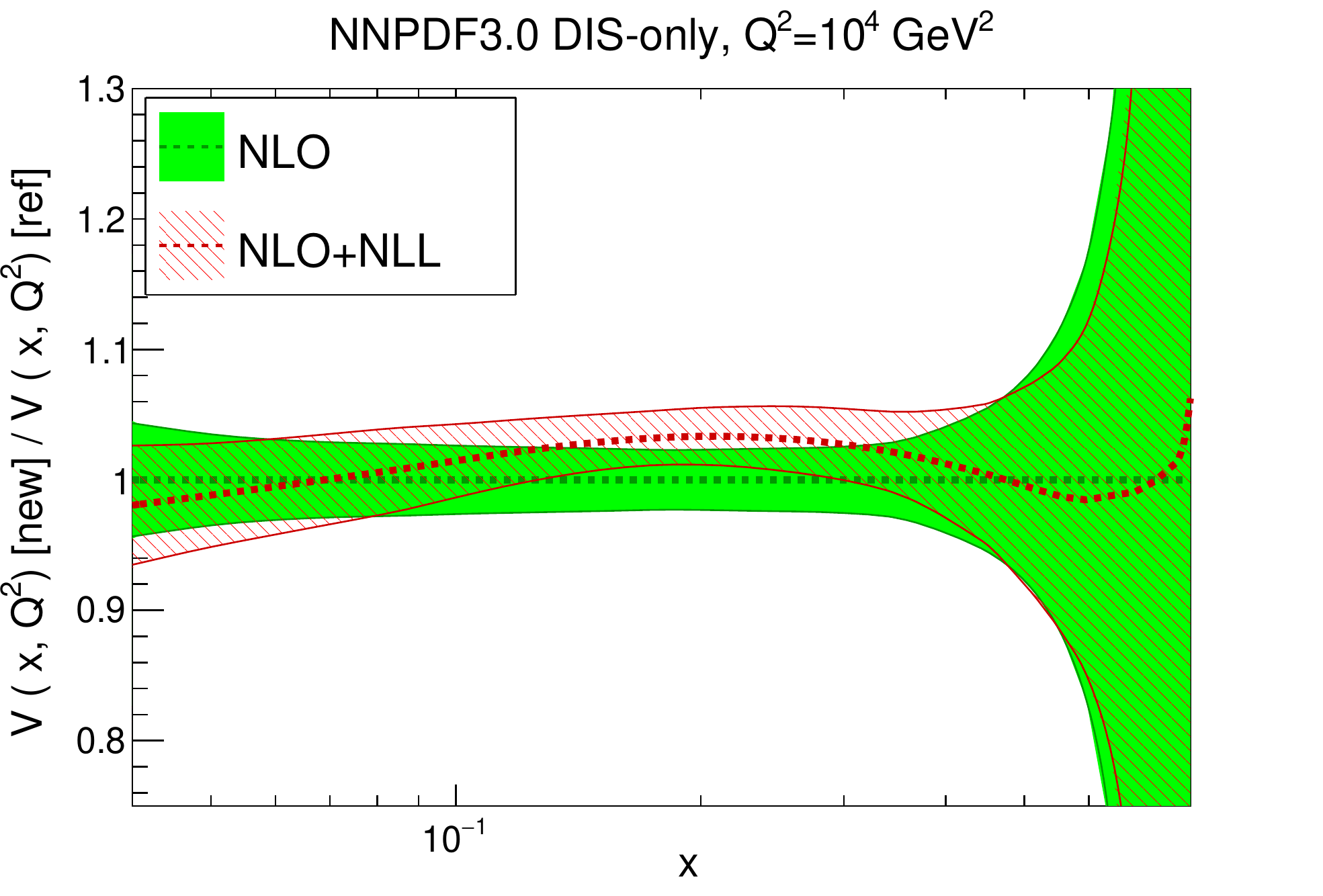}
\includegraphics[width=0.49\textwidth]{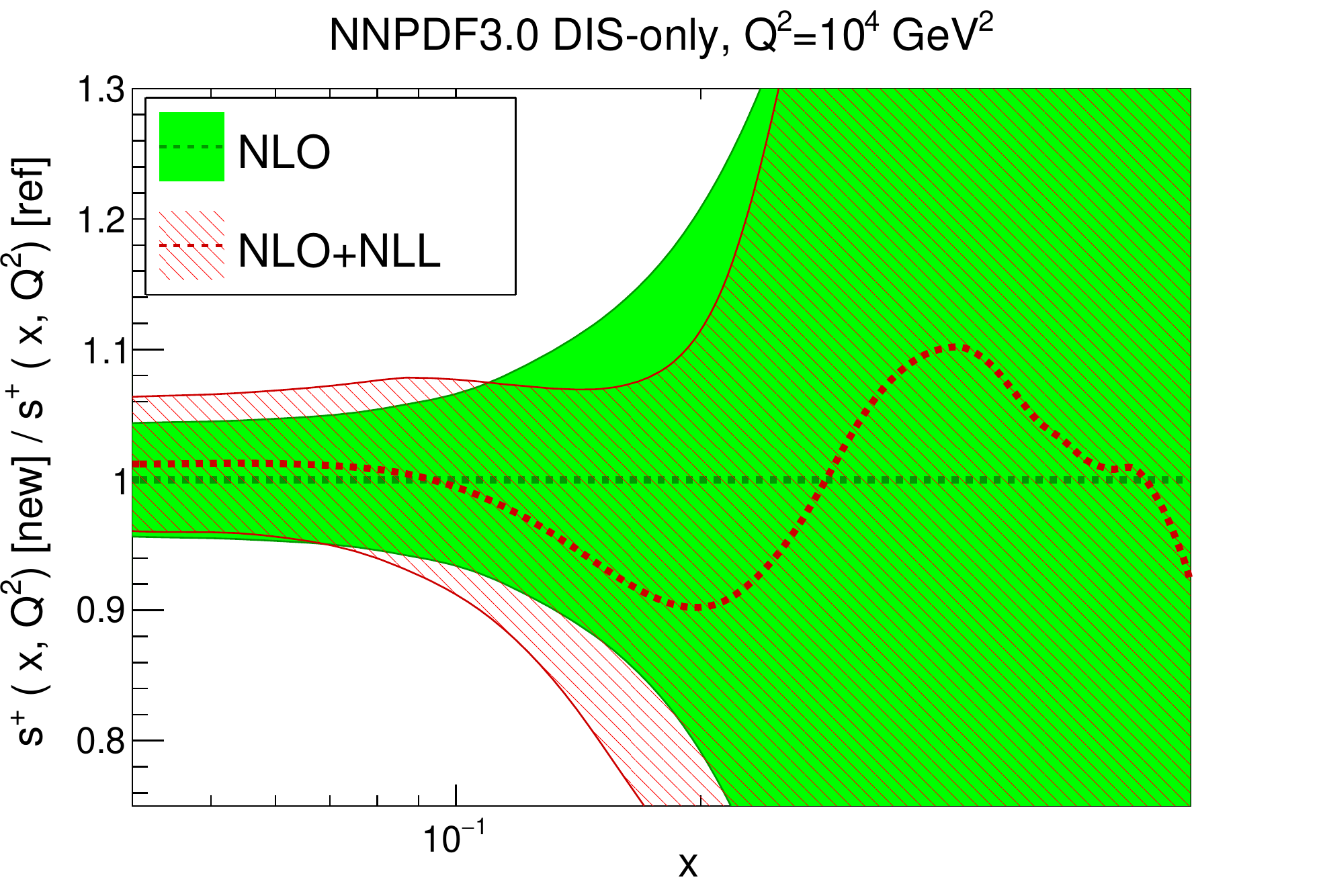}
\end{center}
\vspace{-0.3cm}
\caption{\small
Comparison between the NNPDF3.0 DIS-only NLO fit
and the corresponding NLO+NLL fit,
for  $\as(m_Z^2)=0.118$, at a typical LHC scale
of $Q^2=10^4$ GeV$^2$.
PDFs are normalised to the central value of the fixed-order
fit.
}
\label{fig:dis-nlo-vs-nll-highQ}
\end{figure}
%%%%%%%%%%%%%%%%%%%%%%%

Now we turn to study the impact of the resummation on the PDFs themselves.
In Fig.~\ref{fig:dis-nlo-vs-nll-highQ} (for the NLO) and in
Fig.~\ref{fig:dis-nnlo-vs-nnll-highQ} (for the NNLO) we compare
the  NNPDF3.0 DIS-only (N)NLO set with $\as(m_Z^2)=0.118$, with the
  corresponding (N)NLO+(N)NLL threshold resummed PDFs, respectively.

First of all, we note that as expected the inclusion of threshold resummation
affects only PDFs at large $x$, for $x \ge 0.1$, which is consistent
with the modifications that resummation induces on the DIS structure functions.
We also see that the impact on the PDFs is more important at NLL than at NNLL,
again as expected since NLL captures part of the
NNLO corrections to the
DIS structure functions.
For the quark PDFs, the effect of resummation is a suppression of the central values
for quite large $x$.
For example, for the NLO+NLL fit, the total quark singlet $\Sigma(x,Q^2)$
is suppressed by $\sim$ 5\% at $x\sim 0.6$.
One also observes a small enhancement of the valence PDF for $x\sim 0.2$, 
presumably due to a compensation for the suppression at very large $x$ through the valence sum rules.
Therefore, we expect resummation to have phenomenological impact for the
calculation of quark-initiated heavy production processes in BSM scenarios, which probe rather
large values of $x$.
The large-$x$ gluon is also suppressed at large-$x$, though PDF
uncertainties are too large in a DIS-only fit to make
this suppression significant.

%%%%%%%%%%%%%%%%%%%
\begin{figure}[t]
\begin{center}
\includegraphics[width=0.49\textwidth]{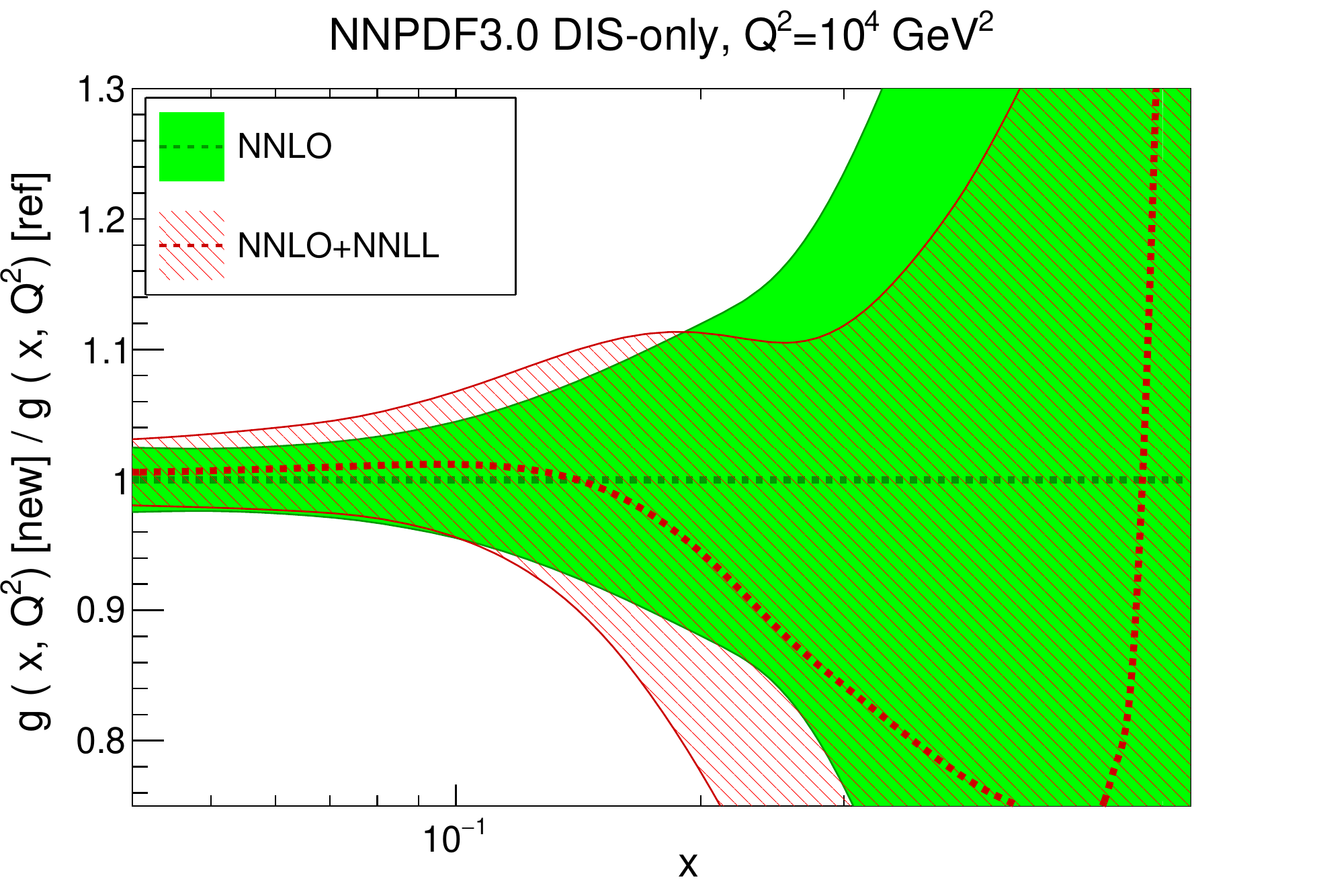}
\includegraphics[width=0.49\textwidth]{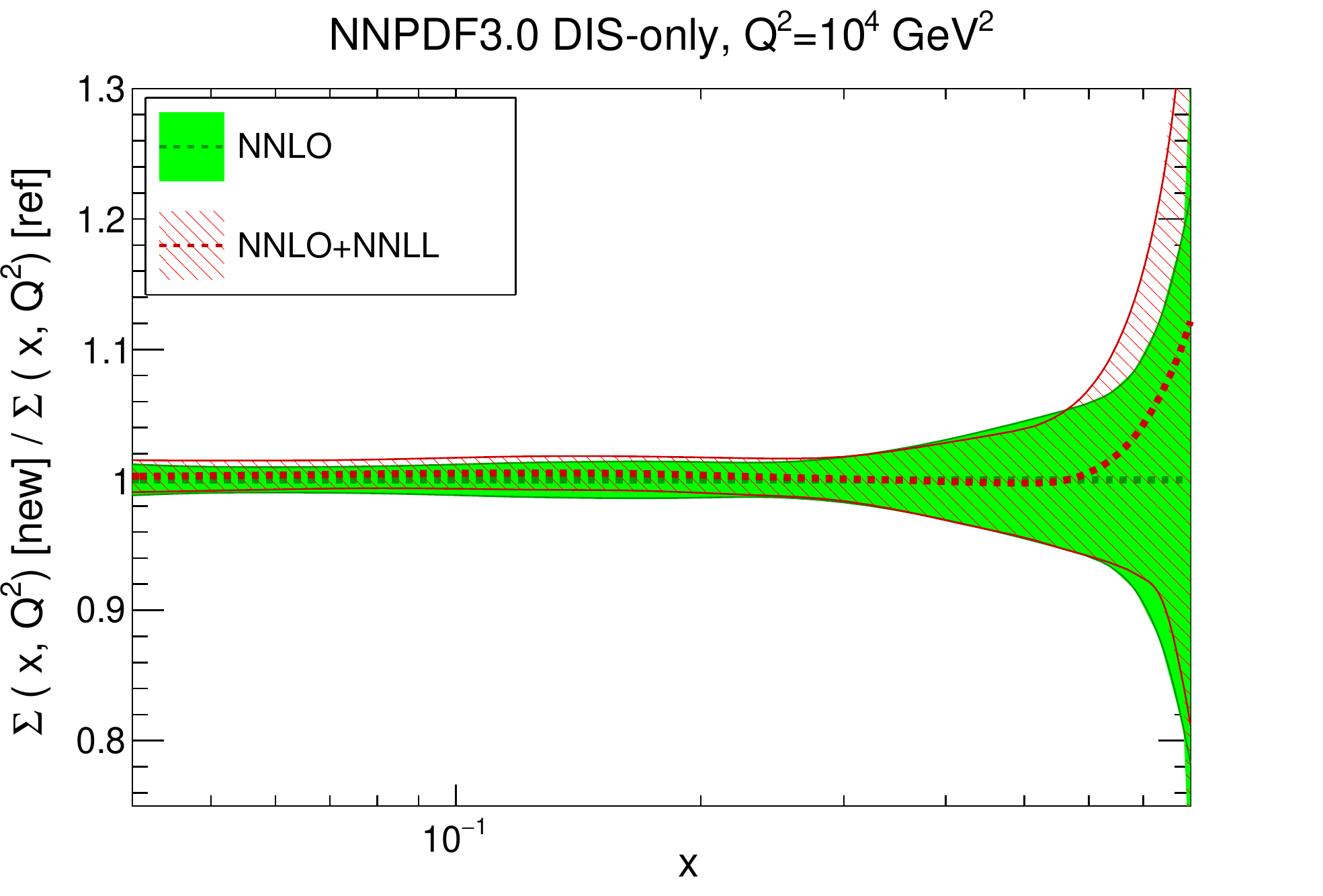}
\includegraphics[width=0.49\textwidth]{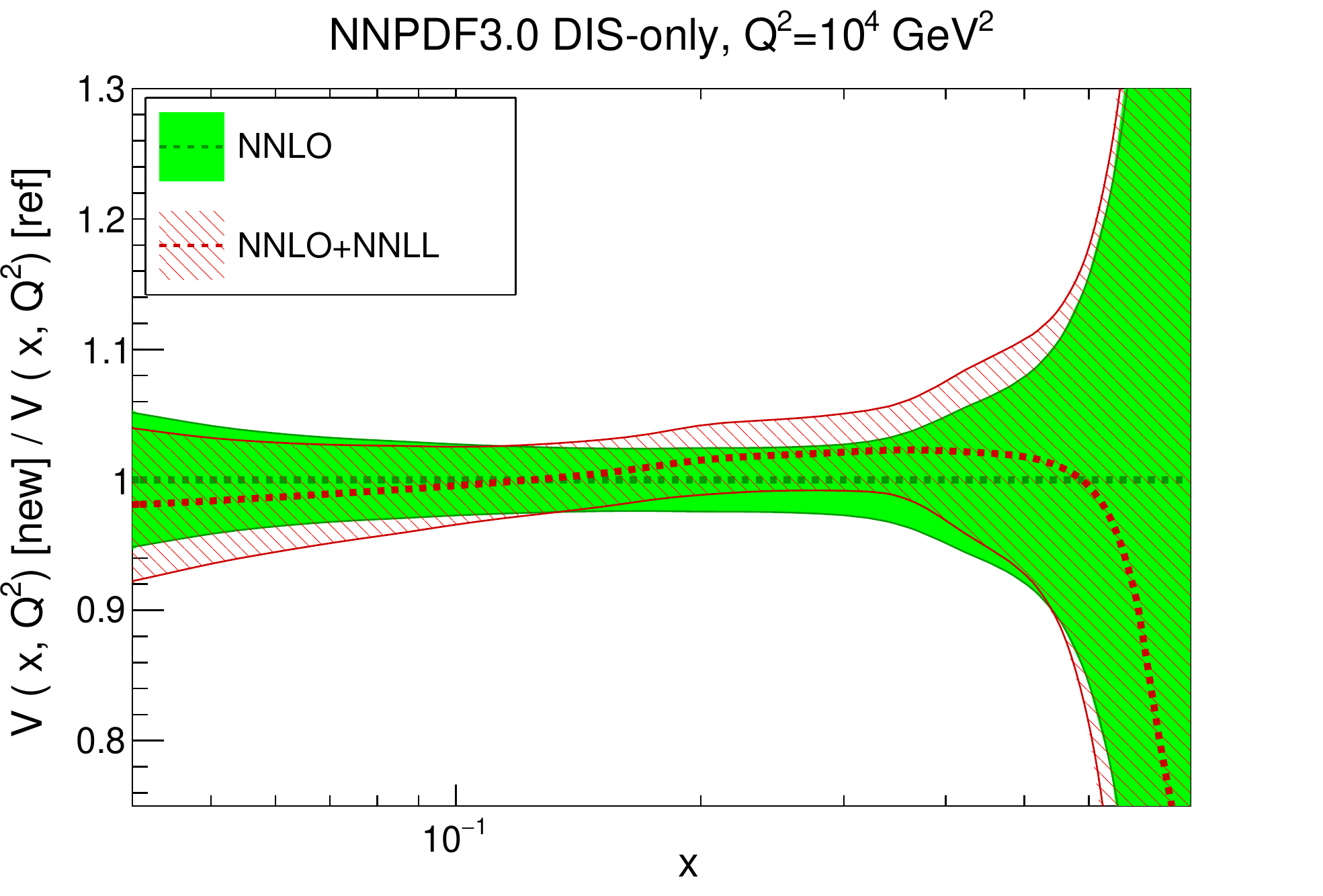}
\includegraphics[width=0.49\textwidth]{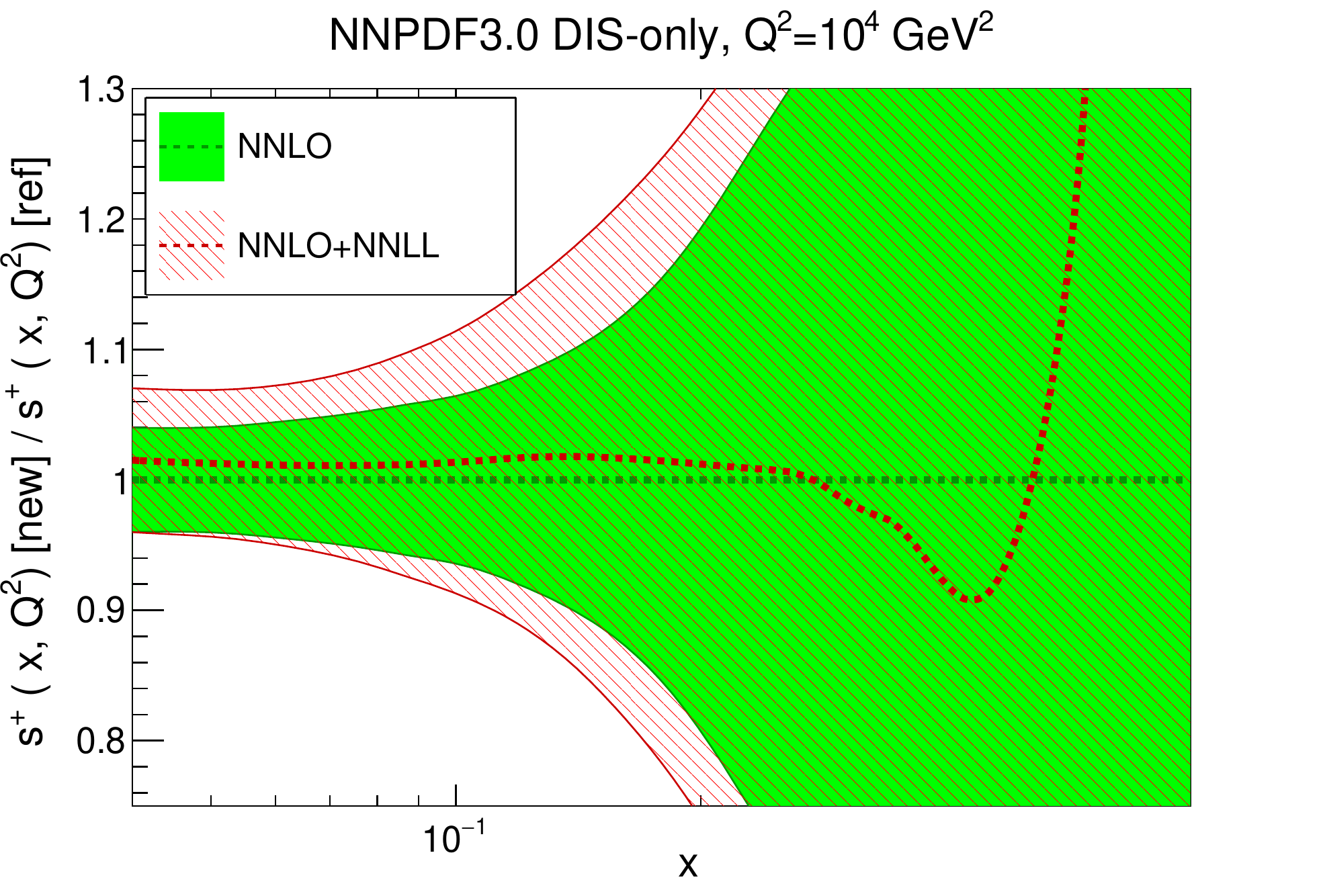}
\end{center}
\vspace{-0.3cm}
\caption{\small
  Same as Fig.~\ref{fig:dis-nlo-vs-nll-highQ} now comparing the NNLO
  DIS-only fit with the corresponding NNLO+NNLL fit.
 }
\label{fig:dis-nnlo-vs-nnll-highQ}
\end{figure}
%%%%%%%%%%%%%%%%%%%%%%%

\subsection{DIS+DY+top resummed PDFs}

Now we turn our discussion to the impact of threshold resummation
in the case of the
fits based on the DIS+DY+top dataset.
In Table~\ref{tab:chi2disdyfit} we provide the
$\chi^2$, again using the
experimental definition,
for all experiments
included in the DIS+DY+top threshold resummed fits, at NLO+NLL and
NNLO+NNLL, compared with their fixed-order counterparts.
%

%%%%%%%%%%%%%%%%%%%%%%%%%%%%%%%%%%%%%%%%%%%%%%%%%%%%%%%%%%%%%%%%%
%%%%%%%%%%%%%%%%%%%%%%%%%%%%%%%%%%%%%%%%%%%%%%%%%%%%%%%%%%%%%%%%%
\begin{table}[h]
\footnotesize
\begin{centering}
\begin{tabular}{|c|C{1.9cm}|C{1.9cm}|C{1.9cm}|c|}
\hline
  Experiment & 
    \multicolumn{4}{c|}{NNPDF3.0 DIS+DY+top} \\
    \hline
    & NLO & NNLO &  NLO+NLL &  NNLO+NNLL \\
    \hline
    \hline
    NMC            &  1.39 & 1.34 &   1.36    &  1.30     \\
      SLAC         &  1.17 & 0.91 &   1.02    &  0.92     \\ 
BCDMS              &  1.20 & 1.25 &   1.23    &  1.28     \\ 
  CHORUS           &  1.13 & 1.11 &   1.10    &  1.09     \\ 
  NuTeV            &  0.52 & 0.52 &   0.54    &  0.44     \\ 
\hline
  HERA-I           &  1.05 & 1.06 &   1.06    &  1.06     \\
ZEUS HERA-II       &  1.42 & 1.46 &   1.45    &  1.48     \\
H1 HERA-II         & 1.70  & 1.79 &   1.70    &  1.78     \\
HERA charm         & 1.26  & 1.28 &   1.30    &  1.28     \\
\hline
 DY E866           & 1.08 & 1.39  &   1.68    &  1.68     \\
 DY E605           & 0.92 & 1.14  &   1.12    &  1.21     \\
\hline
 CDF $Z$ rap       & 1.21 & 1.38  &   1.10    &  1.33     \\
 D0 $Z$ rap        & 0.57 & 0.62  &   0.67    &  0.66     \\
 \hline
ATLAS $Z$ 2010     & 0.98 & 1.21  &   1.02    &  1.28     \\
ATLAS high-mass DY & 1.85 & 1.27  &   1.59    &  1.21     \\
       \hline
CMS 2D DY 2011     & 1.22 & 1.39  &   1.22    &  1.41      \\
   \hline
LHCb $Z$ rapidity  & 0.83 & 1.30  &   0.51    &  1.25      \\
\hline
ATLAS CMS top prod & 1.23 & 0.55  &   0.61    &  0.40      \\
\hline
\hline
Total              &  1.233  &  1.264   &  1.246  &   1.269    \\
\hline
\end{tabular}
\par\end{centering}
\caption{\small Same as Table~\ref{tab:chi2disfit} for
  the DIS+DY+top fits.}
\label{tab:chi2disdyfit}
\end{table}
%%%%%%%%%%%%%%%%%%%%%%%%%%%%%%%%%%%%%%%%%%%%%%%%%%%%%%%%%%%%%%%%%%%%%%%%%
%%%%%%%%%%%%%%%%%%%%%%%%%%%%%%%%%%%%%%%%%%%%%%%%%%%%%%%%%%%%%%%%%%%%%%%%%
%
As we can see from Table~\ref{tab:chi2disdyfit},
in the NLO+NLL fits the fit quality of most of the experiments is improved
as compared to the fixed-order NLO fit.
This is especially marked in the case of SLAC, as in the DIS-only fit, but also
for the CHORUS neutrino structure functions,
the CDF $Z$ rapidity distribution, ATLAS high-mass DY, LHCb $Z$ rapidity
and the top quark pair production.
The only exception is the fixed target Drell-Yan data, where
resummation makes the $\chi^2$ worse.
Note however that in the resummed fit the overall balance between experiments
in the global fit is modified compared to the fixed-order fit, so this does
not necessarily imply that resummation degrades the internal fit quality for
this particular observable.\footnote{We have checked
  that in a fit based only on HERA data and fixed-target Drell-Yan
  data, in both the NLO+NLL and NNLO+NNLL fits we get
  $\chi^2 \sim 1$ for the Drell-Yan data.
  Therefore, the deterioration of the $\chi^2$ of E866
  in the resummed fits can be attributed to tension
  with other datasets, rather than a failure of the resummation
  to correctly describe this dataset.
}
At the level of total $\chi^2$ we see that fixed-order and resummed fits
lead to essentially the same value, since in the resummed case the improvement
in some experiments is compensated by the deterioration of others.

Turning to the NNLO+NNLL fit results in Table~\ref{tab:chi2disdyfit}, we
see that now the effect of resummation is more moderate.
Effects are small, and in the case again for the fixed-target
Drell-Yan data, resummation deteriorates
the fit quality.
Interestingly, the $\chi^2$ for the LHCb $Z$ rapidity data, which, being in the forward region, probe rather
large values of $x$, improves substantially with the inclusion of resummation, even at NNLL.
Given the small differences at the $\chi^2$ level, we also expect smaller differences at the
PDF level, as in the case of the DIS-only NNLO+NNLL fit.

The comparison of the PDFs between the NLO and NLO+NLL DIS+DY+top fits is shown in
Fig.~\ref{fig:disdy-nlo-vs-nll-highQ}, and the corresponding comparison
between the NNLO and NNLO+NNLL fits is found in Fig.~\ref{fig:disdy-nnlo-vs-nnll-highQ}.
These can be compared with the corresponding DIS-only fits, see
Fig.~\ref{fig:dis-nlo-vs-nll-highQ} and Fig.~\ref{fig:dis-nnlo-vs-nnll-highQ}.
In the case of the NLO+NLL fit, the trend is similar to that of the DIS-only fit: softer quarks at very 
large $x$, and a corresponding enhancement of the valence distribution at medium $x$.
At small $x$, the effect of resummation is negligible as expected.
We note that both for the total quark singlet and for the total valence PDF the
effect of the resummation has become more significant than in the DIS-only fit, due
to the reduction of PDF uncertainties.
Indeed, for $\Sigma(x,Q^2)$ for example, the shift of the central value is about $\sim 5\%$ at $x\simeq 0.5$, and the uncertainty bands of the
two fits, although still overlapping, are clearly departing from each other.

%%%%%%%%%%%%%%%%%%%
\begin{figure}[t]
\begin{center}
\includegraphics[width=0.49\textwidth]{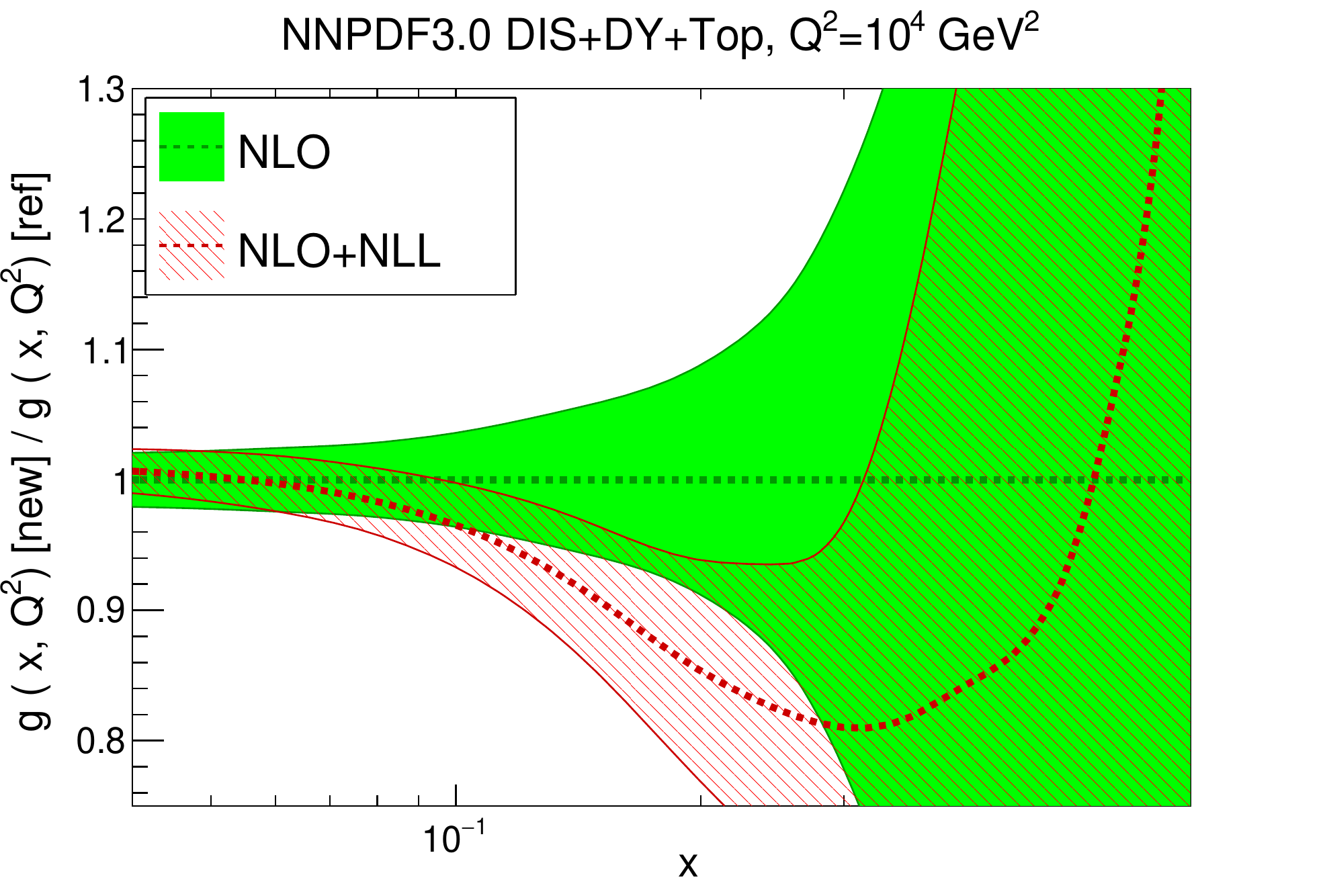}
\includegraphics[width=0.49\textwidth]{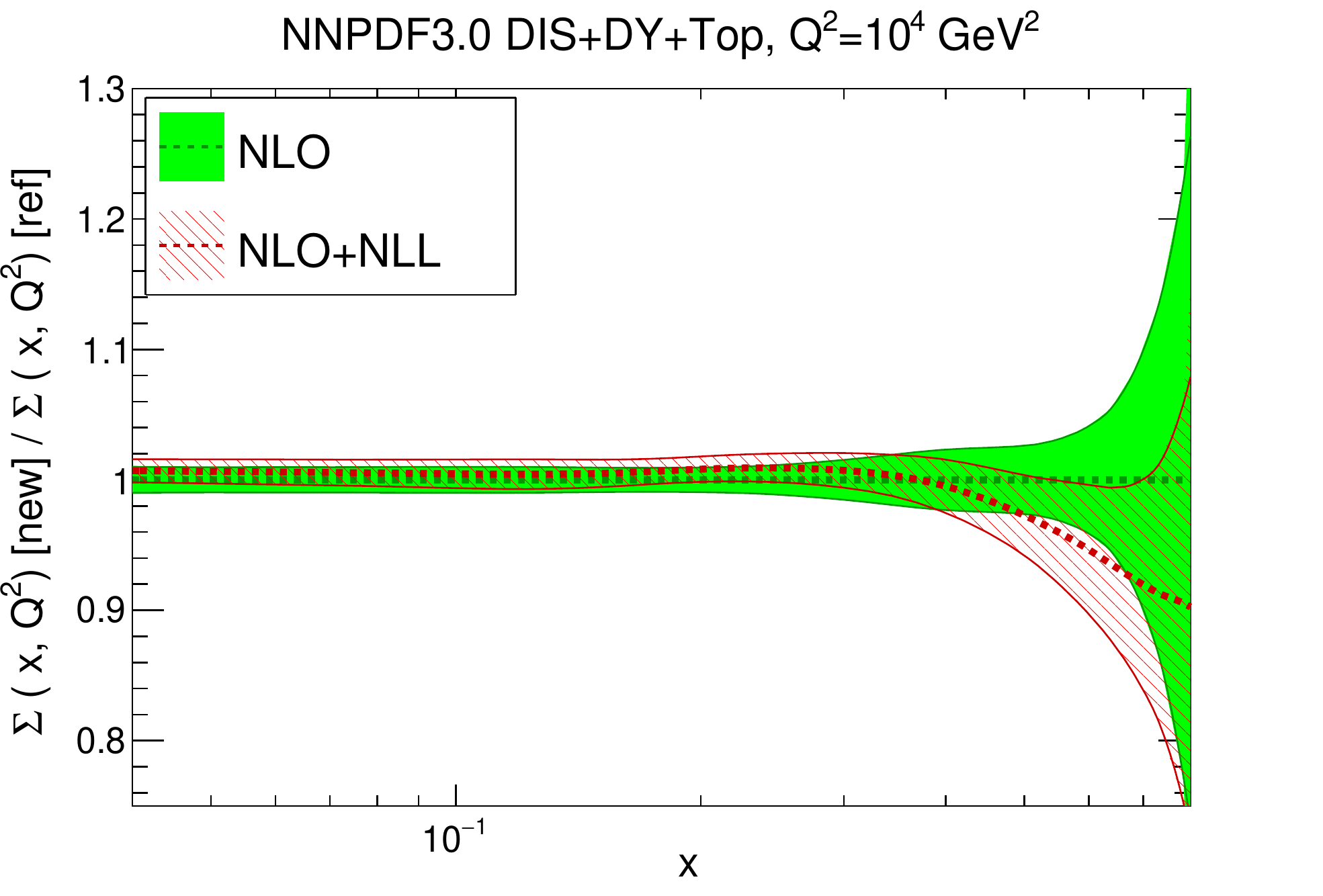}
\includegraphics[width=0.49\textwidth]{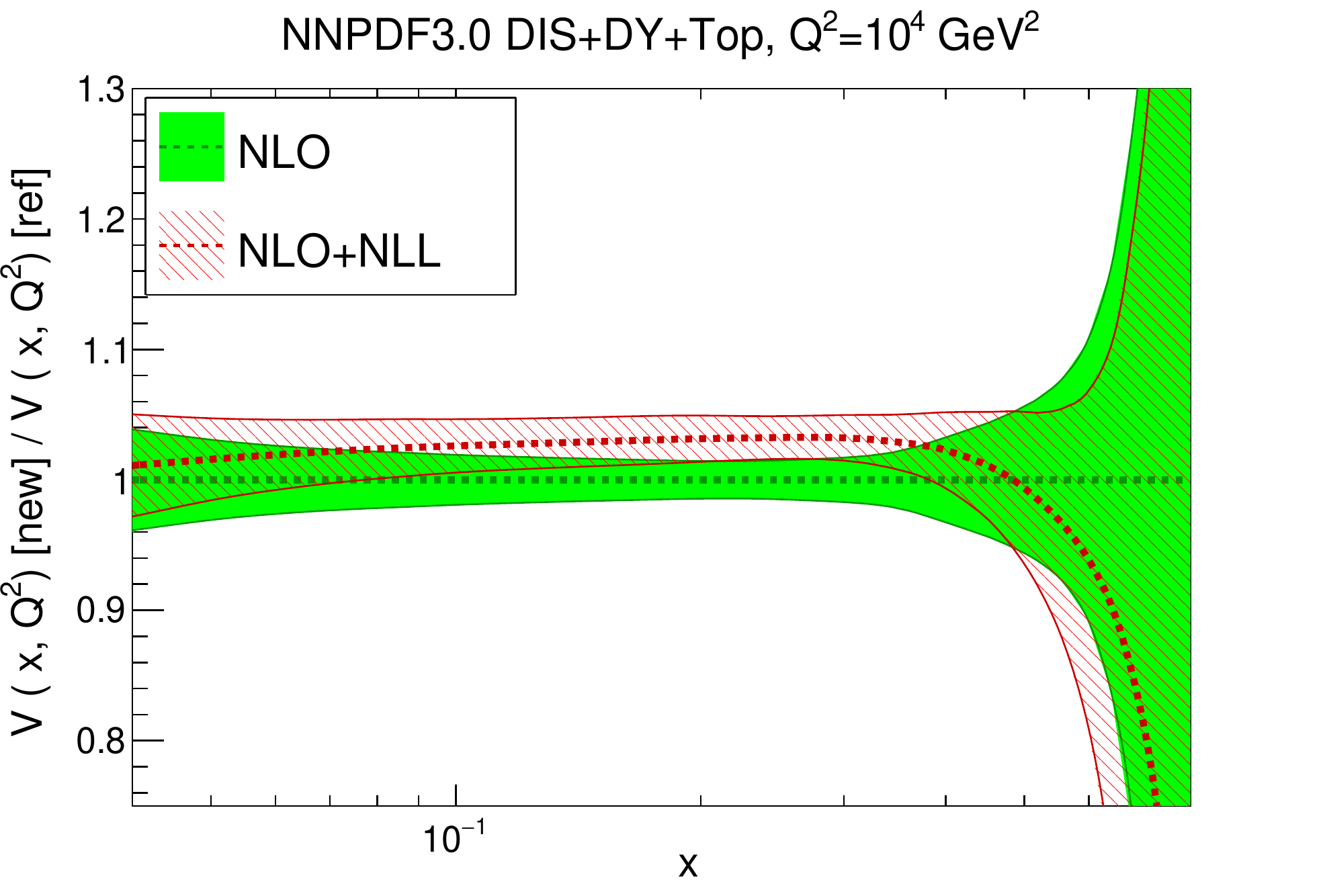}
\includegraphics[width=0.49\textwidth]{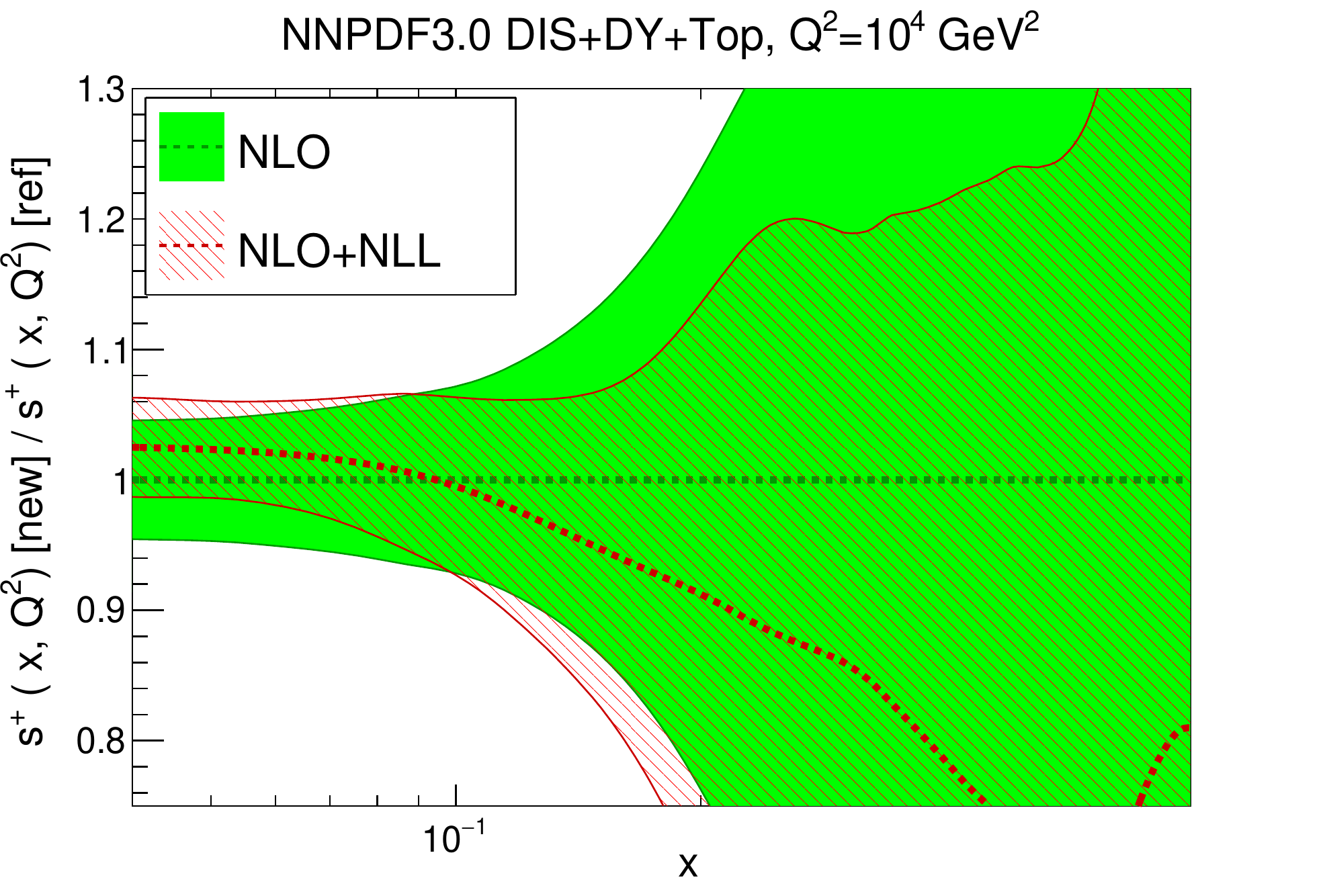}
\end{center}
\vspace{-0.3cm}
\caption{\small
Comparison between the NNPDF3.0 NLO DIS+DY+top fit
and the corresponding NLO+NLL fit, for  $\as(m_Z^2)=0.118$, at a typical LHC scale
 of $Q^2=10^4$ GeV$^2$.
}
\label{fig:disdy-nlo-vs-nll-highQ}
\end{figure}
%%%%%%%%%%%%%%%%%%%%%%%
%
%%%%%%%%%%%%%%%%%%%
\begin{figure}[t]
\begin{center}
\includegraphics[width=0.49\textwidth]{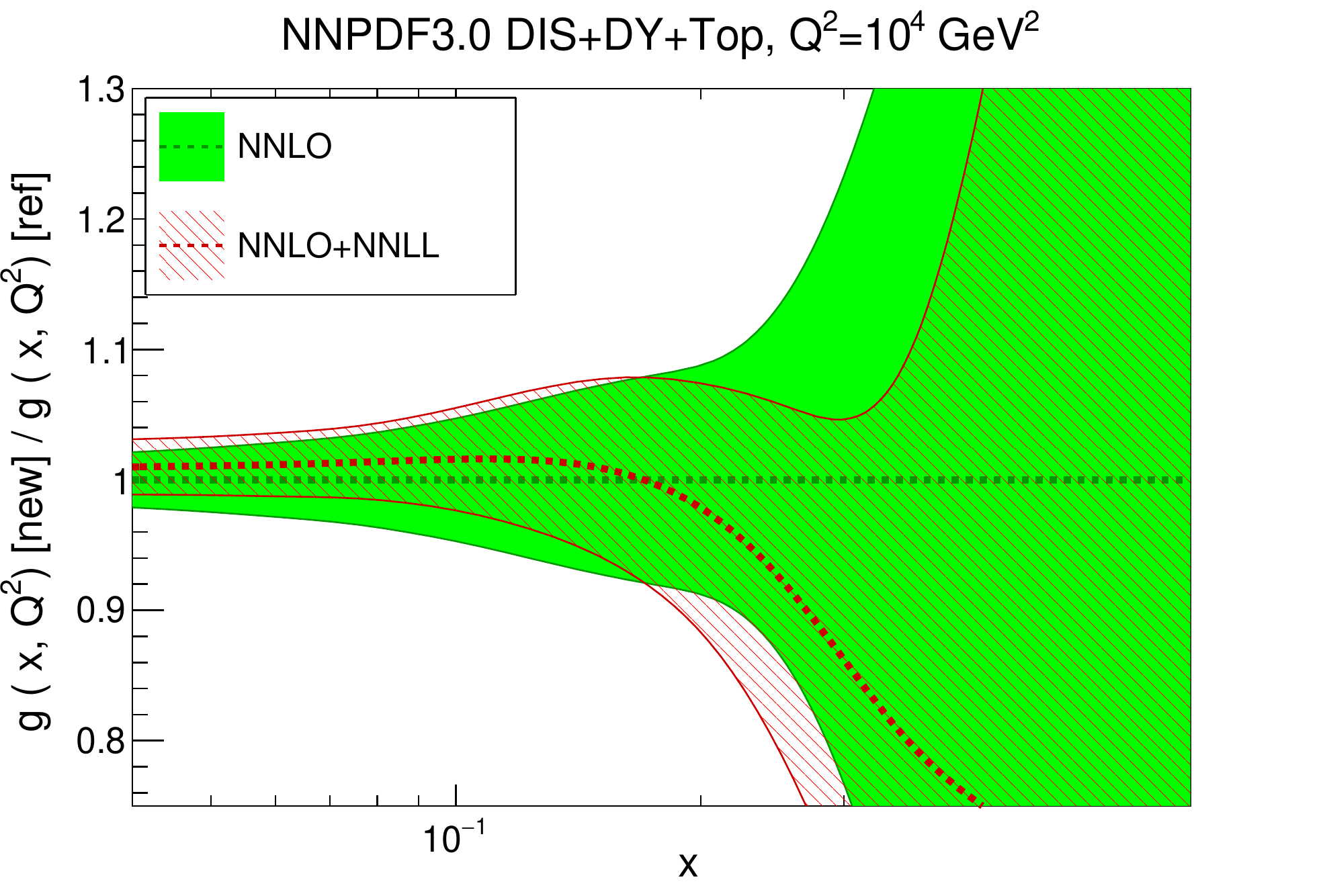}
\includegraphics[width=0.49\textwidth]{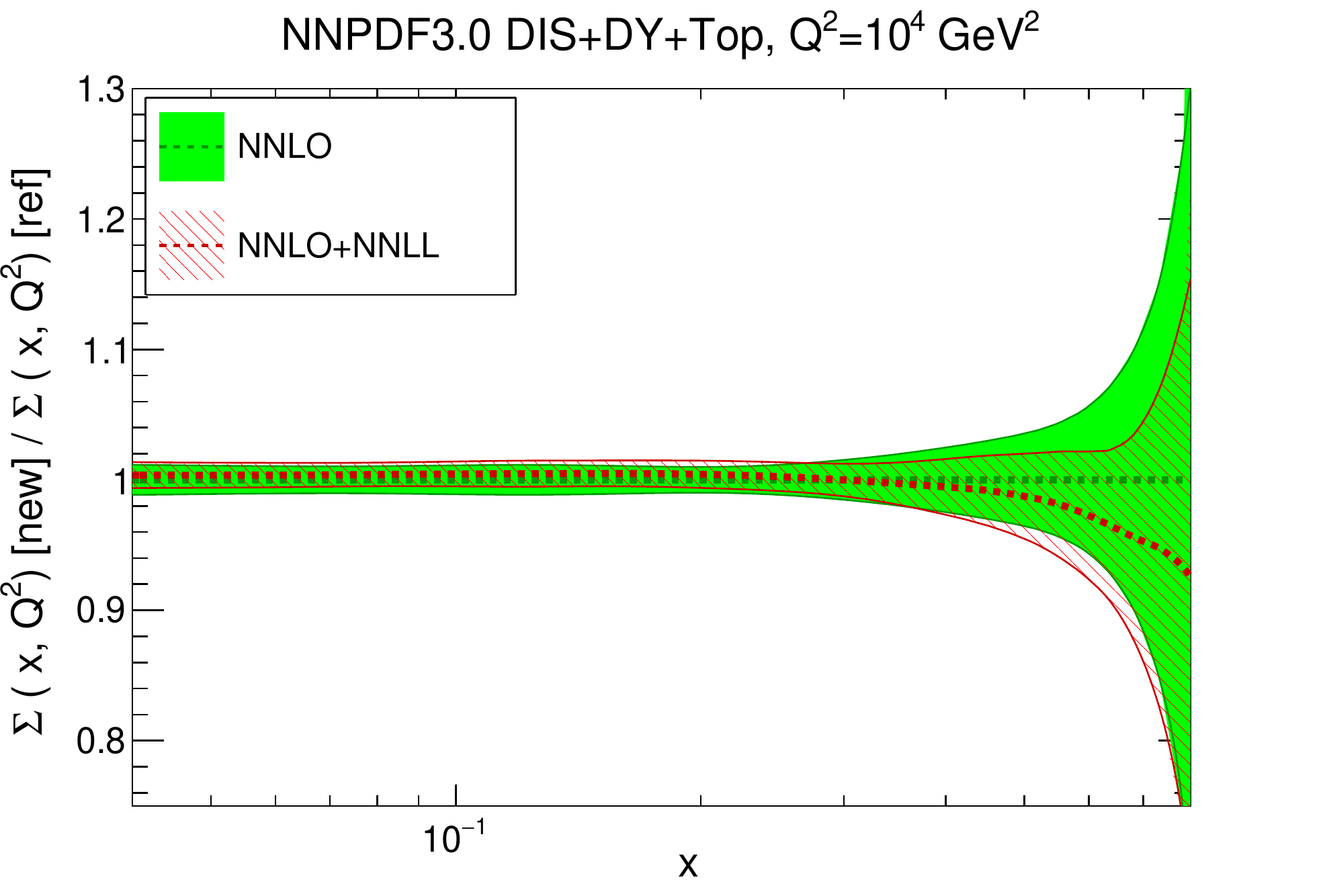}
\includegraphics[width=0.49\textwidth]{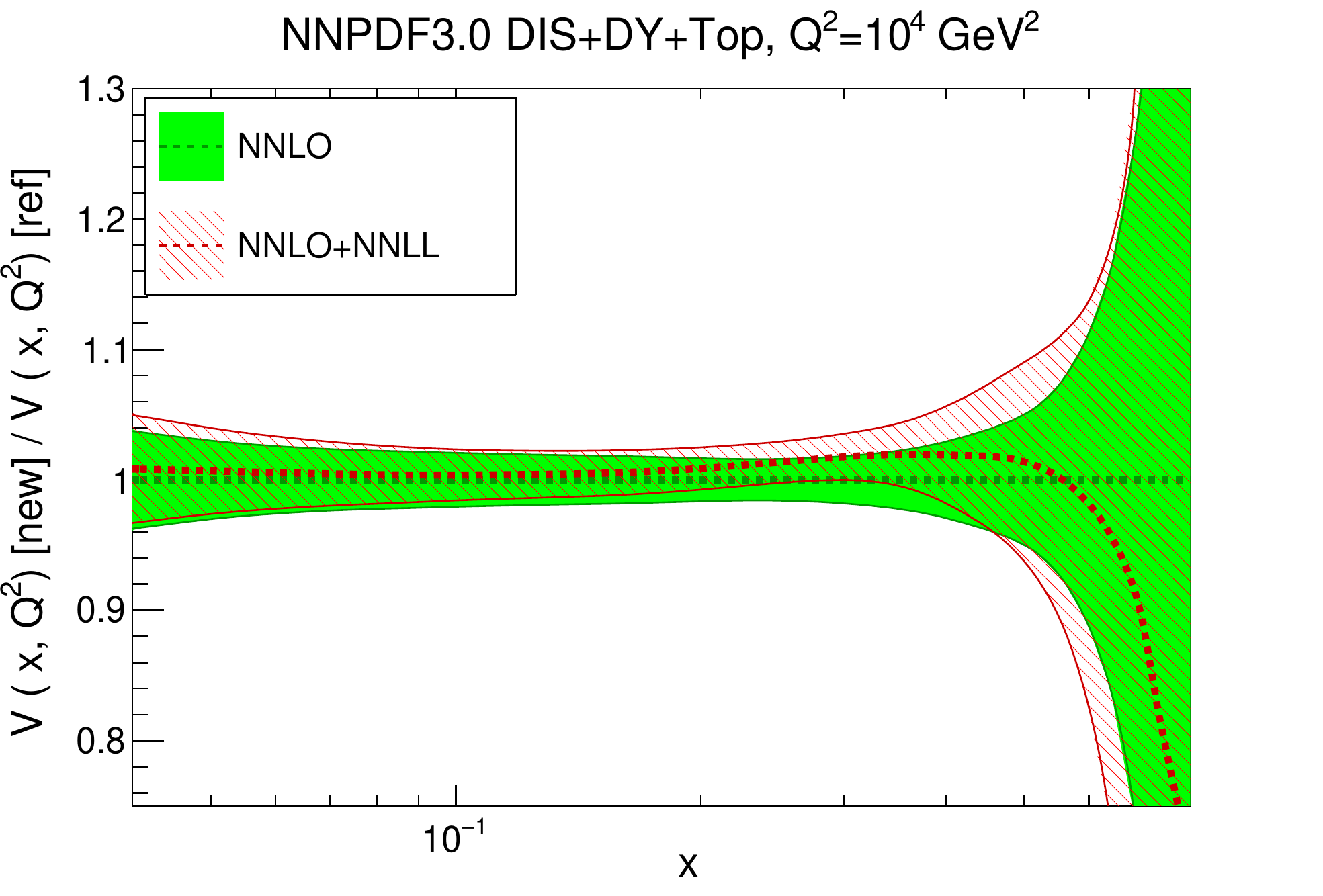}
\includegraphics[width=0.49\textwidth]{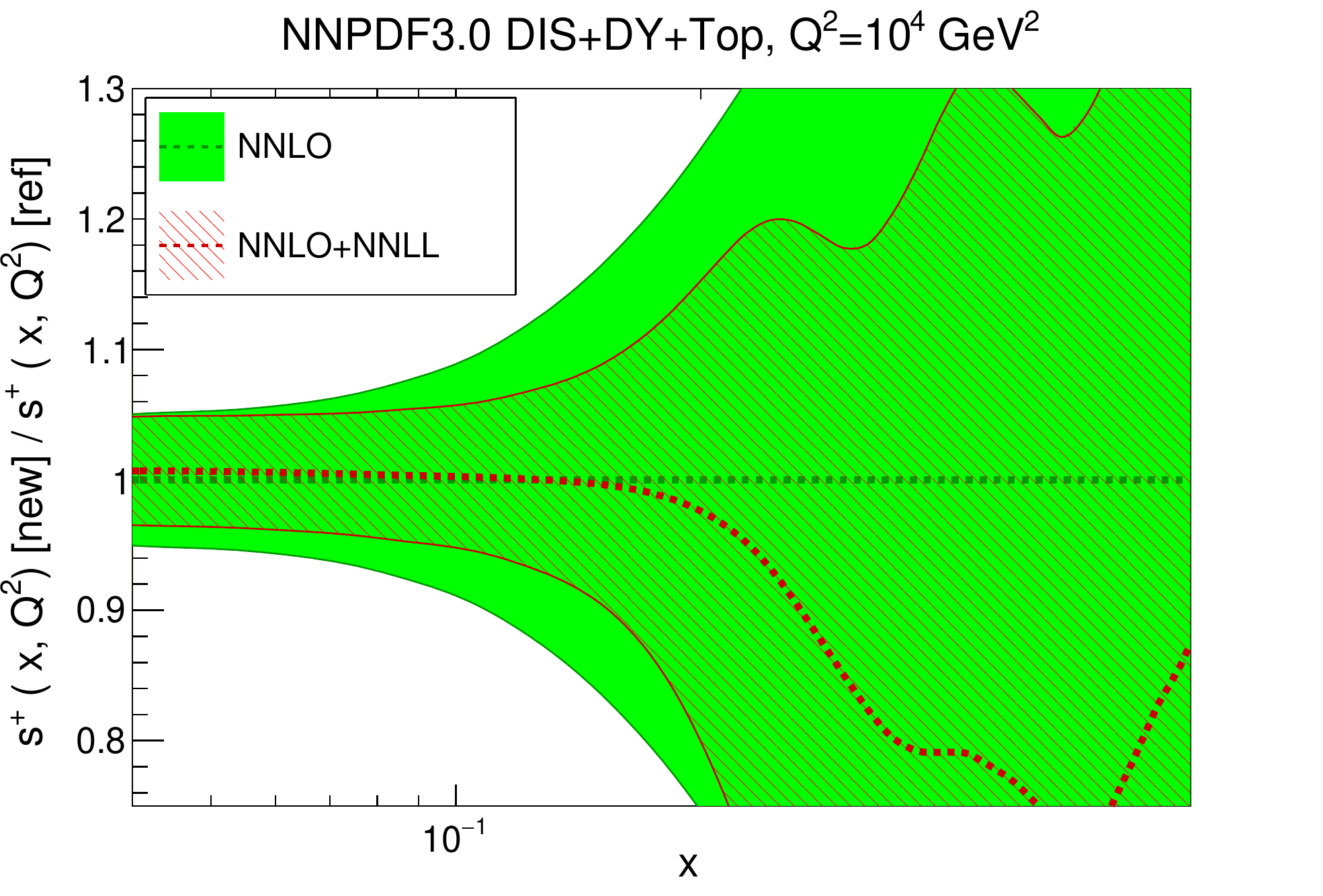}
\end{center}
\vspace{-0.3cm}
\caption{\small
 Same as Fig.~\ref{fig:disdy-nlo-vs-nll-highQ}, now comparing the NNLO
  DIS+DY+top fit with the corresponding NNLO+NNLL fit.
}
\label{fig:disdy-nnlo-vs-nnll-highQ}
\end{figure}
%%%%%%%%%%%%%%%%%%%%%%%
%
At the NNLO+NNLL level, Fig.~\ref{fig:disdy-nnlo-vs-nnll-highQ}, the impact of the resummation
is as expected even smaller.
In particular, the PDF uncertainty bands of the fixed order and resummed fits are quite similar and they overlap in the entire $x$ range.
Large shifts in the central values only occur in regions where the PDF uncertainties are large.
%This said, the central value still shifts in the resummed fit, by an amount which reaches up to
%half the PDF error band at some value of $x$. 
%
For example, for the large-$x$ gluon, at  $x\sim 0.3$ the resummed central value is
$\sim 15\%$ smaller than the fixed order one, however the PDF uncertainty in this region is substantially larger than the central-value shift.
At very large $x$, a similar trend can be seen for the total valence PDF, which however exhibits an enhancement at $x\sim 0.3$, which is as big as the PDF uncertainty.

\subsection{Partonic luminosities}

We now study the impact of the inclusion
of threshold resummation on PDFs at the level of partonic luminosities.
This comparison is useful because it provides direct information
on how the cross sections for the production of a given final state
with invariant mass $M_X$ will be affected by the inclusion
of resummation in the PDFs.
It should be emphasised however that in a consistent calculation
the 
impact of the resummation in the PDFs may be compensated by 
a similar sized effect of the resummation in the partonic matrix elements of the
process under consideration.
The consistent comparison of LHC cross sections with resummation
included both at the PDFs and in the matrix elements
is performed in the next section.

We begin by estimating the effect on the PDF
luminosities of the reduced dataset used in our baseline fits,
as compared to the NNPDF3.0 global fit.
Thus in Fig.~\ref{fig:lumiBaseline} we compare
the NNPDF3.0 NLO partonic luminosities for $\as(m_Z^2)=0.118$,
in the global fit and in the DIS+DY+top baseline fit.
In the upper plots we show the quark-antiquark and quark-quark
luminosities, and in the lower plots the gluon-gluon and gluon-quark luminosities.
The calculation has been performed for the LHC 13 TeV, as a function of the
mass of the final state $M_X$, and results are normalised to the
central value of the global fit.

  As we can see in Fig.~\ref{fig:lumiBaseline}, there
  are some important differences between the global and DIS+DY+top fits.
  For the $qq$ luminosity, the impact of varying the dataset is small,
  both in terms of central values and of PDF uncertainties,
  except at very large values of $M_X$.
  For the $q\bar{q}$ luminosity, the differences are again only
  sizeable at large $M_X$, where the central value of the DIS+DY+top
  fit is softer than that of the global fit, for instance by
  10\% at $M_X\simeq 3$~TeV.
  PDF uncertainties are similar in the two cases, and the two
  fits agree within one-sigma.
  The missing jet data have a stronger impact on the $gg$ and $qg$
  luminosities. For instance for the $gg$ luminosity above 0.5 TeV, PDF
  uncertainties increase by a factor two or more.
  Therefore, in order to consistently assess the impact of
  the resummation, one should compare the
  resummed and fixed-order DIS+DY+top fits,
  rather than the NNPDF3.0 global fit, with the resummed
  fits presented here.

  %%%%%%%%%%%%%%%%%%%%%%%%%%%%%%%%%%%%%%%%%%%%%
\begin{figure}[t]
\begin{center}
\includegraphics[width=0.45\textwidth]{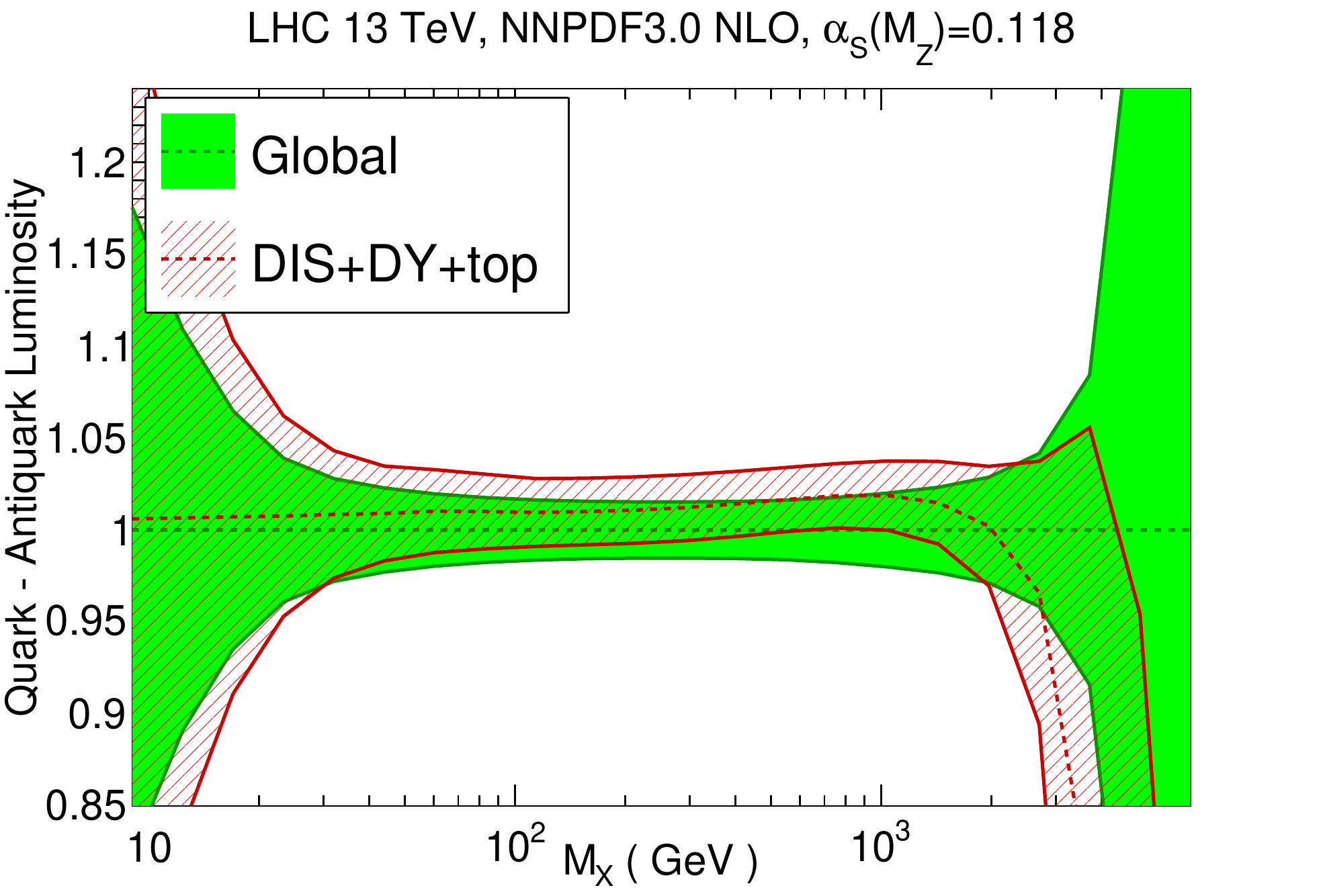}
\includegraphics[width=0.45\textwidth]{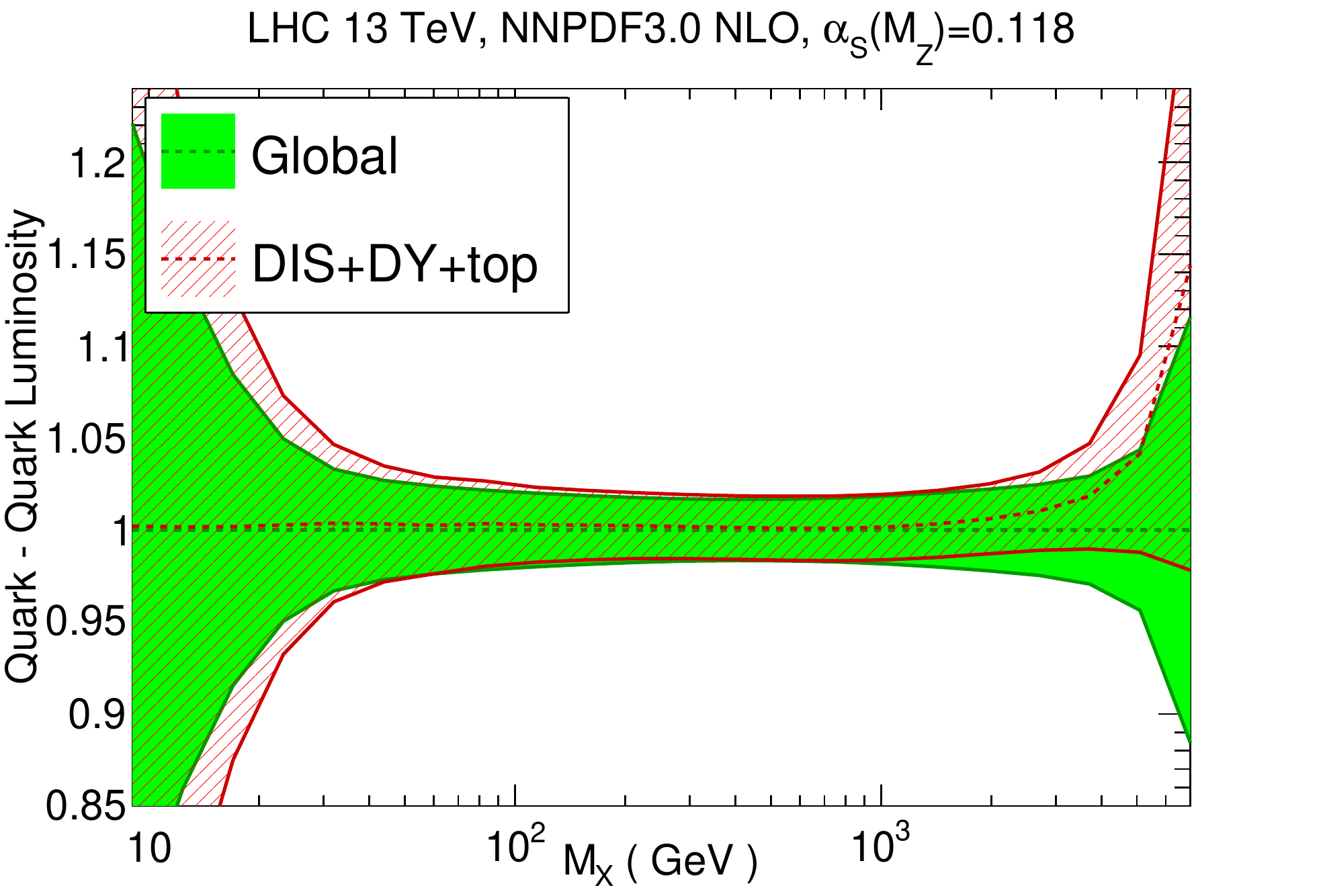}
\includegraphics[width=0.45\textwidth]{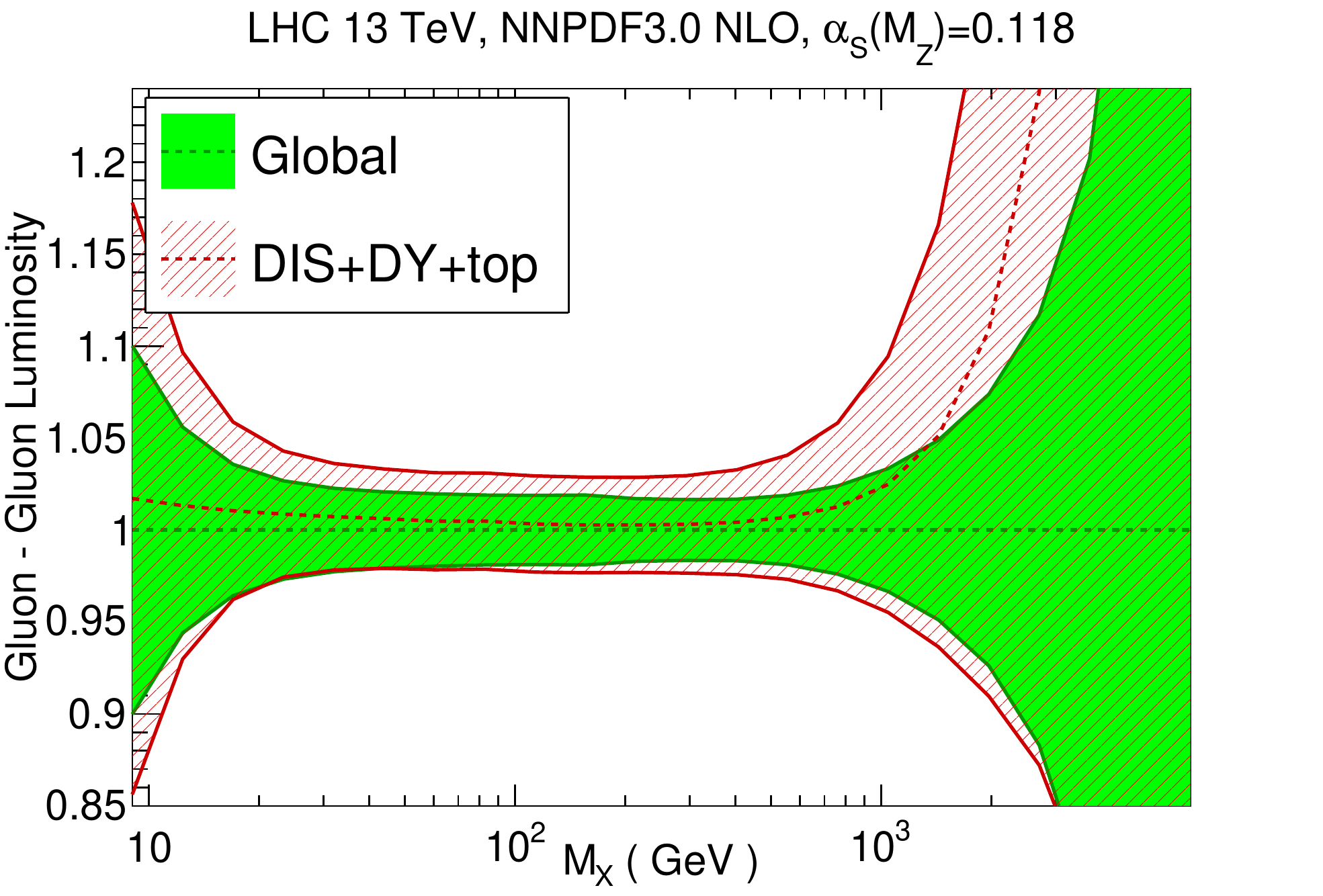}
\includegraphics[width=0.45\textwidth]{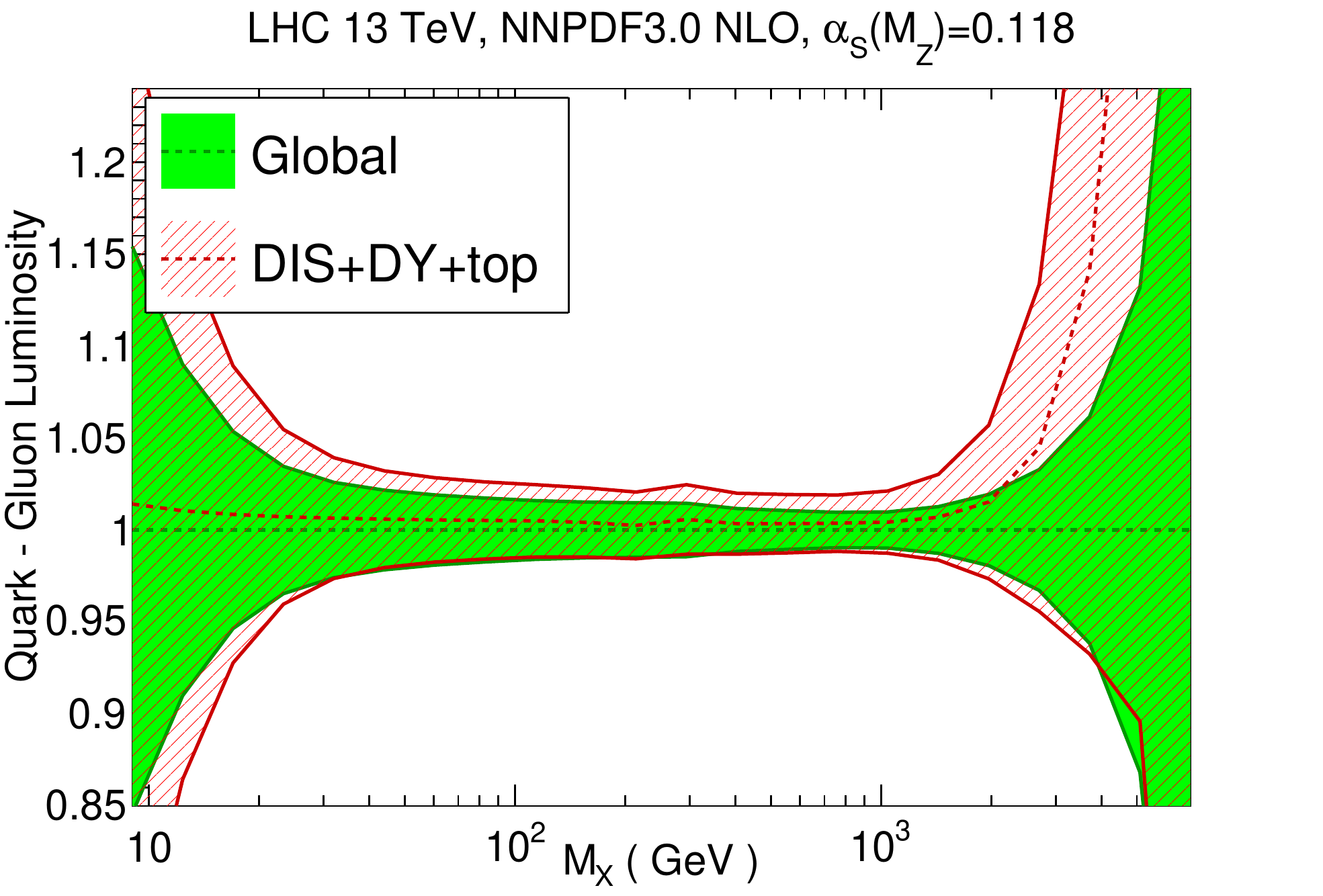}
\end{center}
\vspace{-0.3cm}
\caption{\small  Comparison of
  the NNPDF3.0 NLO partonic luminosities for $\as(m_Z^2)=0.118$,
  in the global fit and in the DIS+DY+top fit which is used as fixed-order
  baseline for the resummed fits.
  In the upper plots we show the quark-antiquark and quark-quark
  luminosities, and in the lower plots the gluon-gluon and gluon
  quark luminosities.
  The calculation has been performed for the LHC 13 TeV, as a function of the
  mass of the final state $M_X$, and results are normalised to the
  central value of the global fit.
}
\label{fig:lumiBaseline}
\end{figure}
%%%%%%%%%%%%%%%%%%%%%%%%%%%%%%%%%%%%%%%%%%%%%%%%%%%%

The comparisons between the DIS+DY+top fixed-order and resummed
fits are displayed in Figs.~\ref{fig:lumi3} (at NLL) and~\ref{fig:lumi4}
(at NNLL).
We see that in all cases the fixed-order and resummed fits agree
at the level of one sigma, and that the effect
of resummation is as expected smaller at NNLL than at NLL.
In the comparison between NLO and NLO+NLL, the $qq$ and $q\bar{q}$
luminosities are enhanced by about one sigma for
$M_X\lsim 1$~TeV, while they are suppressed at larger values of
$M_X$.
This behaviour follows from the corresponding PDF comparisons,
where quarks are slightly enhanced at $x\simeq 0.1-0.4$ but
suppressed for larger values of $x$.
This suppression can be sizeable: for $M_X\simeq 3$~TeV the
$q\bar{q}$ luminosity in the NLO+NLL fit is reduced by $\sim 15\%$.
In the $qq$ channel, this suppression instead is small unless very large values of $M_X$ are probed.
The $gg$ and $gq$ luminosities are also suppressed at large invariant masses,
for instance for $gg$ the suppression is already $\sim 10\%$ at 1~TeV,
though still consistent with the fixed-order fit within the large
PDF uncertainties.

%%%%%%%%%%%%%%%%%%%%%%%%%%%%%%%%%%%%%%%%%%%%%
\begin{figure}[t]
\begin{center}
  \includegraphics[width=0.45\textwidth]{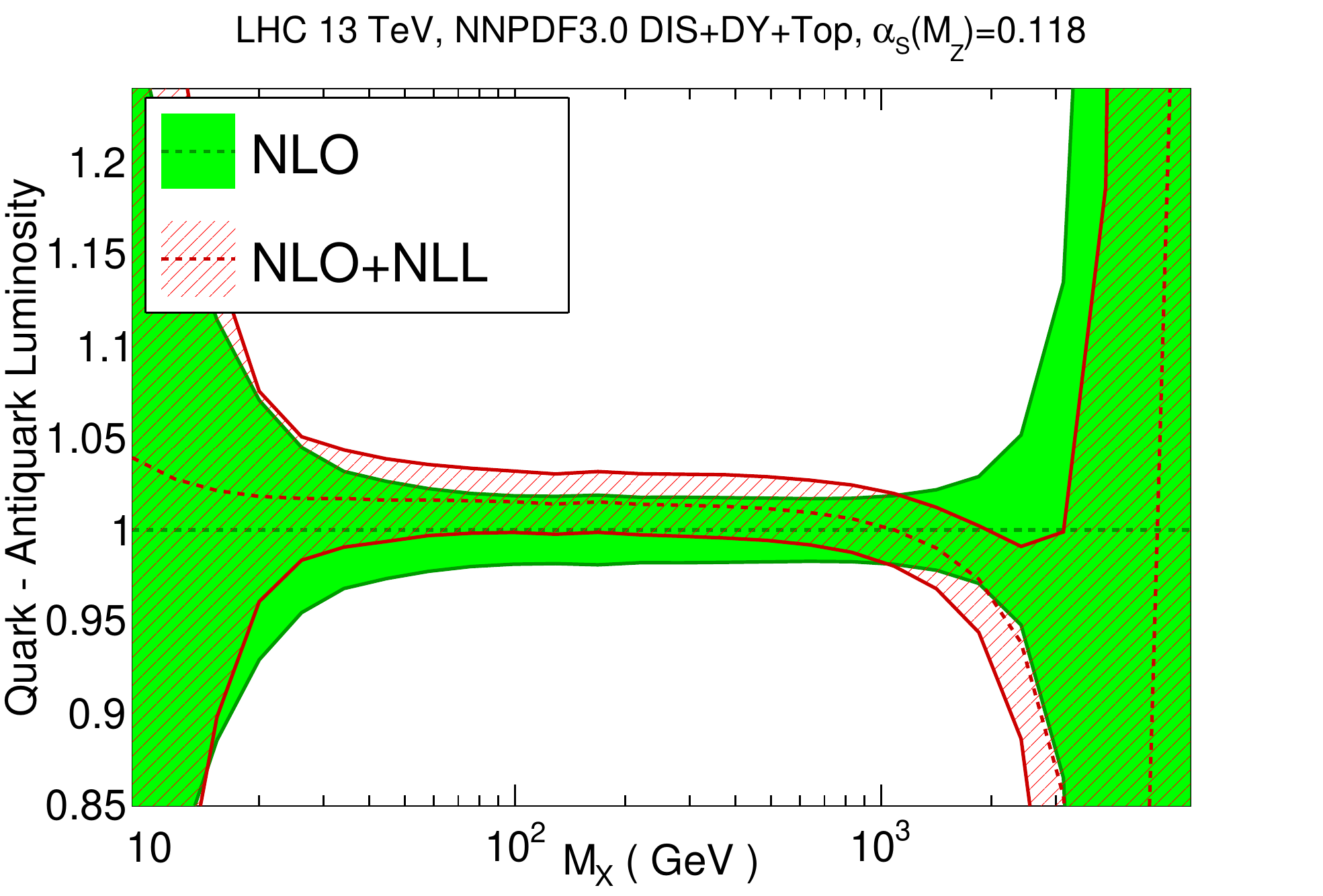}
  \includegraphics[width=0.45\textwidth]{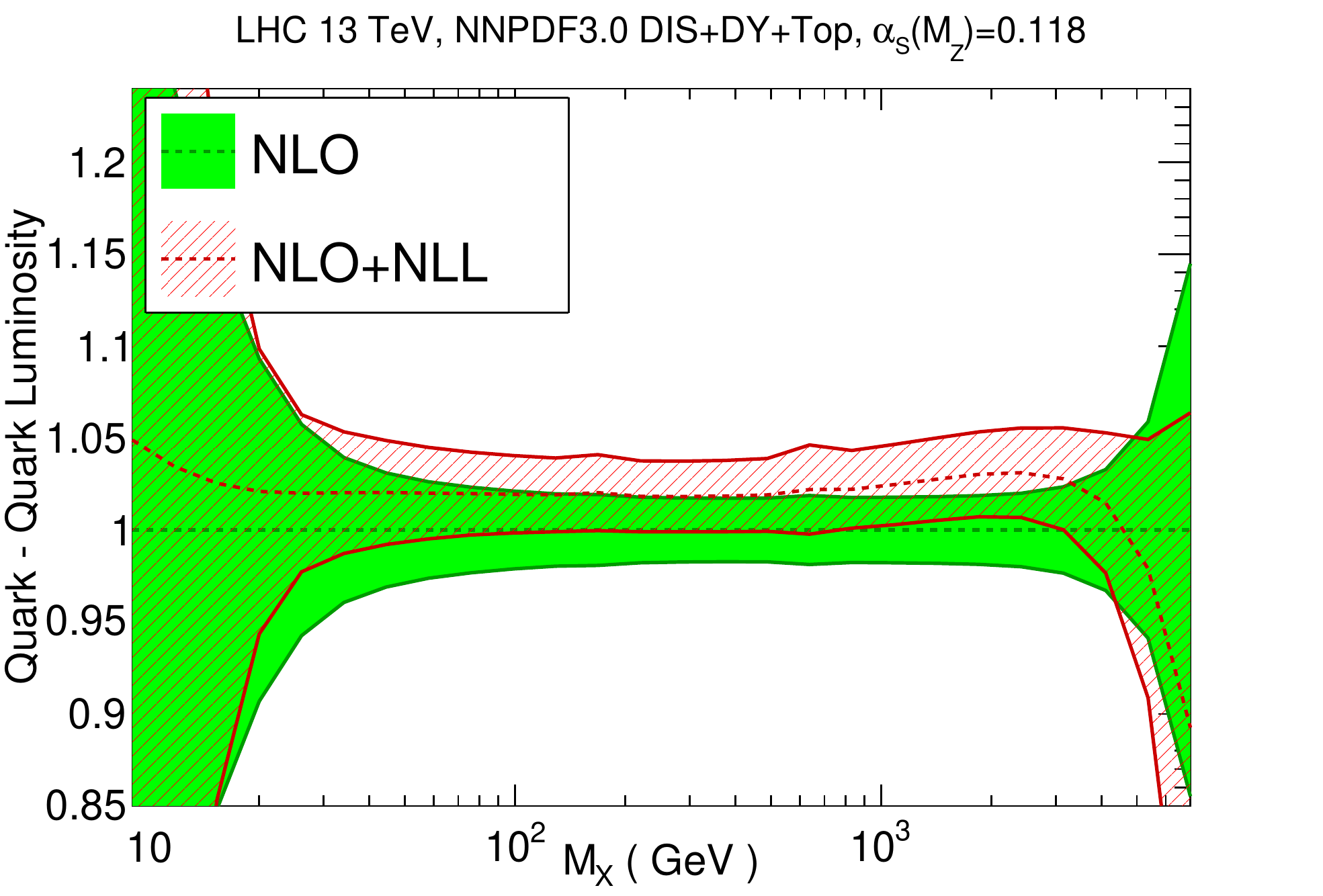}
\includegraphics[width=0.45\textwidth]{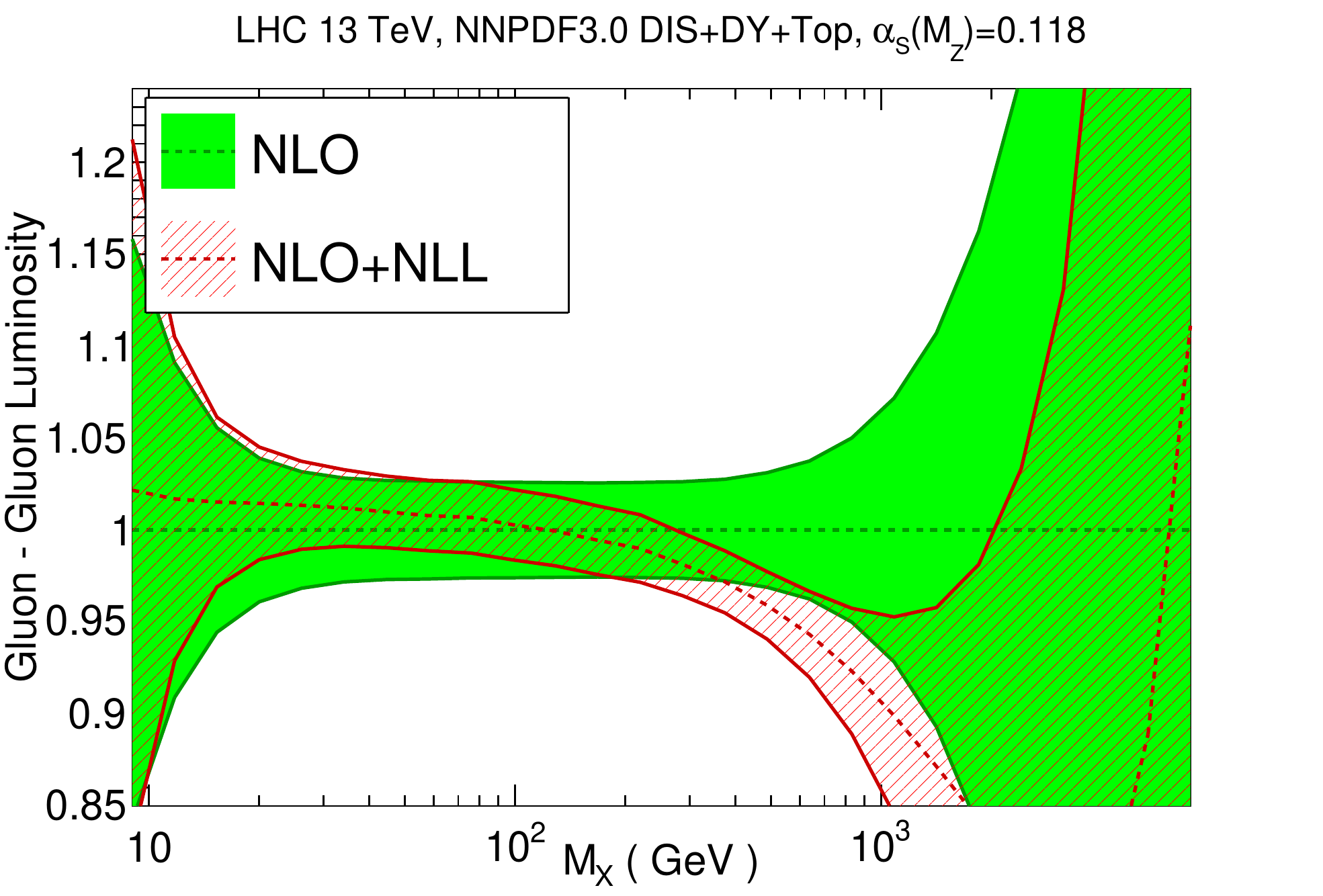}
\includegraphics[width=0.45\textwidth]{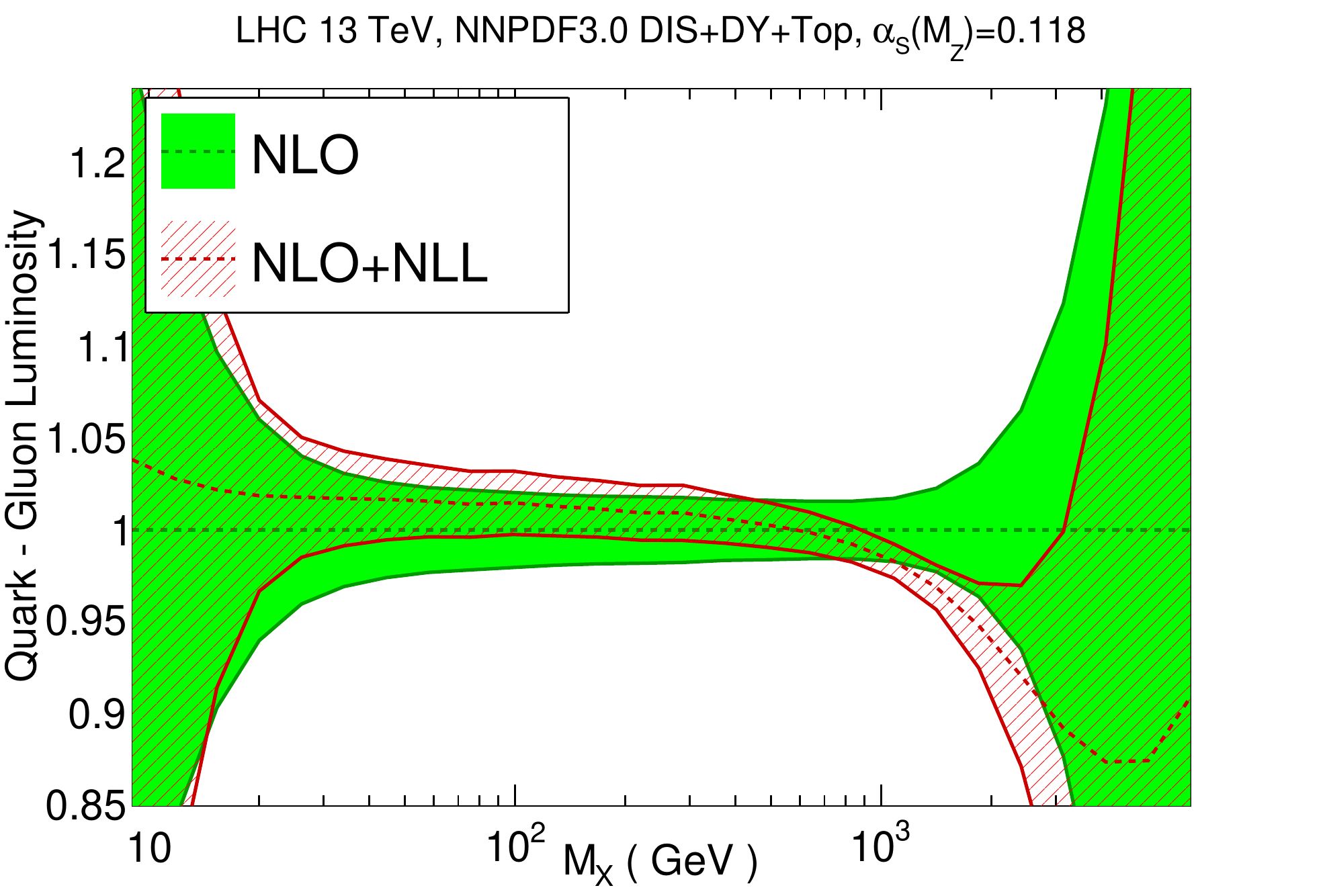}
\end{center}
\vspace{-0.3cm}
\caption{\small  Same as Fig.~\ref{fig:lumiBaseline}, now comparing
  the results of DIS+DY+top fits using either NLO or NLO+NLL
  calculations.
}
\label{fig:lumi3}
\end{figure}
%%%%%%%%%%%%%%%%%%%%%%%%%%%%%%%%%%%%%%%%%%%%%%%%%%%%

From the corresponding comparison between the
NNLO and NNLO+NNLL fits, shown in Fig.~\ref{fig:lumi4},
we see that the effects of resummation are very small
everywhere except for the largest values of $M_X$.
The central values of the $q\bar{q}$, $gg$ and $qg$ luminosities
exhibit some suppression at very large $M_X$, but this suppression
is not relevant when compared to the PDF uncertainties.
From the comparison in Fig.~\ref{fig:lumi4} we thus conclude that the
impact of threshold resummation in a global PDF analysis
is only relevant at NLO, while at NNLO it
appears to be negligible, at least with the current
PDF uncertainties.
If future data leads to substantial reduction of
PDF uncertainties at large-$x$, threshold resummation
could be relevant even for NNLO fits.

%%%%%%%%%%%%%%%%%%%%%%%%%%%%%%%%%%%%%%%%%%%%%
\begin{figure}[t]
\begin{center}
  \includegraphics[width=0.45\textwidth]{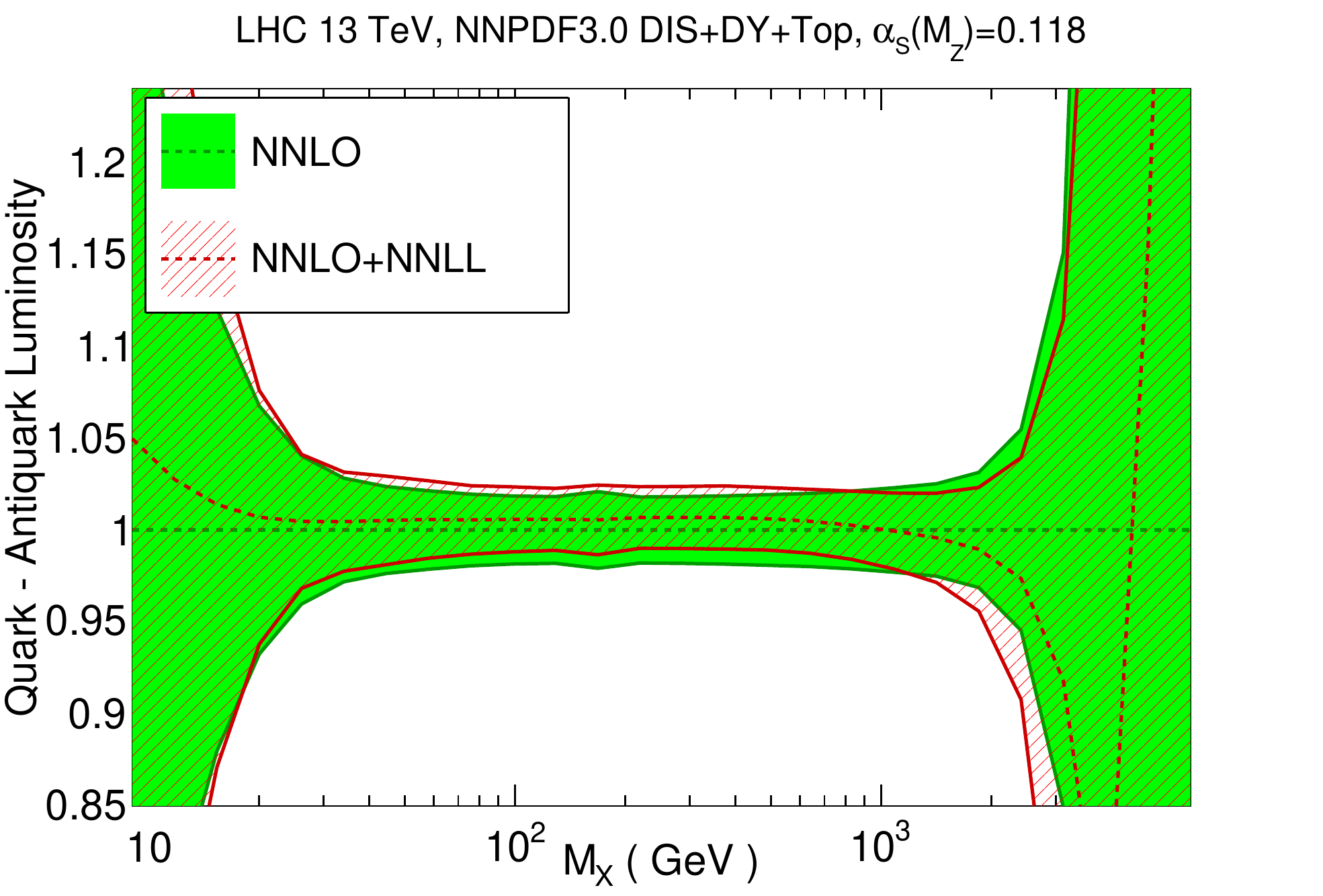}
  \includegraphics[width=0.45\textwidth]{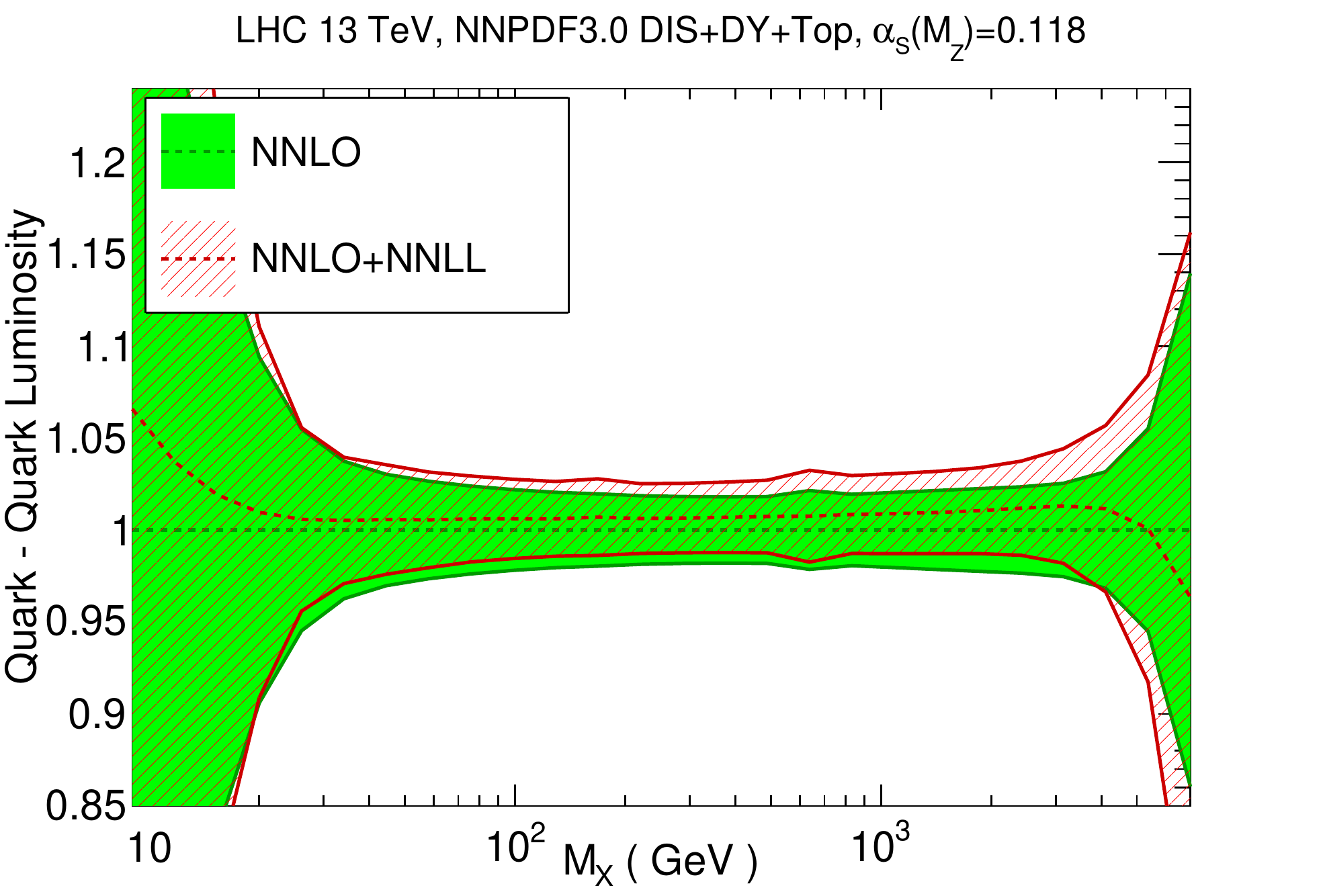}
\includegraphics[width=0.45\textwidth]{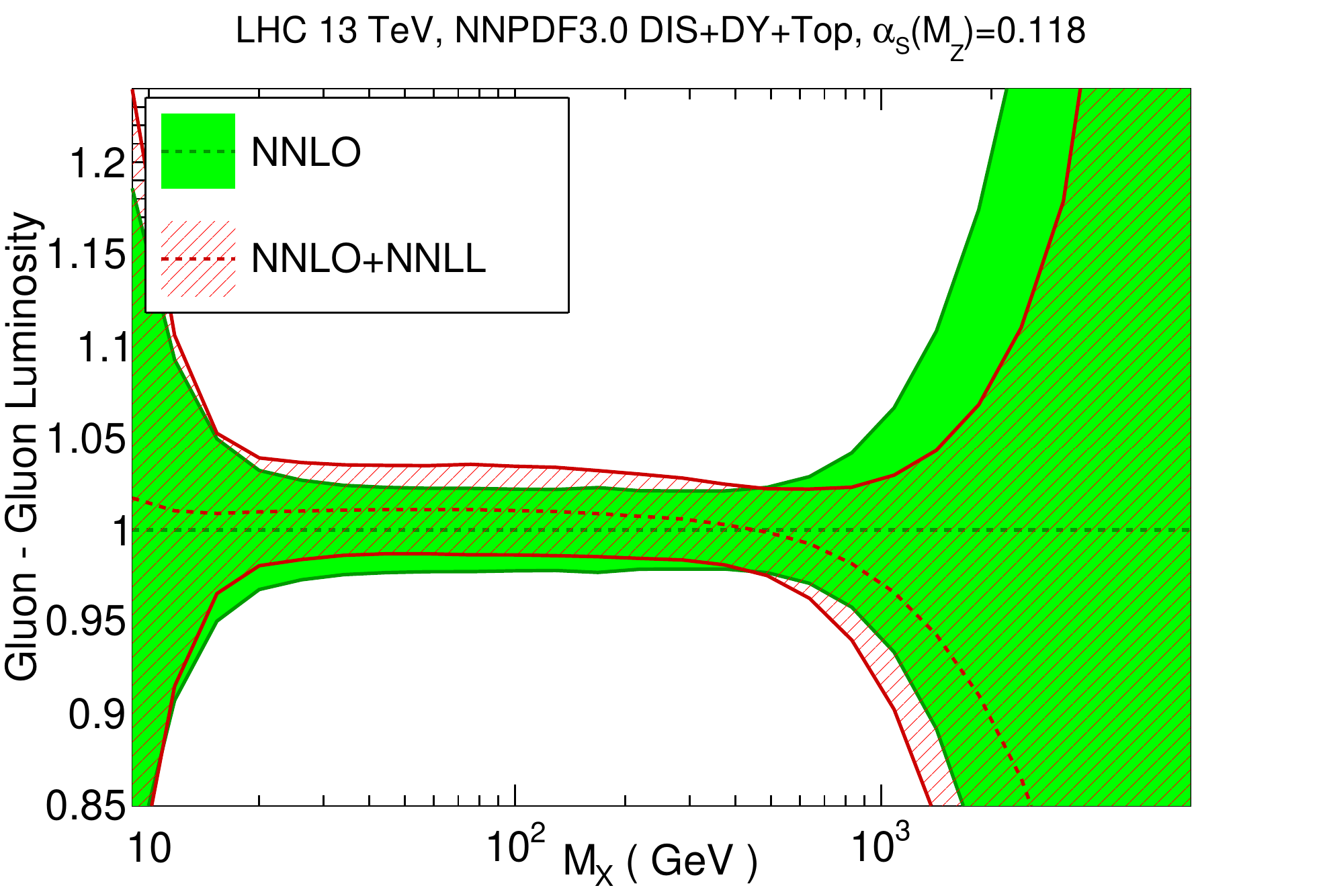}
\includegraphics[width=0.45\textwidth]{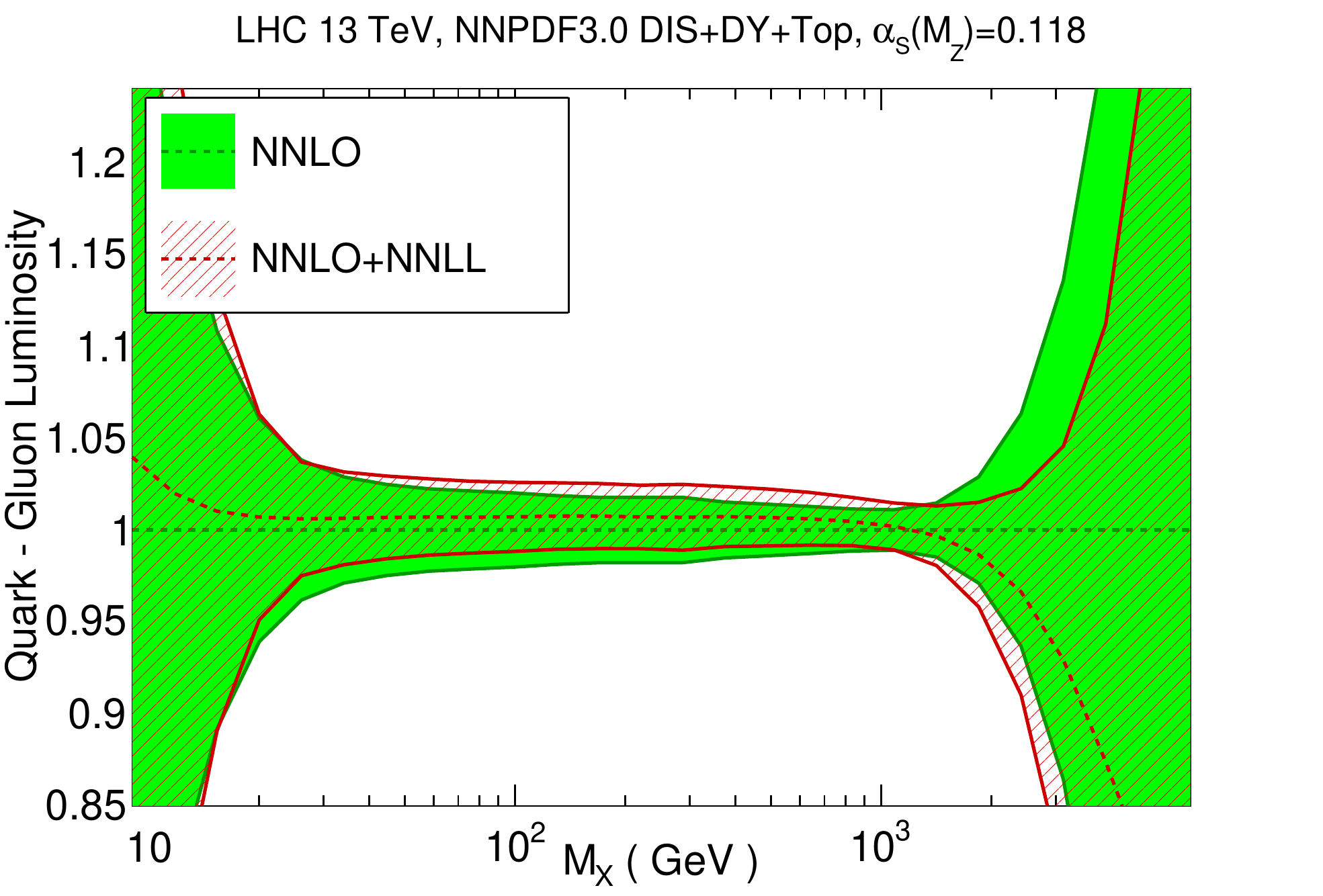}
\end{center}
\vspace{-0.3cm}
\caption{\small Same as Fig.~\ref{fig:lumi3},
  this time comparing the NNLO and NNLO+NNLL fits.
  }
\label{fig:lumi4}
\end{figure}
%%%%%%%%%%%%%%%%%%%%%%%%%%%%%%%%%%%%%%%%%%%%%%%%%%%%

\section{Resummed PDFs: implications for LHC phenomenology}

\label{sec:lhcpheno}

In this section we discuss the implications of the NLO+NLL
and NNLO+NNLL resummed PDF sets for LHC phenomenology.
Our aim is to quantify, for a variety of processes, the difference
between using consistently NLO+NLL and NNLO+NNLL calculations at the
level of both PDFs and matrix elements, and the usual (but inconsistent)
approach of using resummed
partonic cross sections with fixed-order PDFs.

For illustration, we consider three representative
LHC processes for which resummed calculations are
publicly available, either
at the level of total cross sections or of differential
invariant-mass distributions.
We start by considering Higgs production in gluon fusion,
both for $m_H=125$ GeV and for a heavy BSM Higgs-like neutral scalar.
Note that the current recommendation of the Higgs Cross Section Working group
for inclusive Higgs production in gluon fusion is based
on the NNLO+NNLL calculation~\cite{deFlorian:2009hc,Dittmaier:2012vm}.
We then consider threshold resummation for the
invariant mass distributions of
dileptons in the high-mass Drell-Yan process, which is important in
many New Physics searches, for example for $Z'$ searches.
Finally we study the invariant mass distribution of supersymmetric
lepton (slepton) pair production, a typical final state analysed in electroweak
SUSY searches.
While Higgs production is driven by the $gg$ luminosity,
both high-mass Drell-Yan and slepton pair production are driven
by the quark-antiquark luminosity, which is reasonably well
constrained even with the reduced dataset used in the present fits.

A variety of other interesting processes are available
in which resummed PDFs should be relevant, including top quark
differential
distributions~\cite{Ahrens:2011mw,Guzzi:2014wia,Ferroglia:2013awa},
squark and gluino pair production~\cite{Beenakker:2011sf,Beenakker:2014sma}
or stop quark pair production~\cite{Broggio:2013cia}. However for
most of these processes the corresponding resummation
codes are not publicly available.

\subsection{SM and BSM Higgs production in gluon fusion}

The accurate calculation of Higgs production via gluon fusion is
an essential component of the LHC program, since it is required
in order to extract Higgs couplings from the ATLAS and CMS measurements.
As a result of the recent calculation of the inclusive cross section
at N$^3$LO~\cite{Anastasiou:2015vya},
PDF uncertainties are now one of the dominant theory
uncertainties.
In addition to the characterisation of the SM Higgs boson,
many New Physics scenarios
predict heavy Higgs-like
bosons~\cite{Chen:2013rba,Djouadi:2013vqa,Arbey:2013jla},
and thus it is also important to provide accurate
predictions for heavy Higgs production for these BSM searches.

%%%%%%%%%%%%%%%%%%%%%%%%%%%%%%%%%%%%%%%%%%%%%%%%%%%%%%%%%%%%%%%%%
\begin{figure}[t]
\centering
\includegraphics[width=0.495\textwidth,page=2]{plots/higgs_xsec_res.pdf}
\includegraphics[width=0.495\textwidth,page=1]{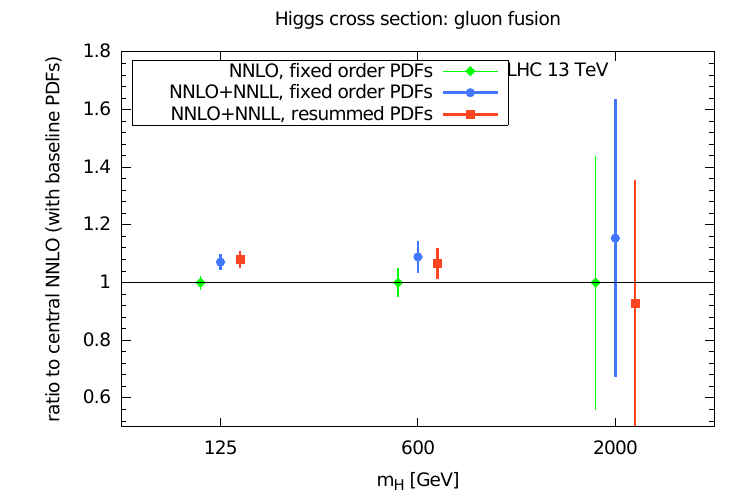}
\caption{\small Left: the total cross section for Higgs production
  in gluon fusion at the LHC 13 TeV for different values of the Higgs mass,
  comparing the predictions of NLO fixed-order with that of NLO+NLL resummed
  calculations (using either fixed order or resummed PDFs).
  Results are shown normalised to the
  central prediction of the fixed order NLO calculation.
  Right: the same comparison now performed at NNLO.
  The calculation has been performed using the {\tt ggHiggs} code.
}
\label{fig:higgs}
\end{figure}
%%%%%%%%%%%%%%%%%%%%%%%%%%%%%%%%%%%%%%%%%%%%

Using the {\tt ggHiggs} code~\cite{gghiggs}, in  Fig.~\ref{fig:higgs} we show
the predictions for the total cross section for Higgs production
  in gluon fusion at the LHC 13 TeV,
  comparing  the (N)NLO fixed-order results
  with those of the (N)NLO+(N)NLL resummed
  calculations, using either fixed-order or resummed PDFs.
  The calculation has been performed
  in the $m_{\rm top}\to
  \infty$ limit and neglecting finite-width effects, which is
  sufficient for current purposes.
All results are normalised to the central value of the fixed-order (N)NLO
calculation, and we provide three different values of the
Higgs mass: $m_H=125$~GeV, $600$~GeV and $2$~TeV.

The comparisons in Fig.~\ref{fig:higgs} are interesting because
they show that for the production of heavy final states that probe
large values of $x$ in the gluon PDF, including resummation in the PDFs can cancel
out the effect of the resummation in the matrix element.
In the case of the NLO calculation, the SM Higgs cross section
is not affected by resummation of the PDFs, but
already for $m_H=600$~GeV, the inclusion of resummed
PDFs cancels almost half of
the enhancement
in the hadronic cross section that arises
from resummation of the matrix element.
For an even heavier Higgs, with
$m_H= 2$~TeV, the consistent NLO+NLL calculation is essentially
identical to the NLO result.
The trend is similar at NNLO, though of course in this case
the effect of perturbative corrections beyond the
fixed-order NNLO calculation is smaller.
Note also that PDF uncertainties are substantial at large Higgs masses,
partly because of the lack of jet data in the baseline and resummed fits.

Our results demonstrate that using consistently resummed PDFs for SM Higgs production at the LHC
has no effect, and therefore puts on a more solid ground the current HXSWG recommendation, which is based on
fixed-order PDFs.
This observation is in agreement with the findings of Ref.~\cite{Forte:2013mda}
regarding the (lack of) need of N$^3$LO PDFs for the SM Higgs production cross section at N$^3$LO.

\subsection{High-Mass Drell-Yan dilepton mass distributions}

At the LHC, high-mass Drell-Yan is one of the most important processes when 
looking for new physics, in particular for new electroweak
sectors.
For instance, ATLAS and CMS have explored a number of BSM
signatures in the high-mass tail of neutral-current
Drell-Yan production~\cite{Aad:2014wca,Aad:2014cka,Khachatryan:2014fba},
such as $Z'$ bosons which appear in several new physics scenarios.
It is therefore interesting to assess the effect of including
consistently threshold resummation both in the PDFs
and in matrix elements, compared to including it
only in the matrix element while using fixed-order PDFs.

In Fig.~\ref{fig:DYnnll} we show the 
dilepton invariant mass distribution for
  high-mass neutral current Drell-Yan production at the LHC 13 TeV,
  comparing the predictions of fixed-order and resummed
  calculations. 
  The fixed-order NLO and NNLO predictions have been computed
  with the code {\tt Vrap}
  supplemented with threshold resummation as provided by
  \troll.
  In Fig.~\ref{fig:DYnnll} we show
the predictions for the dilepton invariant mass distribution,
   comparing  the (N)NLO fixed-order results
  with those of the (N)NLO+(N)NLL resummed
  calculations, using either fixed-order or resummed PDFs.
  The latter comparison quantifies the mismatch when
  resummed calculations are used with fixed-order PDFs. 

%%%%%%%%%%%%%%%%%%%%%%%%%%%%%%%%%%%%%%%%%%%%%%%%%%%%%%%%%%%%%%%%%
\begin{figure}[t]
  \centering
  \includegraphics[width=0.495\textwidth,page=2]{plots/DY_vrap.pdf}
  \includegraphics[width=0.495\textwidth,page=1]{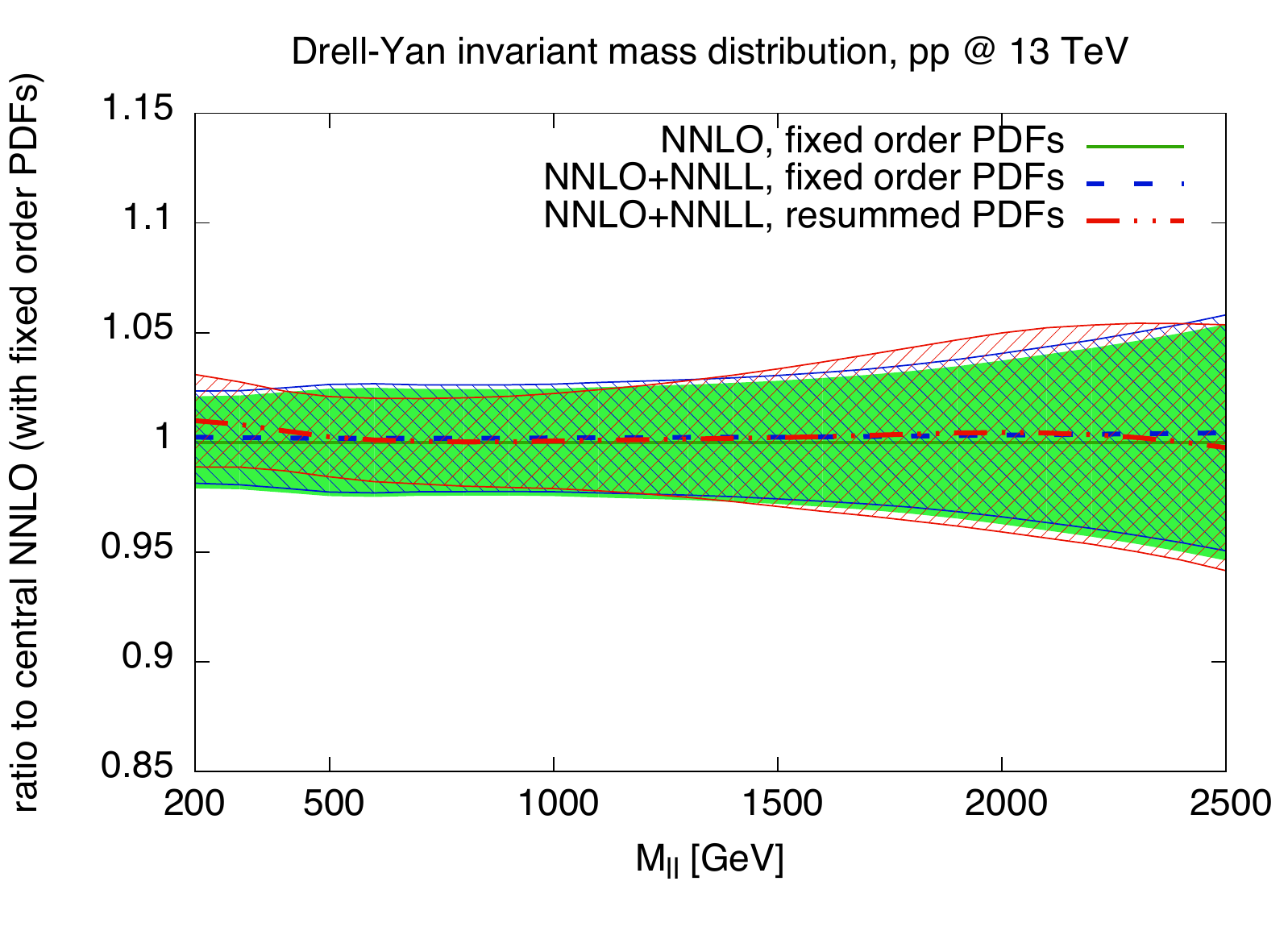}
  \caption{\small Left: dilepton invariant mass distribution for
  high-mass neutral current Drell-Yan production at the LHC 13 TeV,
  comparing the predictions of fixed-order with that of resummed
  calculations. Results are shown normalised to the
  central prediction of the fixed-order NLO calculation.
  Right: the same comparison at NNLO.
}
\label{fig:DYnnll}
\end{figure}
%%%%%%%%%%%%%%%%%%%%%%%%%%%%%%%%%%%%%%%%%%%%

 The results are qualitatively consistent with those of the
  Higgs cross sections in Fig.~\ref{fig:higgs}.
  First, we see that even at NLO and at large invariant masses
  the effect of threshold resummation is moderate: the NLL
  correction amounts (for fixed-order NLO
  PDF) to about 4\% at $M_{ll}=2.5$~TeV, which is within the current PDF uncertainty.
  Including the effect of the resummation in the PDFs
  consistently cancels this effect, for example in the range
  for $M_{ll}\in \lc 1.5, 2.5\rc$~TeV the central value
  of the NLO+NLL calculation
  agrees with the fixed-order NLO result by less than one percent.
  At NNLO the impact of resummation is completely negligible,
  both at the level of the PDFs and of the matrix elements.

\subsection{Supersymmetric particle production}

The theoretical predictions for high-mass
supersymmetric pair production at hadron colliders are currently made at 
NLO, supplemented with either NLL or NNLL resummation of
threshold logarithms.
In particular, the NLO+NLL resummed calculations of
Refs.~\cite{Borschensky:2014cia,Kramer:2012bx} have been
used to produce the benchmark production cross sections at $\sqrt{s}=7$~TeV
and $13$~TeV that are used as by ATLAS and CMS
in the theoretical interpretation of their
searches for
supersymmetry~\cite{Aad:2015pfx,Khachatryan:2015wza,Aad:2015wqa,Khachatryan:2015vra}.

An important limitation of these predictions is the mismatch
between the fixed-order PDFs and the resummed partonic cross sections,
which should be more important at high-masses, precisely the crucial
region for New Physics searches.
Thanks to the availability, for the first time, of general-purpose
resummed PDFs, it is now possible to consistently combine resummed
PDFs and matrix elements into a single calculation.
It is beyond the scope of this work to present a comprehensive
study of the impact of NLO+NLL PDFs for generic supersymmetric
processes.
However, for illustrative purposes,
in this section we will use the public code {\tt Resummino}~\cite{Fuks:2013vua,Fuks:2013lya,Bozzi:2007qr}
to compare the effect of resummed PDFs in the context of NLO+NLL predictions for electroweak supersymmetric
particle pair production at the LHC, in particular for slepton
pair production.

{\tt Resummino} computes resummed and matched predictions for supersymmetric particle production at hadron colliders up to the NLO+NLL level.
Currently the processes implemented include gaugino-pair production and slepton-pair production.
These final states are characteristic signatures in electroweak
SUSY searches at the
LHC~\cite{Aad:2014vma,Aad:2012pra,Khachatryan:2014qwa,Khachatryan:2014mma}.
{\tt Resummino}  is able to compute total cross sections as well as invariant-mass and transverse-momentum distributions.
In this study we focus on the invariant-mass distribution for slepton pair production.
Note that the production of sleptons (like many other electroweak SUSY processes)
is mostly sensitive to the $q\bar{q}$ luminosity;
other processes, such as squark and gluino pair production,
would be sensitive to other PDF combinations such as $qg$ and $gg$.

%%%%%%%%%%%%%%%%%%%%%%%%%%%%%%%%%%%%%%%%%%%%%
\begin{figure}[t]
\begin{center}
\includegraphics[width=0.70\textwidth]{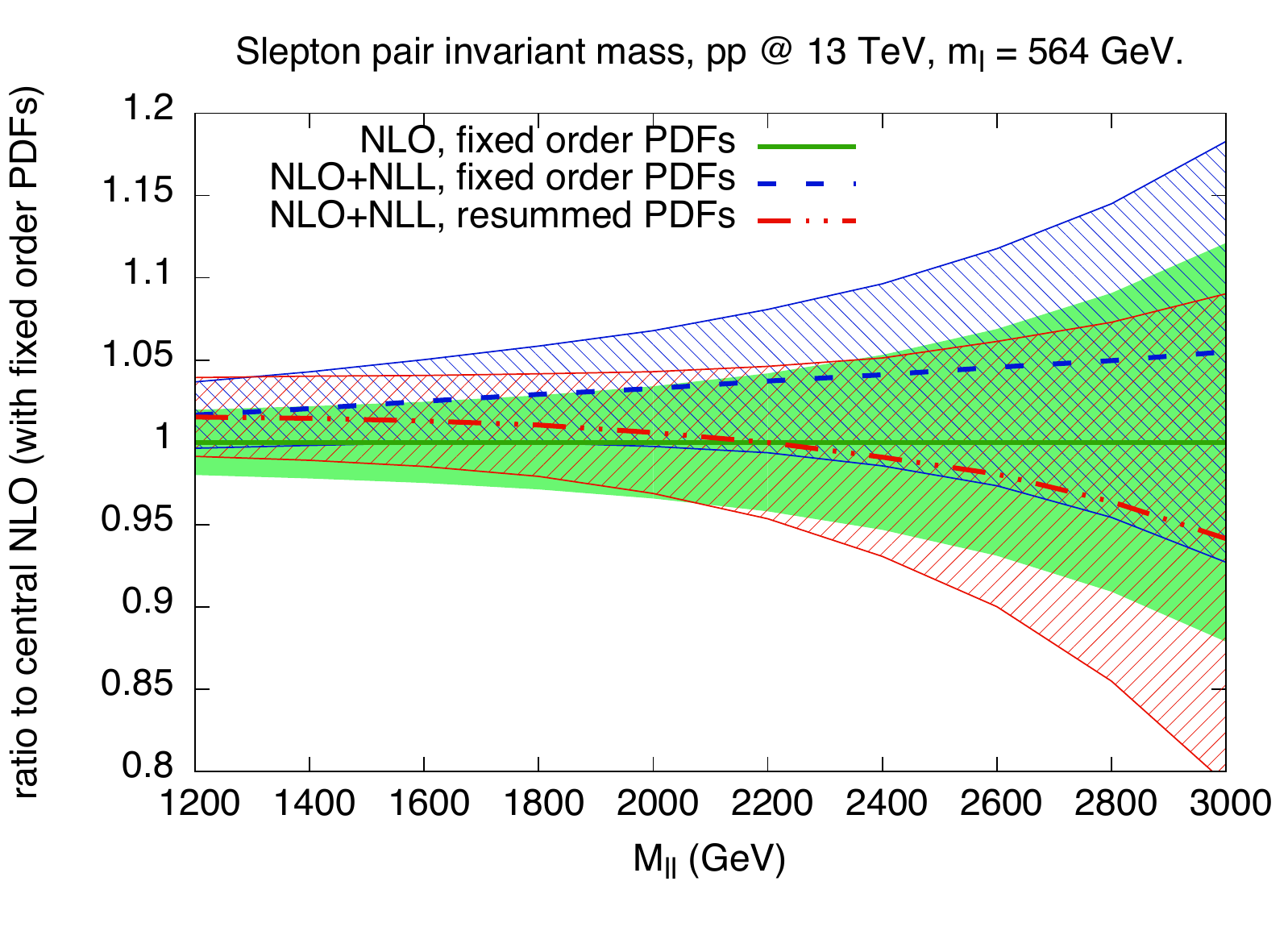}
\end{center}
\vspace{-0.3cm}
\caption{\small The NLO+NLL
  calculation of the invariant mass distribution for 
  slepton pair production at the
  LHC 13 TeV using the {\tt Resummino} program,
  using both the  NLO and NLO+NLL NNPDF3.0 DIS+DY+top PDFs
  as input.
  Results are shown as ratios with respect to the NLO calculation,
  using consistently the NLO baseline PDFs.
  The settings of the SUSY calculation are the default ones in
  {\tt Resummino}. We use a slepton mass of $m_{\tilde{l}}=564$ GeV.}
\label{fig:resummino}
\end{figure}
%%%%%%%%%%%%%%%%%%%%%%%%%%%%%%%%%%%%%%%%%%%%%%%%%%%%

In Fig.~\ref{fig:resummino} we show the results
of the NLO+NLL
  calculation of the invariant mass distribution for
  slepton pair production at the
  LHC 13 TeV obtained with {\tt Resummino},
  using both the  NLO and NLO+NLL NNPDF3.0 DIS+DY+top PDFs
  as input.
  Results are shown as ratios with respect to the NLO calculation,
  using consistently the NLO baseline PDFs.
  The settings of the SUSY calculation are the default ones in
  {\tt Resummino}. We use a slepton mass of $m_{\tilde{l}}=564$~GeV.

The comparison displayed in Fig.~\ref{fig:resummino} is interesting
for a variety of reasons.
First of all, we see that, using the NLO PDFs as input, the NLO+NLL calculation
(i.e.\ with resummation included only in the matrix element) enhances the cross section
by several percent, from 2\% at $M_{\tilde{l}\tilde{l}}\sim 1.2$~TeV up to
5\% at $M_{\tilde{l}\tilde{l}}\sim 3$~TeV.
On the other hand, in the consistent calculation in which resummation
is included both at the level of PDFs and of matrix elements, this increase
is only seen around $M_{\tilde{l}\tilde{l}}\sim 1.2$~TeV. For higher invariant masses up to
$M_{\tilde{l}\tilde{l}}\sim 2.5$~TeV or so the effect of resummation in the PDFs
cancels the one originating from the matrix elements, and the consistent NLO+NLL calculation is
essentially the same as the NLO one.
For even higher masses, the NLO+NLL calculation is suppressed compared to the
NLO calculation by up to $5\%$ at $M_{\tilde{l}\tilde{l}}\sim 3$~TeV, though in this region
PDF uncertainties are very large.

The results of Fig.~\ref{fig:resummino} are consistent with the behavior
of the $q\bar{q}$ luminosity shown in Fig.~\ref{fig:lumi3}.
In particular,
for an invariant mass of $M\simeq1$~TeV, the NLO and NLO+NLL PDF luminosities
are essentially the same, while for $M\simeq 3$~TeV the NLO+NLL luminosity is
suppressed by a factor of approximately $10\%$,
a similar amount as that
inferred from the {\tt Resummino} plot
of Fig.~\ref{fig:resummino}.
This illustrates that, for those processes which are dominated by a single
partonic luminosity, one can approximately correct a NLO+NLL matrix element
calculation using the ratio of PDF luminosities.
This is also consistent with the high-mass Drell-Yan results
of Fig.~\ref{fig:DYnnll}, which are also driven by the
$q\bar{q}$ luminosity.

In summary, even though we have been able to explore only a limited
number of resummed calculations for LHC processes, a consistent
trend appears.
When the produced final state has an invariant mass far from
threshold, the use of resummed PDFs has a rather small effect. 
However, for heavy final states, the main effect of the resummation
of the PDFs is to compensate the effect of the resummation in the matrix element,
so that the consistent (N)NLO+(N)NLL
calculation is rather closer to the fixed-order (N)NLO result.
This shows that using resummation only in the matrix
element but not in the PDF can be misleading, since it may overestimate cross sections and invariant mass distributions.
This is particularly the case for NLO+NLL calculations, because
at NNLO+NNLL the effect of the resummation is much smaller, since much of it 
has already been accounted for in the NNLO fixed-order corrections.

In conclusion, one should in general always use resummed PDFs with threshold
resummed matrix elements.
This said, even if the central value of the consistent
(N)NLO+ (N)NLL calculation is reasonably close to the
original fixed-order (N)NLO result, it is in general still
better to use the resummed calculation, since these
benefit for instance from reduced scale dependence, and
thus smaller theoretical uncertainties.

\section{Summary}
\label{sec:delivery}

In this paper we have presented for the first time global
fits of parton distributions extracted at NLO+NLL and
NNLO+NNLL accuracy, where the fixed-order partonic cross sections
have been systematically improved using soft-gluon
threshold resummation.
We find that the main effect of threshold resummation is to suppress 
the PDFs in the large-$x$ region, as expected
given that the fit compensates from the resummation-induced increment 
in the partonic cross sections used
in the PDF fit.
This suppression is important for
all PDF flavors for $x\gsim 0.1$, while at intermediate values of
$x$, $0.01 \lsim x \lsim 0.1$, the quark PDFs are instead
somewhat enhanced due to the sum rules.
For smaller values of $x$, $x \lsim 0.01$, the effect of resummation becomes completely
negligible.

At the level of PDF luminosities at the LHC 13 TeV, we find that at the NLO+NLL level
the suppression induced by resummation in the PDFs starts to become important for
$M_{X}\gsim$ 400 GeV in the $gg$ channel, $M_{X}\gsim$ 1 TeV in the $q\bar{q}$ and
$qg$ channels, and $M_{X}\gsim$ 5 TeV for the $qq$ channel.
The trend is similar at NNLO, but in this case
differences between fixed-order
and resummed PDFs are much smaller.
We also find that fixed-order and resummed PDFs differ by at most one sigma 
throughout all
the range of $M_X$.

We have investigated the corresponding implications at the level of resummed LHC 
cross sections for three different processes:
 SM and BSM Higgs production in gluon fusion, high-mass Drell-Yan pair
 production and slepton pair production.
 We find that 
 the effect of consistently including resummation in the
PDFs can compensate the enhancement from resummation of the partonic cross sections, 
if $M_X$ is large enough.
For the production of final states with lower $M_X$,
the effect of PDF resummation is negligible.
This trend is likely to be general: when fitting to data, PDFs 
adjust to absorb the effect of the resummation in the partonic 
cross sections, and this compensation inevitably persists 
when extrapolating to predictions for new processes.

Our results illustrate the importance of using the same perturbative order in
all the components that enter hadronic
cross sections: the use of fixed-order PDFs with resummed matrix elements can lead to misleading
results, especially at high invariant masses,
a region crucial for new physics searches.
The partial cancellation between resummation in PDFs and in matrix elements indicates that
consistent resummed calculations can be closer to fixed-order results.
This said,
even in the case of a complete cancellation,
use of resummation would still be advantageous, because of the reduced scale
uncertainty.
For these reasons, we expect that the resummed NNPDF3.0 sets will provide a crucial ingredient,
missing so far,
to improve the precision of all-order resummed calculations for the LHC, and in particular those used 
in searches for new physics.

The main limitation of the present work is that, since resummed calculations are not available
for all processes included in the NNPDF3.0 global fit, we have restricted the dataset in the resummed
fits (and the corresponding baseline) to those processes that can be consistently resummed.
In particular, we have had to exclude the inclusive jet production data and $W$ lepton
rapidity distributions.
Hence, our resummed sets generate larger PDF uncertainties than the
NNPDF3.0 global PDFs, especially for gluon-initiated processes.
It is thus important in the future to provide resummed calculations for these missing
processes, in order to produce a truly global resummed PDF analysis.

With the recent start-up of the LHC Run II, the need for precision calculations is even
more pressing than in Run I, since precision could be the key to uncovering new physics.
The results of this paper offer for the first time fully consistent threshold
resummed calculations, which constitute state-of-the-art accuracy for a number of important
LHC processes, from Higgs to supersymmetric particle production.
Therefore, the resummed NNPDF3.0 sets presented here achieve a new milestone in the program
of precision phenomenology at the LHC.

\subsection*{Delivery}

The resummed calculations for DIS structure function and Drell-Yan distributions
used in this work
have been obtained with the new code \troll, version \texttt{v3.0}.
This code is publicly available from
\begin{center}
\href{http://www.ge.infn.it/~bonvini/troll/}{\texttt{http://www.ge.infn.it/$\sim$bonvini/troll/}}
\end{center}
and can be used by any interested parties to compute their own predictions for
resummed observables.

Our resummed sets are available in the {\tt LHAPDF6} format from the authors upon request.
The available sets are:
\begin{center}
  \tt NNPDF30\_nlo\_disdytop\\
  \tt NNPDF30\_nnlo\_disdytop\\
  \tt NNPDF30\_nll\_disdytop\\
  \tt NNPDF30\_nnll\_disdytop
\end{center}
which stand for the NLO and NNLO baseline fits, and their NLO+NLL and NNLO+NNLL
resummed counterparts.
All these PDF sets are provided for $\as(m_Z^2)=0.118$ with a maximum of
$n_f=5$ active flavors.

\medskip

\acknowledgments
%%%%%%%%%%%%%%%%%%%%%%%%%%%%%%%%%%%%%%%%%%%%%%%%%%%%%%%%%%%%%%%%%%%%%

%
We thank all the members of the NNPDF Collaboration, and especially
S.~Forte and G.~Ridolfi, for discussions and encouragement during this project.
We thank B.~Pecjak and A.~Kulesza for discussions. 
S.~M. and M.~U. would like to thank the Rudolf Peierls Centre for Theoretical Physics, at Oxford University for hospitality during the course of this work.
The work of S.~M. is supported by the U.S.\ National Science Foundation, under grant PHY--0969510, the LHC Theory Initiative.
J.~R. is supported by an STFC Rutherford Fellowship ST/K005227/1.
M.~B., J.~R. and L.~R. are
supported by an European Research Council Starting Grant "PDF4BSM".
S.~C. is supported in part by an Italian PRIN2010 grant and by a European Investment Bank EIBURS grant.
V.B. is supported by the ERC grant 291377, LHCtheory: \emph{Theoretical
  predictions and analyses of LHC physics: advancing the precision
  frontier}.

%%%%%%%%%%%%%%%%%%%%%%%%%%%%%%%%%%%%%%%%%%%%%%%%%%%%%%%%%%%%%%%%%%%%%

%%%%%%%%%%%%%%%%%%%%
\phantomsection
\addcontentsline{toc}{section}{References}

\bibliographystyle{jhep}
\bibliography{biblio}
%\input{DISsoftres.bbl}

%%%%%%%%%%%%%%%%%%%%%

\end{document}